\documentclass{aa}  

\usepackage{graphicx}
\usepackage{txfonts}
\usepackage[colorlinks]{hyperref}
\let\orgautoref\autoref
\renewcommand{\autoref}
        {\def\equationautorefname{Eq.}%
         \def\figureautorefname{Fig.}%
         \def\sectionautorefname{Sect.}%
         \def\subsectionautorefname{Sect.}%
         \def\subsubsectionautorefname{Sect.}%
         \orgautoref}

\usepackage{xcolor}
\definecolor{dark-red}{rgb}{0.9,0.0,0.0}
\definecolor{dark-blue}{rgb}{0.15,0.15,0.9}
\definecolor{dark-green}{rgb}{0.15,0.8,0.15}
\definecolor{medium-blue}{rgb}{0,0,0.9}
\hypersetup{linkcolor={dark-blue},citecolor={dark-blue}, urlcolor={medium-blue}
}
\usepackage{placeins}
\usepackage{float}

\usepackage{natbib}

\begin{document}

   \title{Precise radial velocities of giant stars}

   \subtitle{XVII. Distinguishing planets from intrinsically induced radial velocity signals in evolved stars}

   \author{Dane Spaeth\inst{1}\fnmsep\thanks{Fellow of the International Max Planck Research School for Astronomy and Cosmic Physics at the University of Heidelberg (IMPRS-HD)}
          \and
          Sabine Reffert\inst{1}
          \and
          Trifon Trifonov\inst{1}\fnmsep\inst{2}
          \and
          Adrian Kaminski\inst{1}
          \and
          Simon Albrecht\inst{3}
          \and 
          Frank Grundahl\inst{3}
          \and 
          Mads Fredslund Andersen\inst{3}
          \and 
          Andreas Quirrenbach\inst{1}
          \and
          Pere L. Pall\'{e}\inst{4}\fnmsep\inst{5}}

   \institute{ Landessternwarte, Zentrum für Astronomie der Universit\"at Heidelberg, K\"onigstuhl 12, 69117    Heidelberg, Germany \\
   \email{dane.spaeth@lsw.uni-heidelberg.de}
   \and 
   Department of Astronomy, Faculty of Physics, Sofia University ``St Kliment Ohridski'', 5 James Bourchier Blvd, BG-1164 Sofia, Bulgaria
   \and
   Stellar Astrophysics Centre, Department of Physics and Astronomy, Aarhus University, Ny Munkegade 120, 8000 Aarhus C, Denmark
   \and
   Instituto de Astrofísica de Canarias, 38200 La Laguna, Tenerife, Spain
   \and
   Universidad de La Laguna (ULL), Departamento de Astrofísica, 38206 La Laguna, Tenerife, Spain}

   \date{Received 18 December 2024 /
   Accepted 11 March 2025 }

   \abstract
  {From a long-term Doppler monitoring campaign of 373 giant stars, we have identified ten giants with periodic radial velocity variations that are challenging to associate with planets. Similar cases in the literature are attributed to poorly understood intrinsic processes.}
  {Our goal is to confirm or refute the presence of planets around these ten evolved stars. Additionally, we evaluate the reliability and sensitivity of planet-confirmation metrics when applied to giant stars and present cases of intrinsically induced radial velocity variations, aiming to enhance the physical understanding of the phenomenon.}
  {We combined 25 years of radial velocity data from the Hamilton/Lick, SONG, and CARMENES spectrographs. To assess consistency with Keplerian models, we examined the residuals and tracked changes in statistical significance as new data were incorporated. Additionally, we compared radial velocity amplitudes across optical and infrared wavelengths, searched for periodic variations of activity indicators, and examined their correlations with radial velocities.}
  {Seven of the ten giants exhibit intrinsically induced radial velocity variations. The strongest arguments against planets orbiting the giants are guided by long-term radial velocity monitoring that detects changing periodicity on long timescales or detects systematics close to the original period in the radial velocity residuals. While activity indicators offer some support, their signals are generally weak. Comparing optical and infrared radial velocity amplitudes also proves insufficient for confirming or refuting planets. We find HIP~64823 remains a promising candidate for hosting a giant exoplanet with orbital period $P\sim 7.75\,\mathrm{yr}$. For two stars, the evidence remains inconclusive.}
  {Long-term radial velocity monitoring is essential for distinguishing planetary companions from intrinsic variations in evolved stars.}


   \maketitle

\section{Introduction}
\label{sec:intro}
The majority of the over 5750 exoplanets confirmed to date have been detected via the transit technique ($\sim 75\%$) and the radial velocity (RV) method ($\sim 19\%$).\footnote{\url{https://exoplanetarchive.ipac.caltech.edu/}} Each of the exoplanet detection methods, however, has intrinsic biases that leave a large part of the exoplanet parameter space still to be explored. 

For instance, main-sequence stars more massive than about $ 1.5\,M_\sun$ are challenging to target in both transit (due to the larger radii) and RV surveys. For the latter, the achievable RV precision is limited due to high surface temperatures resulting in a small number of absorption lines, which are significantly broadened by high rotation rates \citep{Sato2003, Galland2005, Johnson2007, Lagrange2009, Assef2009}. An alternative to studying the population of planets around these stars is to target their evolved counterparts, which have significantly cooled down and slowed their rotation velocities, resulting in numerous narrow absorption lines that contain valuable Doppler information. So far, around 150 planets orbiting giant stars have been detected,\footnote{\url{https://www.lsw.uni-heidelberg.de/users/sreffert/giantplanets/giantplanets.php}} the vast majority of which were found using the RV method. 

A recent study by \citet{Wolthoff2022} summarizes the planet occurrence rate around giant stars through a combined analysis of three large RV surveys. They show a positive planet-metallicity correlation, as previously found also for main-sequence stars \citep{Fischer2005, Udry2007} and for giants \citep{Reffert2015}. Furthermore, \citet{Wolthoff2022} report a peak in the planet-occurrence rate at host masses $M = 1.68\pm0.59\,M_\sun$ and at orbital periods of around $2\,\mathrm{yr}$.

However, the RV method applied to giant stars faces two important challenges. On the one hand, short-term p-mode (solar-like) oscillations lead to intrinsic RV jitter with amplitudes around 10 to 20 $\mathrm{m\,s^{-1}}$, even for relatively stable giant stars \citep{Hekker2006a}. The amplitude of these p-mode oscillations increases as the stars become more evolved and luminous \citep{Hekker2008}, and thus the oscillations present more of a challenge for such giants. However, as the periods of these oscillations are typically on scales of hours to days, they can be dealt with as a white noise component in the context of exoplanet surveys targeting companions with orbital periods of several hundred days. Nevertheless, while this short-term RV jitter complicates the detection of low-mass ($M\lesssim 1\,M_\mathrm {Jup}$) planetary companions, the detection of planets of higher masses is not impacted. 

On the other hand, what is more problematic is the suspicion that the RV signals of some of the most luminous evolved planet hosts are in fact caused by poorly understood intrinsic stellar processes instead of planetary companions. The first identified case was \object{$\gamma$ Dra} \citep{Hatzes2018}, which showed stable large-amplitude RV variations with a period of $P \sim 700\,\mathrm{d}$ during nearly eight years of RV monitoring before amplitude changes and phase shifts became apparent. Other stars with RV signals with similar periods have been found in subsequent studies and include \object{Aldebaran} \citep{Reichert2019}, \object{$\epsilon$~Cyg} \citep{Heeren2021}, \object{42 Dra} \citep{Dollinger2021}, Sanders~364\footnote{SIMBAD identifier: \object{BD+12~1917}} \citep{Zhou2023}, \object{HD~135438} \citep{Lee2023a}, and four evolved stars in open clusters: NGC~2423~No.~3, NGC~2345~No.~50, NGC~3532~No.~670, and NGC~4349~No.~127\footnote{SIMBAD identifiers: \object{NGC 2423 3}, \object{NGC 2345 50}, \object{HD 96789}, and \object{NGC 4349 127}} \citep{DelgadoMena2018, DelgadoMena2023}. For several of the stars, the RV variations had previously been attributed to planetary companions. We stress that all of these stars are very luminous ($L>100\,L_\sun$) and that false-positive detections seem to occur frequently only for stars with radii $R \gtrsim 21\,R_\sun$ and RV periods between $300\,\mathrm{d}$ and $800\,\mathrm{d}$ \citep{Dollinger2021}. Planets around less luminous giants are generally not contested. 

The arguments to refute orbital companions as the origin of the RV signals of these luminous giants are quite varied. The orbital companions for $\gamma$~Dra, Aldebaran, $\epsilon$~Cyg, and 42~Dra have been ruled out based on arguments of the coherence and stability of the RV signal, which are only possible after extensive, long-term RV monitoring. The planet around Sanders 364 was ruled out, as periodicity at multiple periods was detected \citep{Zhou2023}. On the other hand, for the cluster giants presented by \citet{DelgadoMena2018, DelgadoMena2023}, orbital companions have been ruled out based on variations of the activity indicators identified with the High Accuracy Radial Velocity Planet Searcher (HARPS; \citealt{Mayor2003}). Unfortunately, such additional spectral diagnostics are mostly only available from pressure and temperature stabilized spectrographs with stable instrumental profiles. Furthermore, since not all published planets have been monitored on sufficiently long timescales or have a large set of reliable activity indicators available, there could, in principle, be more as yet unidentified false-positive planets around the most luminous giants, especially around those that have been announced on the basis of sparse RV sampling.

Furthermore, the origin of these intrinsic RV variations is still uncertain. The long RV periods typically rule out radial pulsations \citep{Hatzes1993, Cox1972, Hatzes2018, Reichert2019}. The modulation of magnetic surface spots is one possibility, but it would in most cases lead to a larger photometric variability than detected \citep{Reffert2015}. More exotic magnetic processes, potentially linked to convection, that could lead to RV variations without associated photometric variations have been suggested \citep{DelgadoMena2023, Rolo2024}, but they remain poorly studied. 
For the binary star $\epsilon$~Cyg, the heartbeat phenomenon was put forth \citep{Heeren2021}, but it cannot explain the variations for single stars. Finally, non-radial oscillations have been discussed for a long time as being a potential source of intrinsic RV variations \citep{Hatzes1994, Hatzes1996, Hatzes1998a}. 

Oscillatory convective modes have been proposed as the origin of non-radial oscillations in very luminous photometrically variable giant stars \citep{Saio2015}, although it is unclear whether these would be applicable for the less luminous giants within the group of false positives. In a recent study, we have shown that simulations of an $l=1, m=1$ non-radial oscillation can reproduce the variations of the RVs and the activity indicators of NGC 4349~No.~127 \citep{Spaeth2024}, an evolved cluster giant reported to host a brown-dwarf companion \citep{Lovis2007} but the finding has since been refuted by \citet{DelgadoMena2018, DelgadoMena2023}. However, it remains to be seen if other stars can be shown to have similar signatures.

The goal of this study is three-fold. (i) Our initial motivation was to investigate whether a number of so-far unpublished planet candidates identified in the Lick RV survey of giant stars are exoplanets or false-positives by intrinsic variations. These candidates were followed up using spectrographs from the Stellar Observations Network Group (SONG) and the Calar Alto high-Resolution search for M dwarfs with Exoearths with Near-infrared and optical Echelle Spectrographs (CARMENES). (ii) With nearly 25 years of RV data and a set of activity indicators commonly used to target planets orbiting main-sequence stars, we further aimed to test which metrics that determine the difference between planets and intrinsic variations are sensitive to intrinsic phenomena in evolved stars. (iii) Last, we aim to present further intrinsically RV-variable stars as well as their observational fingerprints, hoping to contribute toward a better physical understanding of the intrinsic mechanism present in giant stars that can mimic planetary companions quite convincingly in RV data. For this purpose, we present ten such planet candidates along with two secure exoplanet systems. 
For simplicity, we include all orbital companions of substellar minimum mass within the term ``planet'' unless stated otherwise. 

The paper is structured as follows. In \autoref{sec:stellar_params}, we give an overview of the sample and the stellar parameters followed by a summary of the observations in \autoref{sec:observations}. In \autoref{sec:results}, we discuss the RV time series data along with several metrics aimed at differentiating between planets and intrinsic processes causing the (semi-)periodic RV variations. These metrics are discussed star-by-star in \autoref{sec:results:individual_stars}. In \autoref{sec:hip64823}, we explore the most promising planet candidate, HIP~64823, remaining after the activity analysis. Finally, we discuss potential astrophysical origins of the intrinsic variations in \autoref{sec:discussion} and summarize our findings in \autoref{sec:summary}.

\begin{table*}
\caption{Stellar parameters derived by \citet{Stock2018}.}
\label{tab:params}
\renewcommand{\arraystretch}{1.15}
\resizebox{\textwidth}{!}{
\begin{tabular}{lllllllllllll}
\hline \hline
HIP   & HD     & Id.    & Evol. state & $P(\mathrm{state})$ & \phantom{(}$M$                    & \phantom{(}$R$                    & \phantom{0(}$L$                        & $T_\mathrm{eff}$   & \phantom{$-$}$\mathrm{[Fe/H]}$ & \phantom{(}$\log g$               & $B-V$          & Hpscat \\
      &        &                &             &                     & ($M_\sun$)              & ($R_\sun$)              & \phantom{0}($L_\sun$)                  & ($\mathrm{K}$)       & \phantom{--}($\mathrm{dex}$)    & ($\mathrm{cm\,s^{-1}}$)   & ($\mathrm{mag}$) & (mag)    \\ \hline
7607  & 9927   & $\upsilon$ Per & HB          & $0.70$              & $1.54^{+0.25}_{-0.16}$ & $21.6^{+0.3}_{-0.5}$   & \phantom{0}$154.6^{+2.3}_{-3.4}$      & $4385^{+16}_{-17}$ & \phantom{$-$}$0.07\pm0.10$     & $1.96^{+0.07}_{-0.08}$ & $1.28$         & 0.004  \\
7884  & 10380  & $\nu$ Psc      & RGB         & $0.56$              & $1.40^{+0.18}_{-0.17}$ & $34.5^{+1.7}_{-1.6}$   & \phantom{0}$331.4^{+18.7}_{-16.7}$    & $4197^{+52}_{-53}$ & $-0.27\pm0.10$    & $1.51^{+0.08}_{-0.09}$ & $1.35$         & 0.005  \\
16335 & 21552  & $\sigma$ Per   & RGB         & $0.54$              & $1.53^{+0.18}_{-0.12}$ & $35.9^{+0.8}_{-0.7}$   & \phantom{0}$357.6^{+13.5}_{-15.2}$    & $4184^{+9}_{-9}$   & $-0.20\pm0.10$    & $1.51^{+0.04}_{-0.05}$ & $1.37$         & 0.006  \\
38253 & 63752  &                & HB          & $0.88$              & $2.43^{+0.21}_{-0.63}$ & $71.7^{+10.8}_{-10.2}$ & $1374.9^{+889.6}_{-776.4}$ & $4058^{+7}_{-7}$   & $-0.35\pm0.10$    & $1.06^{+0.06}_{-0.05}$ & $1.45$         & 0.006  \\
46390 & 81797  & $\alpha$ Hya   & HB          & $1.00$              & $2.40^{+0.23}_{-0.11}$ & $58.8^{+0.4}_{-0.5}$   & \phantom{0}$868.0^{+7.4}_{-17.9}$     & $4086^{+8}_{-3}$   & $-0.05\pm0.10$    & $1.29^{+0.06}_{-0.02}$ & $1.44$         & 0.007  \\
47959 & 84561  & 18 Leo         & RGB         & $0.98$              & $0.93^{+0.05}_{-0.03}$ & $36.3^{+1.6}_{-1.9}$   & \phantom{0}$296.5^{+24.7}_{-29.5}$    & $3981^{+2}_{-2}$   & $-0.32\pm0.10$    & $1.27^{+0.11}_{-0.01}$ & $1.49$         & 0.008  \\
64823 & 115478 &                & RGB         & $0.90$              & $1.42^{+0.16}_{-0.13}$ & $18.6^{+0.4}_{-0.6}$   & \phantom{0}$109.0^{+4.6}_{-6.4}$      & $4335^{+14}_{-14}$ & \phantom{$-$}$0.10\pm0.10$     & $2.09^{+0.03}_{-0.08}$ & $1.30$         & 0.005  \\
73620 & 133165 & 110 Vir        & HB          & $0.69$              & $1.13^{+0.34}_{-0.12}$ & $12.1^{+0.4}_{-0.6}$   & \phantom{00}$70.3^{+2.2}_{-2.3}$       & $4800^{+79}_{-40}$ & $-0.30\pm0.10$    & $2.43^{+0.05}_{-0.16}$ & $1.03$         & 0.005  \\
75458 & 137759 & $\iota$ Dra    & HB          & $0.88$              & $1.29^{+0.17}_{-0.16}$ & $11.9^{+0.1}_{-0.2}$   & \phantom{00}$57.2^{+0.8}_{-0.5}$       & $4622^{+15}_{-24}$ & \phantom{$-$}$0.11\pm0.10$     & $2.40^{+0.07}_{-0.06}$ & $1.17$         & 0.005  \\
84671 & 156681 & 66 Her          & RGB         & $0.68$              & $1.00^{+0.11}_{-0.09}$ & $47.8^{+1.8}_{-1.7}$   & \phantom{0}$471.8^{+36.5}_{-31.9}$    & $3899^{+8}_{-9}$   & $-0.21\pm0.10$    & $1.09^{+0.05}_{-0.04}$ & $1.54$         & 0.009  \\
88048 & 163917 & $\nu$ Oph      & HB          & $1.00$              & $2.74^{+0.12}_{-0.09}$ & $14.1^{+0.3}_{-0.3}$   & \phantom{0}$107.0^{+2.2}_{-2.3}$      & $4943^{+32}_{-29}$ & \phantom{$-$}$0.06\pm0.10$     & $2.59^{+0.02}_{-0.03}$ & $0.99$         & 0.004  \\
89826 & 168775 & $\kappa$ Lyr   & HB          & $0.91$              & $2.25^{+0.46}_{-0.20}$ & $18.6^{+0.3}_{-0.4}$   & \phantom{0}$138.2^{+2.6}_{-3.2}$      & $4595^{+21}_{-20}$ & \phantom{$-$}$0.15\pm0.10$     & $2.24^{+0.04}_{-0.03}$ & $1.16$         & 0.005  \\ \hline
\end{tabular}}
\tablefoot{Presented are the stellar parameters for the most likely evolutionary state (Evol. state) with its probability $P(\mathrm{state})$. Metallicities were adopted from \citet{Hekker2007}. We also list $B-V$ colors and the photometric scatter present in the \textsc{Hipparcos} data \citep{ESA1997}.}
\end{table*}

\begin{figure}
    \centering
    \includegraphics{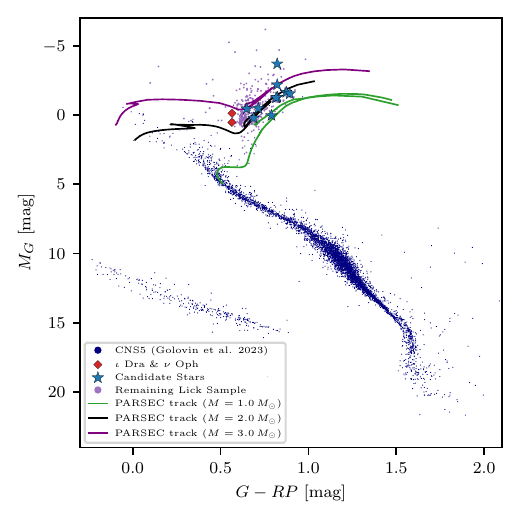}
    \caption{Color-magnitude diagram based on \textit{Gaia} DR3 photometry showing the location of the sample stars using blue star markers as well as red diamonds for the two published planet hosting giants $\iota$~Dra and $\nu$~Oph. We also plot the location of the remaining 361 giant stars within the Lick sample in purple. For comparison, we overplot the fifth catalog of nearby stars (CNS5) \citep{Golovin2023} in dark blue and three adjacent evolutionary tracks for $1M_\sun$ (green), $2M_\sun$ (black), and $3M_\sun$ (purple) taken from PARSEC \citep{Bressan2012}.}
    \label{fig:cmd}
\end{figure}

\section{Sample and stellar parameters}
\label{sec:stellar_params}

The giant stars considered in this work were originally part of the Lick RV survey, which started in 1999 and targeted 373 bright ($V \leq 6$) G and K giants. The Lick sample initially comprised 86 K giants, selected to be photometrically stable and not part of multiple systems from the \textsc{Hipparcos} catalog \citep{Frink2001, ESA1997}. The stellar sample was extended in 2000 and 2004, relaxing the constraints on photometric stability and adding stars with higher masses and bluer colors. However, the actual level of photometric variability remained below 0.01 mag for most stars \citep{Reffert2015}. 

Overall, 18 planets in 15 stellar systems present in the Lick sample of giant stars have been confirmed either by our own or other groups \citep{Frink2002, Hatzes2006, Reffert2006, Sato2007, Liu2008, Schwab2010, Wittenmyer2011, Mitchell2013, Trifonov2014, Lee2014a,  Ortiz2016, Takarada2018, Quirrenbach2019, Luque2019, TalaPinto2020, Hill2021, Teng2023}. A recent overview is also presented by \citet{Wolthoff2022}. Furthermore, a number of stars were identified as periodic RV-variable stars, making them planet-host candidates. 20 stars of the latter group, along with three confirmed planet hosts were monitored using CARMENES starting in 2017 (see \autoref{sec:observations}). Notably, the 20 planet candidates also include HIP~16335, for which a planetary companion was published by \citet{Lee2014a}, but that we classified as a candidate. In this work, we exclude stars with long-period (much longer than the combined Lick-CARMENES baseline) stellar companions as the uncertainty of the outer companions' orbits significantly hinders the analysis. We further exclude two stars that will be the focus of future dedicated publications. The remaining sample of 12 stars thus comprises ten giants that were considered to be candidates to host planets with periods between one and eight years after the Lick survey concluded as well as two published planet hosts, namely $\iota$~Dra (HIP~75458) \citep{Frink2002, Hill2021} and $\nu$~Oph (HIP~88048) \citep{Quirrenbach2011, Quirrenbach2019}, which were added to the CARMENES observations for comparison.

The stellar parameters for the Lick giant star sample have been derived by \citet{Stock2018} using Bayesian inference on a grid of stellar evolutionary models, implemented in the fitting tool \textit{SPOG+}.\footnote{\url{https://github.com/StephanStock/SPOG}} The stellar parameters in the most likely evolutionary state, either on the red giant branch (RGB) or horizontal branch (HB), are presented in \autoref{tab:params}. Many of the stars are relatively luminous. The stars are roughly evenly distributed between the RGB and HB (five vs seven). The solar-like oscillations of HIP~75458 and HIP~89826 have been analyzed as part of asteroseismic studies \citep{Zechmeister2008, Hill2021, Campante2023, Malla2024}. The derived asteroseismic parameters are consistent with those in \autoref{tab:params}. 

Figure~\ref{fig:cmd} shows the loci of the 12 stars, as well as the remaining 361 giants comprising the Lick sample, within a \textit{Gaia} DR3 \citep{GaiaCollaboration2016, GaiaCollaboration2023} color-magnitude diagram. For the stars not present in the \textit{Gaia} catalog due to their brightness (including HIP~46390), we converted the \textsc{Hipparcos} photometric measurements \citep{ESA1997} into the \textit{Gaia} passbands.\footnote{\url{https://gea.esac.esa.int/archive/documentation/GDR2/Data_processing/chap_cu5pho/sec_cu5pho_calibr/ssec_cu5pho_PhotTransf.html}} To provide a visual reference, we overplot the fifth catalog of nearby stars (CNS5; \citealt{Golovin2023}) and three PARSEC evolutionary tracks \citep{Bressan2012} for different stellar masses at solar metallicity. 

\section{Observations}
\label{sec:observations}

\subsection{Lick observations}

The observations at UCO/Lick Observatory started in 1999 using the 0.6 m Coud\'{e} Auxiliary Telescope (CAT) and the Hamilton Echelle Spectrometer \citep{Vogt1987} with a measured resolving power $R\sim50\,000$ at $6000\,\text{\AA}$ \citep{Reffert2015}. The survey used the iodine technique described by \citet{Butler1996} and aimed for an RV precision of $5$ to $8\,\mathrm{m\,s^{-1}}$ \citep{Reffert2015}. The Hamilton spectrograph covers a wavelength range of approximately $3400 \text{--} 9000\,\text{\AA}$ \citep{Fischer2013}. However, the RV determination is limited to the regime spanning roughly $5000 \text{--} 5800\,\text{\AA}$, due to the use of the iodine cell. The observations ended in 2011 when the iodine cell at Lick was damaged \citep{Fischer2013}. A continuation of the survey using the $72\,\mathrm{cm}$ Waltz Telescope located at Landessternwarte Heidelberg \citep{Tala2016} is expected to start in 2025. Overall, 1127 Lick spectra are available for the 12 stars with a time baseline of approximately 12 years. We use the standard Lick RV reduction. 

\subsection{Lick \texorpdfstring{H$\alpha$}{Hα} index}

Since the line spread function of the Lick spectra is comparably unstable \citep{TalaPinto2020}, and due to the inclusion of the iodine cell, activity indicators targeting the shape of the stellar spectral lines only yield low precision (see, for instance, \citealt{Reffert2006}). No such indicators are therefore used in this work. 

An individual spectral line that is present in the Lick wavelength range, lies outside the iodine regime, and is known to be sensitive to chromospheric activity \citep{Kurster2003}, is the H$\alpha$ line ($6562.8\,\text{\AA}$). To quantify the variability of its core, we improved upon the work by \citet{Staudt2020}, who calculates H$\alpha$ indices for the Lick spectra based on the definition by \citet{Boisse2009}. Since the standard wavelength regions used by \citet{Staudt2020} and \citet{Boisse2009} were strongly affected by telluric lines, we searched for relatively unaffected regions within the spectral order containing the H$\alpha$ line and slightly redefine the H$\alpha$ index as
\begin{equation}
    \mathrm{H}\alpha = \frac{F_{\mathrm{H}\alpha}}{\sum_{i=1}^5 F_i},
\end{equation}
with $F_{\mathrm{H}\alpha}$ being the flux in a $0.4\,\text{\AA}$ region centered on the core of the H$\alpha$ line and $F_{1-5}$ being the fluxes in five smaller comparison regions in the continuum. These are defined as $[6538.5\,\text{\AA}, 6540.5\,\text{\AA}]$, $[6565.0\,\text{\AA}, 6567.0\,\text{\AA}]$, $[6576.0\,\text{\AA}, 6578.0\,\text{\AA}]$, $[6582.5\,\text{\AA}, 6584.5\,\text{\AA}]$, and $[6595.0\,\text{\AA}, 6597.0\,\text{\AA}]$. The uncertainties were calculated by quadratically adding the contributions of the estimated intensity error using the $\beta \sigma$ procedure \citep{Stoehr2008, Czesla2018} and the influence of an imperfect wavelength shift by performing a Monte Carlo simulation \citep{Staudt2020}. The H$\alpha$ time series is treated as two separate data sets due to small offsets caused by a CCD camera exchange in April 2004 (some spectra testing the new CCD have already been acquired in 2002/2003).

\subsection{SONG observations}
After the Lick observations concluded, select giants from the Lick survey were monitored using the 1~m robotic SONG telescope located on Tenerife, Spain \citep{Andersen2014, FredslundAndersen2019}. The SONG spectrograph is a high-resolution ($R\sim90\,000$), cross-dispersed \'echelle spectrograph. Similar to the Lick survey, SONG employs the iodine cell technique to derive precise RVs \citep{Grundahl2017}. The SONG spectrograph covers the wavelength range $4400\text{--}6900\,\text{\AA}$ \citep{Grundahl2017}, with RVs derived from the regime $5000\,\text{\AA}$ to $6300\,\text{\AA}$ \citep{Heeren2023}.

Of the 12 stars in the sample, four (HIP~38253, HIP~46390, HIP~47959, and HIP~75458) were included in the SONG observations. Overall 1030 spectra with sufficient S/N are available from SONG between March 2015 and December 2023. We obtained the RVs using the \texttt{pyodine} reduction software \citep{Heeren2023}. No standard activity indicators are available from the SONG observations. We note that some spectra, especially for HIP~75458, were taken as part of other observing campaigns aiming to study short-term variations. In these cases, several (typically less than five, in two nights more than ten) spectra were taken in the same night. Nevertheless, we refrain from binning the spectra, as retaining the full time resolution of the measurements can help to constrain the short-term jitter during the RV modeling.

\subsection{CARMENES observations}
First CARMENES observations for the ten planet candidates were taken between 2017 and 2018 (with a single additional spectrum from 2016). These include archival spectra from other observation programs. The CARMENES observations were continued, after an unfortunate gap, from 2021 to 2023 for the full sample, including the confirmed planet hosts.  

The CARMENES spectrograph \citep{Quirrenbach2014}, installed at the $3.5\,\mathrm{m}$ telescope at Calar Alto, Spain, offers two independent channels acquiring spectra in different wavelength regimes simultaneously. The visual (VIS) channel covers the regime $5200\,\text{\AA}$ to $9600\,\text{\AA}$, with a resolving power $R \sim 94\,600$. The near-infrared (NIR) channel continues the coverage at longer wavelengths from $9600\,\text{\AA}$ to $17\,100\,\text{\AA}$ with a resolving power $R \sim 80\,400$.\footnote{\url{https://carmenes.caha.es/ext/instrument/index.html}} 

We reduced the CARMENES data using two reduction pipelines. The standard SERVAL reduction pipeline \citep{Zechmeister2018} uses a coadded stellar template to derive the RVs using least squares minimization. The pipeline furthermore yields several activity indicators, namely the chromatic index (CRX), differential linewidth (dLW) (see \citealt{Zechmeister2018} for their definitions), and several line indicators: H$\alpha$ ($6562.8\,\text{\AA}$), Na \textsc{i} D1 ($5895.9\,\text{\AA}$), Na \textsc{i} D2 ($5889.9\,\text{\AA}$), and the Ca \textsc{ii} infrared triplet (IRT) lines at $8498.0\,\text{\AA}$, $8542.1\,\text{\AA}$, and $8662.1\,\text{\AA}$. For brevity, we refer to the line indicators as NaD$i$ and CaIRT$i$, respectively. These indicators have been shown to be sensitive to chromospheric activity (see, for instance, \citealt{Kurster2003, Diaz2007, Martinez-Arnaiz2011, Martin2017, Schofer2019, Huang2024, Gehan2024}).

Due to the brightness of the stars ($V \leq 6\,\mathrm{mag}$), they are excellent filler targets and have been observed in variable, sometimes even quite bad, observing conditions. As a consequence, the S/N of the individual observations varies, although it is generally high. Additionally, in some good nights exposure times were increased in an attempt to reach sufficient S/N in the simultaneous Fabry-P\'erot exposures to measure the instrumental drift over the course of the night \citep{Schafer2018}. As a consequence, a few observations have S/N exceeding the standard SERVAL limits of 500 and 400 for the visual and near-infrared channels, respectively. To avoid any non-linear effects due to a saturation of the CCD affecting all observations, we first created a SERVAL template using only spectra below the standard SERVAL S/N limits, which we then used to derive the RVs and activity indicators for all spectra. Other than that, we use the standard settings of the SERVAL pipeline for the two respective CARMENES channels. We note that this includes the selection of only a subset of the NIR spectral orders, as many orders suffer from severe telluric contamination \citep{Reiners2018, Nagel2023}.
 
We furthermore derived the RVs and activity indicators by calculating the cross-correlation function (CCF) with a weighted binary mask by using the RACCOON reduction software \citep{Lafarga2020}. The RACCOON pipeline yields spectral diagnostics such as the full width at half maximum (FWHM) and contrast of the CCF, as well as the bisector inverse slope (BIS). By default, masks are only provided for M dwarfs targeted in the CARMENES GTO sample \citep{Reiners2018, Ribas2023}. We therefore calculated new masks for each star using the functionality provided by RACCOON and the SERVAL templates detailed above. The RACCOON RVs are generally in good agreement with the SERVAL results. The deviation of the RV results between the SERVAL and RACCOON results have a standard deviation $\sigma=3.3\,\mathrm{m\,s^{-1}}$ and $\sigma=11.6\,\mathrm{m\,s^{-1}}$ for VIS and NIR, respectively. The larger deviations in the near infrared are expected due to a lower number of usable lines as reported by \citet{Lafarga2020}. For the analysis presented in \autoref{sec:results}, we always use the RVs as derived by SERVAL, and complement these with the FWHM, contrast, and BIS measurements derived by RACCOON.

Although it was attempted to obtain instrumental drift measurements for the spectra simultaneously, exposure times were often too short to ensure sufficient Fabry-P\'erot exposures in the second fiber of the CARMENES spectrograph. For these spectra, the instrumental drift was modeled over the night with a third-order polynomial using the drift measurements from the sky calibration frames and all stellar spectra (including other programs). The quality of the fits was ensured by eye. The fits were generally found to sufficiently describe the nightly variation of the instrumental drift. From these fits, the drifts for the individual spectra were interpolated. The uncertainties of the drift measurements were derived by quadratically adding the uncertainty of the polynomial fit at the observing time of the spectrum, derived via a Monte Carlo simulation, and the weighted standard deviation of the residuals of the drift measurements (after subtracting the third order polynomial) over the course of the night. The drifts are typically smaller than the stellar jitter of the evolved stars. Nevertheless, there are several nights that show strong instrumental drifts or offsets that lead to large outliers in the RVs if not corrected properly. We further correct the spectra for nightly zero-points measured from CARMENES RV standard stars and provided through the CARMENES consortium. These are typically found to be small.

We carefully cross-checked the spectra with any reported issues in the CARMENES observing logs and the routine analysis carried out by the CARMENES consortium to ensure the data quality. Twelve spectra with reported issues were removed from the analysis. We further removed four spectra for which the peak S/N per pixel is below 30 as well as five overexposed spectra in VIS and NIR each. Finally, we removed two spectra that were taken at very high airmass $\sim 2.9$ in the transition between nautical and civil twilight in eastern observing direction. These showed strong outliers in the activity indicators. Finally, we remain with 337 spectra from the CARMENES VIS and NIR channels each for all 12 stars. We note that not all 337 spectra are pairs of contemporary VIS and NIR observations since readout issues in one detector still allow the use of the other channel. As for the SONG data, we refrain from binning any data points, even though in the earlier CARMENES observation campaigns multiple spectra were taken in the same nights for some stars. 

Overall, we are left with 1127 Lick spectra, 1030 SONG spectra, and 337 spectra from each of the CARMENES channels, totaling 2831 spectra for all 12 stars.\footnote{The RVs and activity indicator measurements are available at the CDS.} Combined, these cover the time frame from 1999 to 2023, albeit with a large gap between 2011 and 2017 for many of the stars. An overview of the number of spectra per instrument, as well as the observational baseline is given in \autoref{tab:nr_of_obs}. Although the number of CARMENES observations for most stars is significantly smaller than that from Lick and SONG, we consider the observational baseline sufficient to reliably detect the large-amplitude RV variations. However, for certain stars, such as HIP~46390, HIP~73620, or HIP~89826, the CARMENES sampling was suboptimal. This could result in smaller-amplitude variations, such as those potentially present in the CARMENES activity indicators, not being reliably detected. We also note that the two secure planets hosts were only added to the CARMENES sample in 2021. However, both were relatively densely sampled, allowing for periodicity close to the main RV period ($P\sim500\,\mathrm{d}$) to be identified with confidence.

\begin{table*}[]
\centering
    \caption{Number of spectra collected per instrument along with the observational time baselines (years in parentheses).}
    \label{tab:nr_of_obs}

    \begin{tabular}{lllll}
         \hline \hline
         HIP & Lick & SONG & CARM-VIS & CARM-NIR \\
         \hline
         7607 & 82 (2000--2011) &  & 39 (2017--2023) & 39 (2017--2023) \\ 
         7884 & 92 (2000--2011) &  & 29 (2017--2023) & 28 (2017--2023) \\ 
         16335 & 71 (2000--2011) &  & 47 (2017--2023) & 47 (2017--2023) \\ 
         38253 & 74 (1999--2011) & 102 (2015--2022) & 23 (2017--2023) & 24 (2017--2023) \\ 
         46390 & 79 (2000--2011) & 289 (2015--2022) & 18 (2016--2023) & 19 (2016--2023) \\ 
         47959 & 70 (2000--2011) & 114 (2015--2022) & 17 (2017--2023) & 18 (2017--2023) \\ 
         64823 & 42 (2000--2011) &  & 25 (2017--2023) & 26 (2017--2023) \\ 
         73620 & 98 (2000--2011) &  & 22 (2017--2022) & 22 (2017--2022) \\ 
         75458 & 193 (2000--2011) & 525 (2015--2023) & 27 (2021--2023) & 26 (2021--2023) \\ 
         84671 & 123 (1999--2011) &  & 34 (2017--2023) & 34 (2017--2023) \\ 
         88048 & 150 (2000--2011) &  & 17 (2021--2023) & 16 (2021--2023) \\ 
         89826 & 53 (2000--2011) &  & 39 (2017--2023) & 38 (2017--2023) \\ 
         \hline
         TOTAL & 1127 & 1030 & 337 & 337 \\
         \hline
    \end{tabular}
\end{table*}

\section{Results}
\label{sec:results}
\subsection{Radial velocity time series and Keplerian fits}
\label{sec:results:keplerian_modelling}

The ten evolved RV-variable stars to be examined in this work were found to be planet candidates after the conclusion of the Lick RV survey in 2011. Each of the stars has larger variations in the RVs than expected from the p-mode oscillations even for these relatively luminous stars. These variations have further been found to be periodic with (mostly) a single dominating peak in a periodogram of the Lick RVs (see, for instance, the Lick panels in Figs.~\ref{fig:mlp_part1} and \ref{fig:mlp_part2}). However, the RVs of each of the systems are challenging to model with orbital companions due to remaining systematics in the RVs; therefore, we refrained from publishing these as planet detections. Thus, a strong selection bias toward (semi-)periodic, but challenging RV variability is inherent to the star sample. On the other hand, with the addition of the more recent SONG and CARMENES data, this makes the stars ideal for studying differences between real planets and intrinsic processes in giant stars. 

For many of the stars, the SONG and CARMENES RVs, which span a much more recent time window than the Lick observations, show periodicity at different periods than present in the Lick RVs. This leads to complex and often multi-periodic combined RV periodograms. We examine this behavior more closely in \autoref{sec:results:mlp} and note that this finding already challenges the existence of most of the potential planets. Nevertheless, in order to provide baseline models to test and treat all stars homogeneously, we performed simple one-planet Keplerian fits to the RV data, motivated by the mono-periodicity present in the Lick RVs and attempting to model the most dominant periodicity.

We performed a dynamic nested sampling \citep{Skilling2004, Skilling2006, Higson2019} of the parameter posterior distributions as provided by the {\sc dynesty} \citep{Speagle2020, Koposov2024, Feroz2009} implementation in the \texttt{Exo-Striker} RV fitting tool \citep{Trifonov2019a}.\footnote{\url{https://github.com/3fon3fonov/exostriker}} We used broad uniform priors, specified in \autoref{tab:Kepler_priors}, and adopted the mode of the posterior distributions for the further analysis. As the RV jitter is typically much larger than the instrumental scatter, we adopted a single RV jitter term $\sigma_\mathrm{jit}$ for all instruments. For the two secure planet systems, HIP~75458 and HIP~88048, we used a double-Keplerian model motivated by the detection of a second long-period brown-dwarf companion in the former system \citep{Hill2021}, and the multi-planetary system comprised of two brown dwarfs for the latter \citep{Quirrenbach2011, Quirrenbach2019}. We stress that the final Keplerian models presented for the ten RV-variable stars merely serve to describe the most dominating RV variability in the data, and should not be interpreted as planet detections. These models are tested in the following sections. 

\begin{figure*}
    \centering
    \includegraphics{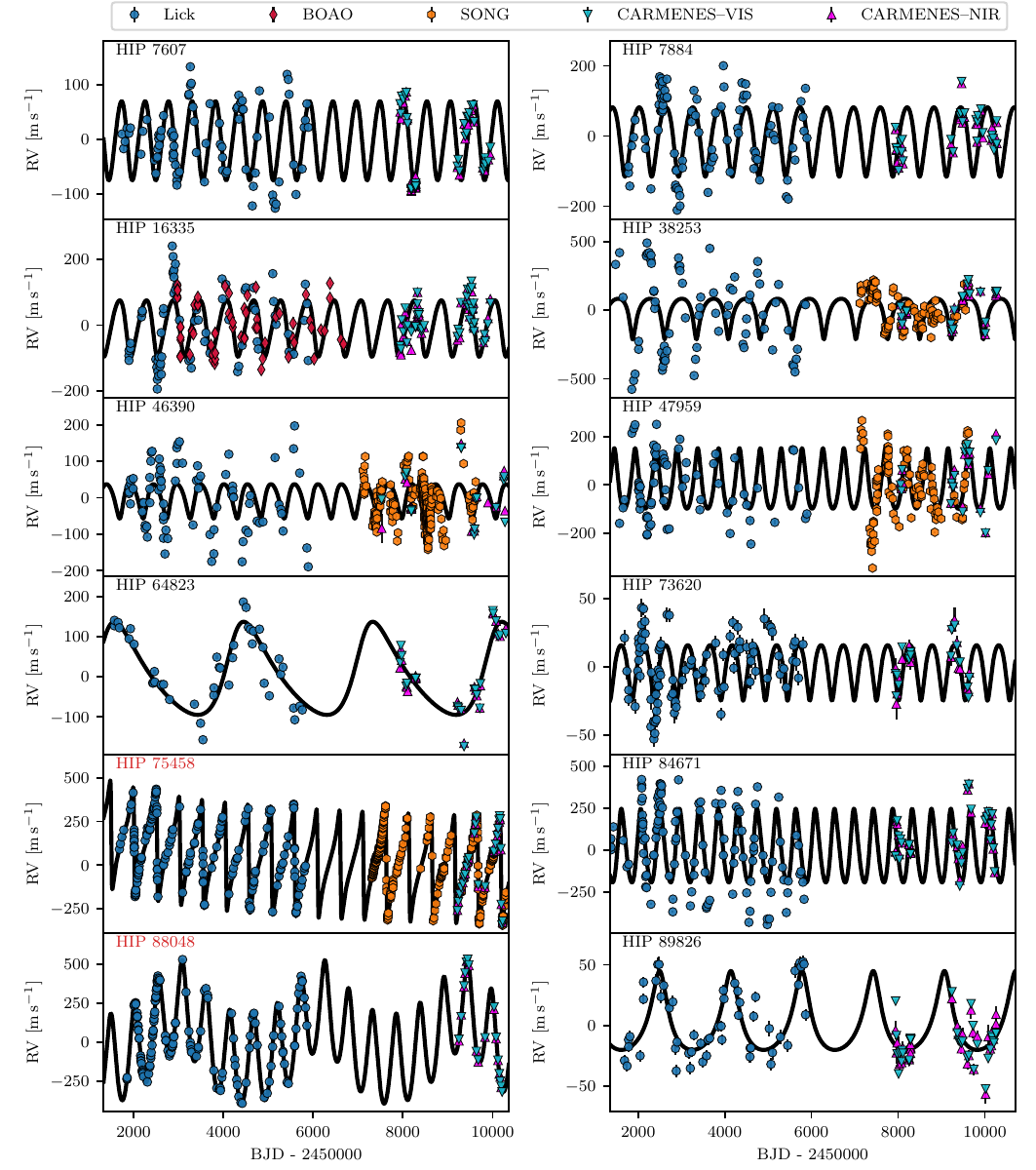}
    \caption{Best Keplerian models for the 12 stars in the sample. The error bars of the RVs, representing the formal uncertainties of the measurements, are typically smaller than the markers. The identifiers of the two planet hosts, HIP~75458 and HIP~88048, are marked in red. Significant intrinsic RV jitter is present for most of the stars.}
    \label{fig:multi-rv}
\end{figure*}

The best model inferred from the nested sampling is portrayed in \autoref{fig:multi-rv} and given in \autoref{tab:nest_samp_params}. We also give estimates of the expected p-mode jitter following the scaling relations of \citet{Kjeldsen1995} and \citet{Kjeldsen2011}. The main difference between the two scaling relations is that the latter, apart from the luminosity $L$ and mass $M$ of the star, additionally depends on the effective temperature $T_\mathrm{eff}$ and the mode lifetime $\tau_\mathrm{osc}$ \citep{Kjeldsen2011}. This is based on the assumption that the power in the velocity fluctuations due to acoustic oscillations behaves similarly to the fluctuations due to granulation. \citet{Muller2019} investigates the applicability of both scaling relations to the Lick sample and finds considerable mismatches especially for cooler stars. Within this study, the modeled jitter for most stars tends to be more consistent with the generally lower estimates by \citet{Kjeldsen1995}, see \autoref{tab:nest_samp_params}, than the updated scaling relations by \citet{Kjeldsen2011}, despite the mentioned shortcomings of the applicability to giant stars of the former.

Several of the stars show varying periodicity in the different data sets. Many of these show multi-modality in the parameter posterior distributions that lead to difficulties in determining the best fit parameters. The most severe of these cases is HIP~46390, for which the default wide prior often picks up a long-period variability at $P\sim1200\,\mathrm{d}$, which is however only present in the SONG data set and coincides with a peak in the SONG window function as can be seen in \autoref{fig:mlp_part1}. We thus constrained the period prior to $\mathcal{U}(100\,\mathrm{d}, 1000\,\mathrm{d})$ for this star.

For the two published planet systems, HIP~75458 and HIP~88048, the derived orbital parameters are consistent with the more detailed models by \citet{Hill2021} and \citet{Quirrenbach2019}, respectively. Revising the orbital parameters is beyond the scope of this publication. 

\begin{table}
\caption{Keplerian parameters obtained through nested sampling.}
\label{tab:nest_samp_params}
\resizebox{\columnwidth}{!}{
\begin{tabular}{llllll}
\hline \hline
HIP & \phantom{00(}$P$ & \phantom{(}$K$ & \phantom{(}$\sigma_\mathrm{jit}$  & $\sigma_\mathrm{jit}$(K\&B95) & $\sigma_\mathrm{jit}$(K\&B11) \\
    & \phantom{00}($\mathrm{d}$) & ($\mathrm{m\,s^{-1}}$) & ($\mathrm{m\,s^{-1}}$) & ($\mathrm{m\,s^{-1}}$) & ($\mathrm{m\,s^{-1}}$) \\
\hline
7607 & \phantom{00}523.3 & \phantom{0}72.7 & \phantom{0}33.0 & $23.5\pm3.5$ & \phantom{$0.$}$61\pm13$\\
7884 & \phantom{00}618.0 & \phantom{0}99.1 & \phantom{0}62.9 & $55.4\pm8.2$ & \phantom{$.$}$181\pm38$\\
16335 & \phantom{00}597.3 & \phantom{0}85.8 & \phantom{0}51.9 & $54.7\pm6.7$ & \phantom{$.$}$173\pm28$\\
38253 & \phantom{00}735.3 & 148.1 & 164.6 & \phantom{$.$}$132\pm84$ & \phantom{$.$}$380\pm250$\\
46390 & \phantom{00}523.8 & \phantom{0}47.3 & \phantom{0}55.2 & $84.6\pm7.9$ & \phantom{$.$}$237\pm29$\\
47959 & \phantom{00}481.3 & 125.7 & \phantom{0}97.2 & $74.6\pm8.8$ & \phantom{$.$}$375\pm48$\\
64823 & \phantom{0}2878.1 & 116.3 & \phantom{0}36.7 & $18.0\pm2.3$ & $50.8\pm8.8$\\
73620 & \phantom{00}512.5 & \phantom{0}20.3 & \phantom{0}13.3 & $14.5\pm3.1$ & $29.9\pm9.5$\\
75458 b & \phantom{00}510.9 & 310.0 & \phantom{0}11.7 & $10.4\pm1.5$ & $23.5\pm4.7$\\
75458 c & 28193.7 & 153.3 &  &  & \\
84671 & \phantom{00}447.2 & 219.6 & 135.0 & \phantom{$.$}$110\pm15$ & \phantom{$.$}$580\pm100$\\
88048 b & \phantom{00}529.9 & 289.1 & \phantom{00}7.5 & \phantom{$0$}$9.1\pm0.7$ & $10.7\pm1.0$\\
88048 c & \phantom{0}3173.4 & 175.0 &  &  & \\
89826 & \phantom{0}1642.6 & \phantom{0}32.5 & \phantom{0}12.8 & $14.4\pm2.3$ & $25.2\pm5.8$\\
\hline
\end{tabular}
} 
\tablefoot{For HIP~75458 and HIP~88048, we present the period $P$ and semi-amplitude $K$ for both orbital companions labeled as ``b'' and ``c'', respectively. $\sigma_\mathrm{jit}$(K\&B95) and $\sigma_\mathrm{jit}$(K\&B11) denote the respective p-mode jitter estimates based on \citet{Kjeldsen1995} and \citet{Kjeldsen2011}.}
\end{table}

\subsection{Radial velocity residual analysis}
\label{sec:results:coherence}

Having modeled the RV variations of the 12 stars using Keplerians, our goal is to examine several standard tools used to confirm the presence of planets in contrast to an intrinsic origin of these RV variations. In this section, we first describe each metric for the whole sample before discussing the results star-by-star in \autoref{sec:results:individual_stars}.

For some false-positive planet detections around giant stars, such as $\gamma$~Dra \citep{Hatzes2018}, Aldebaran \citep{Reichert2019}, or $\epsilon$~Cyg \citep{Heeren2021}, the existence of the proposed planets have been ruled out as changes of the amplitudes, phases, or periods of the RV variability have been detected on long timescales. Similar suspicions arise when attempting to model many of the stars in this sample using Keplerians.

\begin{figure*}
    \centering
    \includegraphics{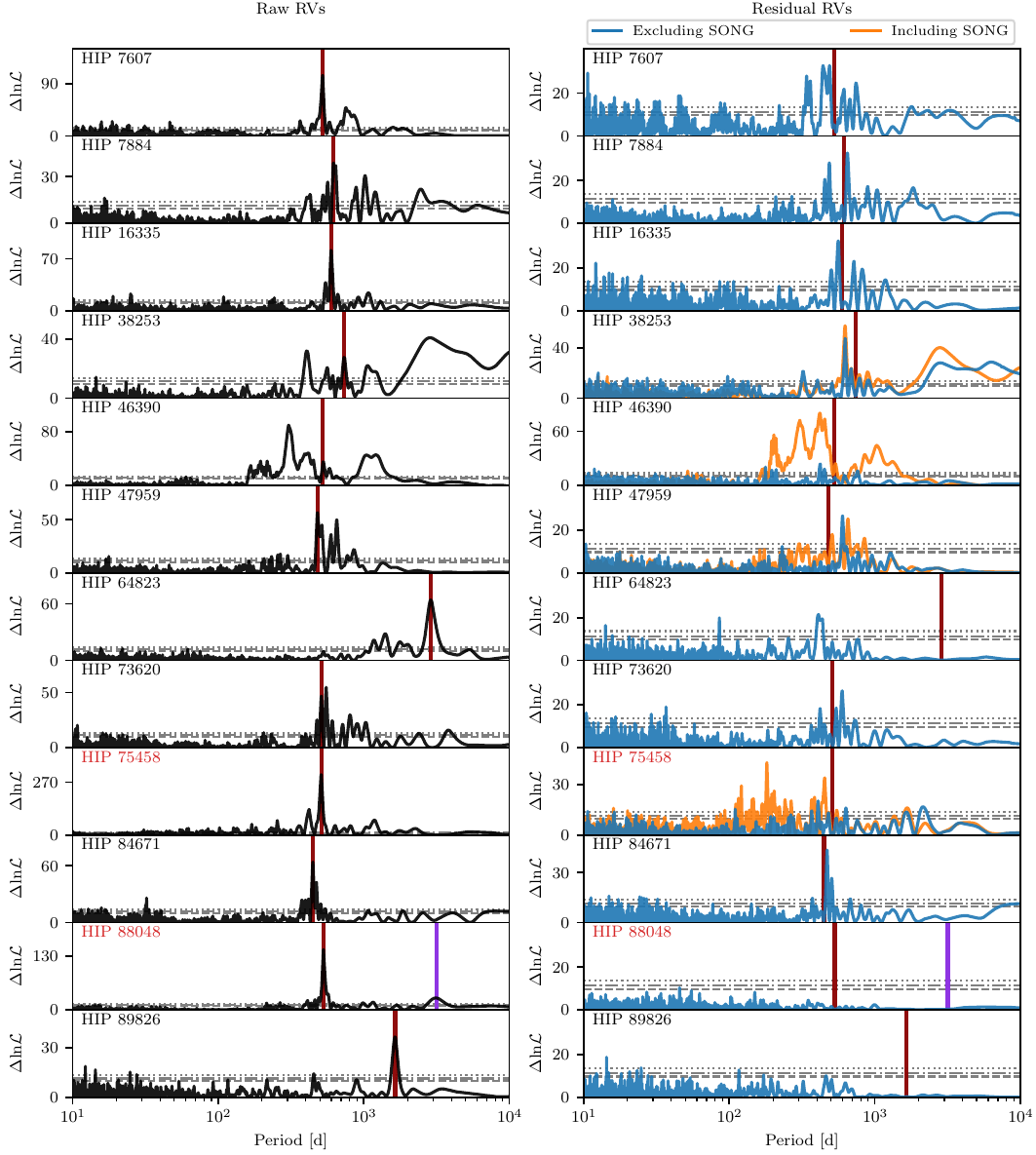}
    \caption{Maximum likelihood periodograms of the RVs (left) and the residuals (right) after subtracting the best Keplerian model. As the SONG RVs are known to show spurious RV periodicity at a one-year period \citep{Heeren2023}, we plot the MLPs of the residual RVs excluding the SONG data in blue, while the periodograms including the SONG data are shown in orange. The FAPs of $5\%$ (dashed), $1\%$ (dash-dotted), and $0.1\%$ (dotted) are plotted as gray horizontal lines. The identifiers of the two planet hosts, HIP~75458 and HIP~88048, are marked in red. Significant periodicity remains close to the strongest RV period after removing the best fit for many of the stars.}
    \label{fig:residuals}
\end{figure*}

One straightforward way to illustrate such problematic cases is to assess the residual
variability in the data through a periodogram search. Figure~\ref{fig:residuals} portrays the maximum likelihood periodograms (MLPs) \citep{Zechmeister2019} of the RVs in the left column, as well as the MLPs of the residuals in the right. We mark the RV period of the best Keplerian model as the red vertical lines. It is known that SONG RVs can suffer from spurious yearly or half-yearly periodicity \citep{Heeren2023}. Therefore, we plot the MLPs of the residuals excluding the SONG data in blue, while we plot the data set including the SONG data in orange.

The main advantage of the MLP, in contrast to the more established generalized Lomb-Scargle (GLS) periodogram \citep{Zechmeister2009}, is that an independent noise term is fitted for each period probed in the periodogram analysis, which is important for the luminous giant stars in our sample (see \autoref{tab:nest_samp_params}). 
We computed the MLPs with an oversampling factor of 50 and derived the false alarm probabilities (FAP) following the independent frequency method summarized by \citet{VanderPlas2018}
\begin{equation}
    \mathrm{FAP} = 1 - (1- \exp(-\Delta \ln \mathcal{L}))^{N_\mathrm{eff}},
\end{equation}
with $\Delta \ln \mathcal{L}$ being the log-likelihood difference to a constant model and $N_\mathrm{eff} = \Delta f \Delta T$ estimating an effective number of independent frequencies. Here, $\Delta f$ is the range of the probed frequencies, while $\Delta T$ is the range of the time interval covered by the data set. In \autoref{fig:residuals}, we overplot the FAPs of $5\%$ (dashed), $1\%$ (dash-dotted), and $0.1\%$ (dotted) as gray horizontal lines.

As long as the star-planet interactions due to tidal forces are small, which is generally to be expected for the relatively long-period variations considered in this work, the RV variations caused by a single planet should occur at a stable, largely unchanging period.
For a well-behaved planetary system, one can therefore expect that no periodicity remains in the RV data in the adjacent period region after removing the best Keplerian model. 

It can be seen from \autoref{fig:residuals} that several of the stars (for instance HIP~7607, HIP~7884, HIP~16335, HIP~38253, HIP~46390, HIP~47959, HIP~73620, and HIP~84671) have significant residual power very close to the best period of the fit, indicating that the RVs are not mono-periodic. Due to the proximity of the residual periods, and given the masses needed to explain the dominant RV variations with planetary companions, which are all Jovian or higher, strong stability constraints rule out additional companions at these periods. From this relatively simple test alone, we can therefore conclude that planetary companions are unlikely explanations in these cases. 

In comparison, for the two published planet systems, HIP~75458 and HIP~88048, the best fits remove the power close to the RV periods almost completely. We note that for HIP~75458 some significant peaks (after removing the SONG data) are left in the proximity of the Kepler period of $\iota$~Dra~b. Several of these peaks are close to alias periods between the Keplerian period and peaks in the combined window function. These peaks are likely a consequence of the large eccentricity of the orbit of $\iota$~Dra~b. All residual peaks for $\iota$~Dra have much lower $\Delta \ln \mathcal{L}$ than the original peak corresponding to the planet period. This is in contrast to the aforementioned challenging cases.  

Two planet candidates, HIP~64823 and HIP~89826, stand out as there is no residual power close to the RV period. Both also show the longest periodicity (see \autoref{tab:nest_samp_params}) among the planet candidates. We conclude that testing the residual RVs for significant periodicity can be a simple check for a multi-periodic (and thus likely intrinsic) behavior of the RV variations. 

\subsection{Infrared test}
Another approach that mostly emerged in the context of RV variability caused by star spots, is to test whether the RV signal is consistent at optical and infrared wavelengths. For star spots, as the contrast ratio between a cool spot and the stellar surface is much smaller in the infrared than in the optical, the RV variation introduced by a spot is generally expected to be smaller in the infrared compared to the optical \citep{Desort2007, Huelamo2008, Reiners2010}. As the photometric amplitudes of pulsating stars are known to be different in optical and infrared passbands as well \citep{Percy2001}, one can also expect the infrared test to be applicable in the context of suspected non-radial oscillations in RV data \citep{Mitchell2013, Trifonov2014, Trifonov2015, Ortiz2016}. In contrast, a planetary companion should induce a consistent RV signal in all wavelength regimes. 

To first order, one can expect that the semi-amplitude $K$ is the only changing parameter when fitting contemporary RVs taken in the VIS and NIR channels of CARMENES. We thus kept all parameters but $K$ fixed from the previous modeling detailed in \autoref{sec:results:keplerian_modelling} and performed another dynamic nested sampling of the posterior distribution of $K$ using only the CARMENES--VIS and CARMENES--NIR data, respectively. We adopted the mode of the distributions and derived the uncertainties by identifying the interval around the mode that encompasses 68\% of the posterior samples, corresponding to a 1 $\sigma$ credibility interval. We ensured the unimodality of each distribution by eye. We stress that we only compare the semi-amplitudes between the two CARMENES channels. These can, in some cases, be inconsistent with the determined semi-amplitudes using the whole data set, indicating that the RV signals are not stable over long time frames but can also be a consequence of sparse sampling. 

\begin{figure}
    \includegraphics{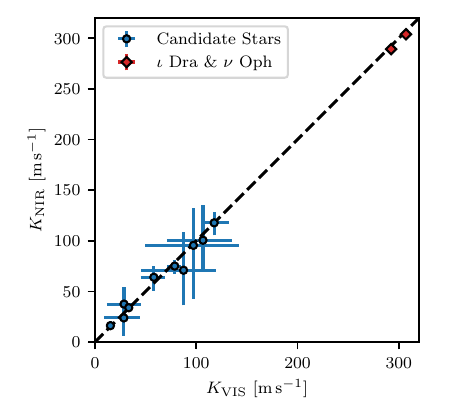}
    \caption{Semi-amplitudes $K$ from the CARMENES--NIR channel plotted against the semi-amplitudes in the VIS channel. We highlight the two secure planet hosts, $\iota$~Dra and $\nu$~Oph as red diamonds, while we show the planet candidates in blue. For $\iota$~Dra and $\nu$~Oph, only the inner orbital components are examined due to the short baseline of the CARMENES observations.}
    \label{fig:IR}
\end{figure}

We plot the results of the infrared test in \autoref{fig:IR}. We find no significant differences between the amplitudes of the variations derived from the CARMENES--VIS and CARMENES--NIR data. For several of the very luminous stars, large error bars due to the large RV jitter and relatively few data points, sampling only two to three phases of the RV variations, reduce the informative value of the infrared test. 

In \citet{Spaeth2024}, we present a simulation of an $l=1, m=1$ non-radial oscillation that can reproduce the variations of the RVs and the activity indicators for the false-positive brown-dwarf host NGC~4349~No.~127 \citep{DelgadoMena2018, DelgadoMena2023}.
Using the same simulation settings as in \citet{Spaeth2024} to derive hypothetical RVs in the CARMENES--VIS and CARMENES--NIR channels leads to semi-amplitudes $K_\mathrm{VIS}=249.6\,\mathrm{m\,s^{-1}}$ and $K_\mathrm{NIR}=257.6\,\mathrm{m\,s^{-1}}$, respectively, a relative difference of about $3\%$. This would, for most of the stars, be insignificant given the derived uncertainties in the presence of large stellar jitter. Thus, we conclude that the absence of significant differences between semi-amplitudes derived in the optical and the infrared cannot rule out a non-radial oscillation model.

We note that these findings are only valid for relatively small temperature variations employed in the simulations (here $\delta T_\mathrm{eff}=2.5\,\mathrm{K}$), for which the chromatic limb-darkening correction dominates the chromatic behavior of the RVs. For larger temperature variations and other oscillation modes, the relative amplitude difference between infrared and visual RVs can change slightly compared to the presented case. However, our test simulations, including more extreme values and different oscillation modes, did not reveal any significant increase in the relative RV difference between the VIS and NIR channels, while maintaining photometric variations comparable to those of the stable stars in this sample.
The infrared test is thus not sensitive enough to validate or disprove potential planets orbiting evolved stars.

\subsection{Analysis of the activity indicator time series}
State-of-the-art, stabilized spectrographs, such as CARMENES, acquire a set of activity indicators simultaneous to the RVs. These activity indicators have proven to be very useful in the study of magnetic activity, as manifested in star spots or plages, in RV surveys that target main-sequence or low-mass stars. \citet{DelgadoMena2018, DelgadoMena2023} have used activity indicators measured by HARPS successfully to rule out a number of exoplanets around cluster giants, showing that these indicators are also applicable to giant stars, for which magnetic activity is not necessarily the major concern. 

\subsubsection{Correlation analysis}
\label{sec:results:corr}
In \citet{Spaeth2024}, we show that the HARPS activity indicators for NGC 4349 No.\ 127, a known false-positive host \citep{DelgadoMena2018, DelgadoMena2023}, show peculiar correlations with the RVs. The CRX and the RVs are slightly positively correlated, while the dLW, FHWM, and contrast of the CCF have ``closed-loop'', Lissajous-like correlations with the RVs. We further show that these are the correlations expected from an $l=1, m=1$ non-radial oscillation mode. 

Motivated by these findings, we scanned the Lick H$\alpha$ measurements and the CARMENES activity indicators for similar correlations with the RVs. By eye, we did not find any reliable closed-loop correlations in the sample. These could potentially be obscured by variations due to short-term p-mode oscillations, or the superposition of different modes \citep{Spaeth2024}. We further note that the RV amplitudes of the planet candidates in this sample are typically smaller than those of NGC 4349 No.\ 127, which makes the identification of such correlations more challenging. 

\begin{figure*}
    \centering
    \includegraphics{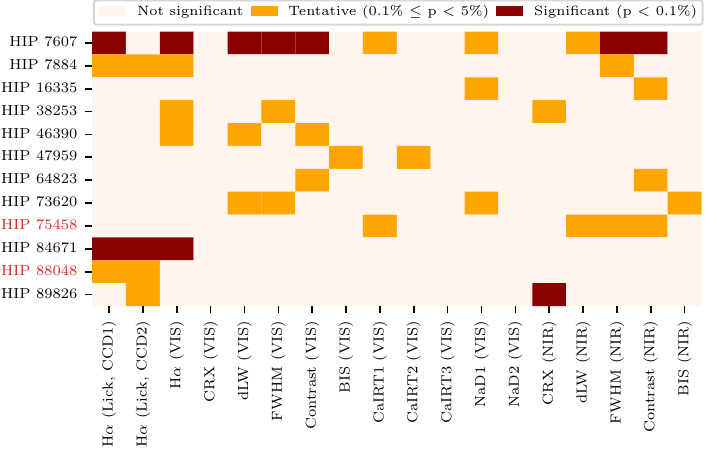}
    \caption{Results of the linear correlation analysis. For each star we show the significance of the linear correlation between the activity indicator labeled on the x-axis and the respective RV data set. Red colors indicate significant ($p<0.1\%$) linear correlations, while orange colors represent tentative ($0.1\% \leq p < 5\%$) correlations. The identifiers of the two planet hosts, HIP~75458 and HIP~88048, are marked in red.}
    \label{fig:corr}
\end{figure*}

Lacking indications for more complex correlations, we systematically tested the activity indicators and RVs for linear correlations. For each of the activity indicators and corresponding RV time series, we fitted for a linear relation. We assessed the significance of the resulting slope by performing a permutation test. That is, we uniquely permuted the pairs of variables $N=10\,000$ times and counted the number of simulated slopes $m$ with $|m| \geq |m_\mathrm{real}|$. The two-tailed $p$-value of the significance of the measured slope is then defined, following \citet{Ernst2004} and \citet{Phipson2010}, as
\begin{equation}
    p = \frac{N_{|m|\geq|m_\mathrm{real}|} + 1}{N+1}. 
\end{equation}
We define correlations with $0.1\% \leq p < 5\%$ as tentative correlations, while we regard $p<0.1\%$ as significant. We computed $p$-values for all correlations without applying any threshold to the slope beforehand. The combined results of the linear correlation analysis can be seen in \autoref{fig:corr}. We observe that only three of the stars, HIP~7607, HIP~84671, and HIP~89826 show any significant correlations. Only the former two show correlations in more than one indicator. That is surprising since we can find strong arguments against planets as the cause of the RV variations for a majority of the stars based on other metrics. In contrast to the case of NGC 4349 No. 127, we also do not find correlations between the CRX and the RVs in the CARMENES VIS channel for any of the stars. 

On the other hand, we observe that all stars show at least one tentative correlation between the indicators and the RVs, even the two secure systems HIP~75458 and HIP~88048. Of course, these two systems could in principle also be contested, but given the extremely well-behaved RV variations of both systems with a large eccentricity in case of HIP~75458 and well-studied dynamical constraints for HIP~88048 \citep{Quirrenbach2019}, there are good reasons to assume that the planetary systems around these stars are real. In fact, it rather reflects the relatively poor performance of the simple linear correlation analysis.

This can be understood since searching for linear correlations in this way is sensitive to different kinds of variations of the activity indicators. These could, for instance, be linked to the known short-term p-mode pulsations or the stellar rotation and might be completely unrelated to any RV periodicity introduced by a planet. Such short-term variations in the activity indicators, combined with long-period RV variations of a different origin, can lead to spurious correlations. We thus conclude that a simple linear correlation analysis is not apt to differentiate between planets and intrinsic variations for the majority of the stars and argue in favor of more sophisticated metrics such as the periodogram analysis discussed in the following. We note that a combination of a well-behaved linear correlation with a finding of clear periodicity at the RV period could be helpful to remove the contribution of the intrinsic signal from the RVs (see, for instance, \citealt{Robertson2014}).   

\subsubsection{Activity periodogram analysis}
\label{sec:methods:mlp}
\begin{figure*}
    \centering
    \includegraphics[scale=1.0]{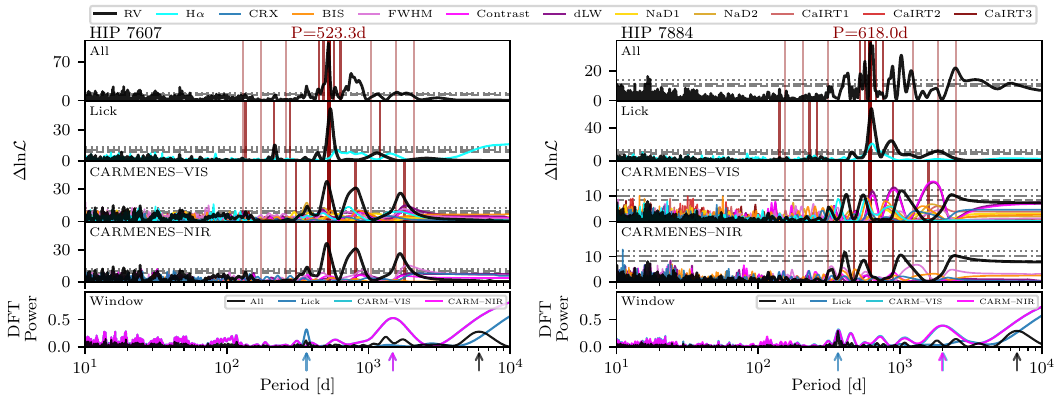}
    \caption{Maximum likelihood periodograms for two example stars. The top panel for each star portrays the combined RV MLP. In the remaining panels, we show the MLPs of the RVs (black) and activity indicators (see legend) for each instrument. The FAPs of $5\%$ (dashed), $1\%$ (dash-dotted), and $0.1\%$ (dotted) are plotted as gray horizontal lines. The bottom panel shows the instruments' window functions derived as discrete Fourier transforms. The thick, red, vertical line indicates the period of the Keplerian fit (see \autoref{tab:nest_samp_params}). The alias periods, corresponding to the most significant peak in the respective window function, are shown as thinner lines. The positions of these window peaks are marked with colored arrows in the bottom panels. Notably, the CARMENES VIS and NIR window functions largely overlap. Additionally, we display the harmonics and sub-harmonics of the period of the Keplerian fit as red lines with lower opacity. The full figures are shown in Figs.~\ref{fig:mlp_part1} and ~\ref{fig:mlp_part2}.}
    \label{fig:mlp_example}
\end{figure*}

To assess whether the activity indicators vary at periods linked to the established RV periodicity, we employ a periodogram search using MLPs. We present the results for two examples stars in \autoref{fig:mlp_example}. The full figures are shown in Figs.~\ref{fig:mlp_part1} and \ref{fig:mlp_part2}. These show the combined MLPs for all time series (RVs and activity indicators) of all instruments available for each star. Again, we overplot as gray horizontal lines the FAPs of $5\%$ (dashed), $1\%$ (dash-dotted), and $0.1\%$ (dotted). We generally refer to peaks with $0.1\% \leq \mathrm{FAP} < 5\%$ to be tentatively significant, while we consider peaks with $\mathrm{FAP} < 0.1\%$ to be significant. We further overplot the period for each Keplerian model (see \autoref{tab:nest_samp_params}) as thick, vertical, dark red lines and label their respective periods at the top of each panel. We also show the aliases of the RV period, computed with the strongest peak in each respective window periodogram, as thinner red lines, while we portray three harmonics and sub-harmonics of the RV period with lower opacity. The periodograms of the window functions were computed as discrete Fourier transforms and are plotted in the bottom panel. The strongest window peak for each instrument was detected in the range between $300\,\mathrm{d}$ and the baseline of the respective time series and is marked as the colored arrows on the $x$-axis of the panel. Following \citet{VanderPlas2018}, periodicity at the alias frequencies or the harmonics of the RV period should be regarded with caution. We note that many of the window functions show multiple peaks. However, we refrain from computing and plotting the aliases for each peak to avoid a cluttering of the plot. The results of the activity periodogram analysis are discussed on a star-by-star basis in \autoref{sec:results:individual_stars}. 

\label{sec:results:mlp}

\subsubsection{DBSCAN cluster analysis}
\label{sec:results:clustering}
A useful way to summarize and detect associated periodicity in multiple CARMENES activity indicators is presented by \citet{Kemmer2024}. Following this approach, we first detected all peaks by their FAPs and periods in the MLPs of all CARMENES activity indicators (both channels). We note that the process allows for multiple peaks of the same indicator time series to be detected. Next, we employed the \texttt{DBSCAN} \citep{Ester1996} clustering algorithm available through \texttt{scikit-learn} \citep{Pedregosa2011} on all identified peaks with $\mathrm{FAP}<50\%$. We tested different settings for the \texttt{DBSCAN} parameter $\epsilon$, which defines the maximum distance between two peaks to be considered neighbors, setting it to one-third of the width of the peaks in the periodogram as estimated by $\delta f = (t_\mathrm{max} - t_\mathrm{min})^{-1}$. We found this value to provide good results at the long periods of interest in this work. We limited the identified clusters to have at least three peaks, with at least one of the peaks having tentative ($\mathrm{FAP<5\%}$) significance. We further limited the cluster search to periods longer than $100\,\mathrm{d}$, to only target periodicity close to the dominating, long-period variations of the RVs.

\begin{figure*}
    \centering
    \includegraphics{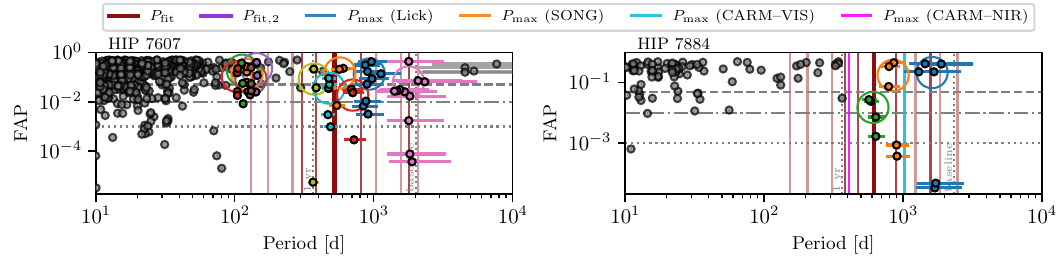}
    \caption{Example results of the DBSCAN clustering algorithm for two stars. In each panel, all identified peaks in the MLPs of the CARMENES activity indicators are shown as filled circles. Arbitrarily colored circles represent activity peaks assigned to clusters by DBSCAN, while gray circles indicate peaks that are not part of any cluster. The larger, empty circles indicate the center periods and FAPs of the clusters. We further overplot the FAP levels of $5\%$ (dashed), $1\%$ (dash-dotted), and $0.1\%$ (dotted) as horizontal gray lines, and the fitted RV period (thick), its aliases (thin, based on the VIS window function), and harmonics (lower opacity) as vertical red lines.
    Furthermore, for stars where individual instruments exhibit a strongest period that deviates from the fitted period by at least $10\%$ (restricted to $100\,\mathrm{d} < P_\mathrm{max} < 5000\,\mathrm{d}$), we overplot the strongest RV period detected in the MLP of the respective instrument using vertical, colored lines specified in the legend. The full figure is shown in \autoref{fig:cluster}.}
    \label{fig:cluster_example}
\end{figure*}

The results of the cluster analysis are plotted for two example stars in \autoref{fig:cluster_example} and for the full sample in \autoref{fig:cluster}. These are discussed in detail on a star-by-star basis in the following. In general, we find that a majority of the stars whose RV variations are likely intrinsically induced show clustered activity periodicity close to the main RV period. These findings are mostly contributed by the VIS channel, which, however, also has several additional line indicators in contrast to the NIR. Somewhat unsurprisingly, we often find that the line shape indicators (dLW, BIS, contrast and FWHM of the CCF) vary at similar periods and thus form activity clusters. However, we often also find contributions to the same clusters from other indicators, including the CRX and the chromospheric indicators targeting the H$\alpha$ and NaD lines.

\section{Combined results for individual stars}
In this section, we discuss the results star-by-star, aiming to determine whether planets or intrinsic variations are causing the observed RV variations. 
\label{sec:results:individual_stars}

\subsection{HIP 7607}
The horizontal branch giant \object{HIP~7607} ($\upsilon$ Per) shows strong and clearly significant RV periodicity close to the fitted period of $P=523.3\,\mathrm{d}$ in the Lick and both of the CARMENES data sets in Figs.~\ref{fig:mlp_example} (left) and \ref{fig:mlp_part1} (top left). For both instruments, periodicity can also be found at the respective alias periods (thinner, red lines). We note that the peaks in the Lick RV MLP ($P=538\,\mathrm{d}$, see also \citealt{Hekker2008}) and the CARMENES RV MLPs (both $P=508\,\mathrm{d}$) are slightly offset. However, given the relatively short baseline of the CARMENES observation and the width of the peaks, this difference cannot be regarded as significant. 

Examining the activity indicators in \autoref{fig:mlp_example}, we find a number of peaks with tentative significance in the vicinity of the RV period in the CARMENES VIS channel. The most significant peak ($\mathrm{FAP}\sim0.1\%$) is contributed by the NaD2 line at $P=491\,\mathrm{d}$, slightly offset from the RV period. Further tentative peaks close-by can be found by the dLW, contrast, and BIS indicators. Further periodicity in the activity indicators is also present close to the alias period of the fitted RV period. The \texttt{DBSCAN} cluster analysis in Figs.~\ref{fig:cluster_example} (left) and \ref{fig:cluster} (top left) reveals 
two clusters of activity variability centered at $P=478\,\mathrm{d}$ (cyan) and $P=572\,\mathrm{d}$ (orange). Together, the activity indicators clearly cast doubts on the planetary nature of the RV signal. 

The doubts are further strengthened by several significant correlations present between the activity indicators and the RVs, as evident from \autoref{fig:corr}. However, as discussed in \autoref{sec:results:corr}, the overall mixed results of the correlation analysis reduce the informative value of such correlations. The final and strong argument against the planetary nature of the RV signal comes from analyzing the residual RVs. As can be seen in \autoref{fig:residuals}, there is significant periodicity left at $P=441\,\mathrm{d}$, $P=489\,\mathrm{d}$, $P=614\,\mathrm{d}$, and other more distant periods. Similar systematics in the residuals of the Lick data of HIP~7607 have been noted by \citet{Quirrenbach2011}. Trying to explain this additional periodicity with another orbital companion requires a minimum mass exceeding $1\,M_\mathrm{Jup}$, which would make the proposed system unstable. It is thus much more likely to be a manifestation of a multi-periodic incoherent type of RV variation that is linked to intrinsic processes. 

\subsection{HIP 7884}
One star that highlights the usefulness of the periodogram analysis of activity indicators is \object{HIP~7884} ($\nu$ Psc), a bright ($V=4.4\,\mathrm{mag}$) K3 giant star likely located on the RGB (see \autoref{tab:params}). Focusing first on the Lick RV MLP in Figs.~\ref{fig:mlp_example} (right) and \ref{fig:mlp_part1} (top right), one can clearly see the dominant and highly significant peak in the RV periodogram at $P=628\,\mathrm{d}$. However, in the CARMENES RVs (third and fourth panels) no significant periodicity can be found at that period, revealing peaks at $P=418\,\mathrm{d}$, $P=549\,\mathrm{d}$, and $P=1028\,\mathrm{d}$ in the VIS instead, with peaks in the NIR at similar periods. As a consequence, the MLP of the combined RVs (top panel) reveals a complex structure, reducing the significance of the most prominent peak compared to the Lick data set alone, and introducing further peaks close-by. These findings likely indicate that the periodic variation present in the Lick data set, is not present in the CARMENES RVs, which cover a more recent time baseline. Analyzing the residual periodogram in \autoref{fig:residuals}, one can see this behavior manifested as a second peak, a mere $30\,\mathrm{d}$ distant in period space.

Examining the available activity data in Figs.~\ref{fig:mlp_example} and \ref{fig:cluster_example}, one can find a significant peak of the Lick H$\alpha$ measurements at the period of the Lick data, as well as peaks of the dLW and the contrast of the CCF of the CARMENES VIS channel, which are both evident at the RV period and at its alias frequencies (thinner, red lines). Thus, a planetary origin of the RV periodicity can be excluded. The activity peaks in the CARMENES data appear to belong to a cluster (including tentative peaks from the CARMENES H$\alpha$ indicator and the BIS) centered very close to the fitted period (\autoref{fig:cluster_example}). One further observes two clusters centered at the two long-period aliases of the RV period. Despite these significant detections of activity variations at the RV period, there are only tentative correlations present in \autoref{fig:corr}. These include the two time series of Lick H$\alpha$ measurements, as well as the H$\alpha$ indicator derived from the visual CARMENES channel. Each of the H$\alpha$ time series shows a tentative positive correlation with the RVs. The FWHM of the CCF of the NIR channel is also tentatively positively correlated with the RVs.

Combining all the above findings, these observations make HIP 7884 another example of a giant star whose measured RVs, when viewed in isolation, seemed to be consistent with an orbital companion for over 12 years. However, by using a much longer RV baseline and acquiring reliable activity indicators, its cause can decisively be concluded to be of intrinsic origin.

\subsection{HIP 16335}
One of the most interesting stars within the sample is \object{HIP~16335} ($\sigma$ Per), which was published to host a planetary companion with minimum mass $m_\mathrm{p}\sin(i)=6.5\pm1.0\,M_\mathrm{Jup}$ in a $P=579.8\pm2.4\,\mathrm{d}$ orbit by \citet{Lee2014a}. The authors used RV data from the Bohyunsan Observatory Echelle Spectrograph (BOES) at the Bohyunsan Optical Astronomy Observatory (BOAO) in Korea, along with activity indicators derived from the Ca\,\textsc{ii}\,H line, and studied the bisector variations of the Ni\,\textsc{i} $6643.6\,\text{\AA}$ line. They find no periodicity or correlations with the RVs. We include the BOAO data set as published by \citet{Lee2014a} in this analysis. 

While the peaks in the BOAO and Lick RV MLPs (\autoref{fig:mlp_part1}), which are roughly contemporary, are clearly apparent and strongly significant, the RVs in the CARMENES visual channel show a much less significant peak, which is further slightly offset to $P=609\,\mathrm{d}$. The infrared RVs of CARMENES show a strong peak at $P=599\,\mathrm{d}$. However, two strong peaks close to the originally published planet period emerge in the CARMENES--VIS activity indicator MLPs: Both the CRX and the BIS of the visual channel have significant peaks at $P=559\,\mathrm{d}$ and $P=565\,\mathrm{d}$, respectively, close to the originally published planet period $P=579.8\pm2.4\,\mathrm{d}$ \citep{Lee2014a}. We further note additional peaks of the VIS H$\alpha$ indicator, the CCF contrast, and the BIS in the vicinity of the alias and harmonic periods of the RVs.

The \texttt{DBSCAN} clustering algorithm presented in \autoref{fig:cluster} shows two adjacent clusters neighboring the RV period, which shows that there is further, tentative periodicity of activity indicators close-by. We note that the two significant peaks in the CRX and BIS are attributed to the cluster centered at $P=527\,\mathrm{d}$ (green).

The strong evidence against the planetary companion is further enhanced when looking at the residuals of the RV fit in \autoref{fig:residuals}. After subtracting the best planet fit, significant variability with a period $P=561\,\mathrm{d}$ remains, a mere $37\,\mathrm{d}$ offset from the fitted period. This additional variability can hardly be explained by a second orbital companion fulfilling stability constraints. Instead, we interpret it to be a sign of a non-constant periodicity or amplitude of the RV signal. Altogether, these findings strongly argue against the planetary nature of the RV signal of HIP~16335 and thus contradict the findings by \citet{Lee2014a}. 

\subsection{HIP 38253}
\object{HIP~38253} (HD~63752) is the most luminous star in the sample and shows relatively large-amplitude RV variations with $K \sim 148.1 \,\mathrm{m\,s^{-1}}$. First hints of an intrinsic origin have already been found by \citet{Hekker2005}. Examining the MLPs in \autoref{fig:mlp_part1} confirms these findings.

The Lick RVs show a dominant peak at $P=619\,\mathrm{d}$, with a small side-lobe and secondary peak at $P=766\,\mathrm{d}$. In contrast, the SONG data set, bridging between the Lick RVs and the most recent CARMENES observations, does not show the same periodicity but instead a most dominant peak at $P=416\,\mathrm{d}$, and further power at longer periods. This shorter-period peak is, however, no longer present in the CARMENES data set, which rather shows two less significant peaks at $P=631\,\mathrm{d}$ and $P=874\,\mathrm{d}$. It is thus apparent that RV variations at inconsistent periods are present in data sets spanning different time windows. This is also evident from the complex structure of the combined MLP in the top panel. The nested sampling algorithm reveals $P_\mathrm{fit}=735.3\,\mathrm{d}$ to be the modal value of the period posterior distribution, which shows a slight bimodal structure. It is clear that this RV variation is not stable and cannot be caused by a planetary companion, which would be expected to be consistently present in all data sets. This is also evident from the rather unsatisfying mono-periodic fit to the RV data in \autoref{fig:multi-rv}. 

Due to this multi-periodicity in the RV data, the activity time series is more challenging to interpret. Taking a look at the clustering results in \autoref{fig:cluster}, it is evident that a strong cluster of significant activity indicator periodicity is present very close to the RV period present in the SONG data set (marked as the orange line). The most dominant peaks stem from the line shape indicators (dLW, FWHM and contrast of the CCF, BIS) in the VIS, but also partially in the NIR. There are further tentative peaks by the H$\alpha$ and CaIRT2 lines in the vicinity, as apparent from \autoref{fig:mlp_part1}.

Close to the strongest period in the Lick RVs ($P=619\,\mathrm{d}$, blue line) is another, less significant, cluster of activity variations, apparent both in the line shape indicators in VIS and NIR, as well as the H$\alpha$ line, and the CRX in the VIS. Finally, there appears to be another cluster of activity variations present at somewhat longer periods. While there is no significant peak exactly at the fitted RV period, it appears that each tentative peak of the CARMENES RVs is accompanied by at least tentative variability in the activity indicators.

As noted for the whole sample, the correlation analysis reveals only tentative correlations between the activity indicators and the RVs. This result is surprising given the strong support for an intrinsic origin of the RV variations. It strengthens our conclusion that a simple linear correlation analysis cannot properly capture the complexities of the relations between activity indicators and RVs for giant stars. It can likely be understood as a consequence of the multi-periodic nature of the variations in both the RVs and the indicators. Looking at the residual RVs in \autoref{fig:residuals}, it is clear that significant residual periodogram power is present in the vicinity of the fitted RV period, confirming this multi-periodicity. We conclude that the RV variations of HIP~38253 are of intrinsic origin. 

\subsection{HIP 46390}
Somewhat similar to HIP~38253 is the behavior of the very bright ($V = 1.97\,\mathrm{mag}$) giant \object{HIP~46390} (Alphard, $\alpha$ Hya, HD\,81797), although the amplitude of the variations is much smaller  ($K\sim47.3\,\mathrm{m\,s^{-1}}$). Again, the strongest peak in the Lick data set (2000--2011) at $P=524\,\mathrm{d}$ is not dominant in the later SONG observations (2015--2022), which have the strongest periodogram peak at $P=306\,\mathrm{d}$ (see \autoref{fig:mlp_part1}). The most recent CARMENES observations (2016--2023) show their most significant periodicity at $P=422\,\mathrm{d}$. As for HIP~38253, this also rules out the presence of a stable planetary system for HIP~46390, which coincidentally is the second most luminous star in the sample. As mentioned in \autoref{sec:results:keplerian_modelling}, the nested sampling algorithm shows multimodal posterior distributions with different periodicities being picked up in different runs of the algorithm. The final model used in this work is mostly dominated by the Lick RVs, with the SONG RVs showing a clear phase shift in a phase-folded RV plot.

Subtracting the best Keplerian model from the RVs and examining the residual periodogram in \autoref{fig:residuals}, we find that there remain several significant peaks when including the SONG data (orange). Removing the SONG data set from the analysis, we find that there is residual periodicity with significant peaks at $P=422\,\mathrm{d}$ and $P=451\,\mathrm{d}$ (blue periodogram). Thus, the best model cannot remove all variability close to its modeled period.

Examining the activity indicator periodograms in \autoref{fig:mlp_part1} and their clustering in \autoref{fig:cluster}, we find 
a significant peak by the CARMENES--VIS H$\alpha$ indicator ($\mathrm{FAP}=0.03\%$) close to the Lick period. This activity peak is attributed to a small cluster, jointly with a tentative peak by the contrast of the CCF in the NIR channel, centered at $P=515\,\mathrm{d}$. Close to the SONG RV period ($P=306\,\mathrm{d}$, orange line in \autoref{fig:cluster}) that dominates the combined RV periodogram, we find a close cluster at $P=325\,\mathrm{d}$ of significant and tentative peaks mostly contributed by the dLW and contrast of the CCF in the VIS, as well as the NaD2.
Thus, the dominant RV periodicity in the SONG RVs also seems to be present in the variations of the spectral lines as measured by CARMENES. Finally, we observe a tentative peak by the dLW close to the most prominent RV peak in the CARMENES data set. As with many other stars, we find only tentative correlations between the H$\alpha$ indicator, dLW, and CCF contrast on the one hand, and the RVs on the other. Taken together, there is overwhelming evidence that the RV variations observed for HIP~46390 are of intrinsic origin, with changing periodicity on timescales of a quarter of a century.

\subsection{HIP 47959}
\object{HIP~47959} (18~Leo, HD~84561) is another relatively luminous star within our sample. Examining \autoref{fig:mlp_part1} (bottom right panel), the RV variations can again be shown to have varying periodicity in different time windows. While the Lick RVs show dominant periodicity at $P=611\,\mathrm{d}$, the SONG data taken at later epochs have a shifted peak at $P=495\,\mathrm{d}$, with a secondary peak at $P=654\,\mathrm{d}$, close to the Lick RVs. The former periodicity dominates the RV period determined through the nested sampling ($P=481.3\,\mathrm{d}$). The CARMENES data do not yet show a significant peak but have their strongest periodicities at $P=580\,\mathrm{d}$ in both channels. The combination of all available RV data then leads to a relatively complex periodogram apparent in the top panel. Following from this multi-periodic behavior, one can find a number of significant peaks close to the removed period in the residuals (\autoref{fig:residuals}).

Adding to these strong arguments against a planetary nature of the RV variations, we can find a number of tentative peaks in the activity indicators of the CARMENES channels (mainly the VIS), with some periodicity (both significant and tentative) present in the Lick H$\alpha$ indicator at somewhat shorter periods. Analyzing these using the \texttt{DBSCAN} clustering algorithm in \autoref{fig:cluster}, we find a strong cluster at $P=478\,\mathrm{d}$, coinciding closely with the fitted RV periodicity that was mainly dominated by the SONG RVs. It is thus evident that the RV variations are again of intrinsic origin. Despite these indications of intrinsic RV variations, only two tentative correlations are found between the activity indicators and the RVs, namely, for BIS and CaIRT2. This illustrates again the shortcomings of the correlation analysis. 

\subsection{HIP 64823}
\object{HIP~64823} (HD~115478) stands out within the sample as the candidate star with the longest-period RV variation, with a period of $P=2878.1\,\mathrm{d}$. It is likely  ($P=90\%$) located on the RGB \citep{Stock2018}. HIP~64823, unfortunately, was not part of the SONG observations, such that only Lick and CARMENES data are available. The baseline of the latter is furthermore a little shorter than one whole period of the RV variations such that an exact determination of the RV periodicity in the CARMENES data is not yet possible. Nevertheless, an increase of the likelihood at the period of the Lick RVs can be seen in \autoref{fig:mlp_part2}. 

Analyzing the activity indicators both in Figs.~\ref{fig:mlp_part2} and ~\ref{fig:cluster} suffers from the same limitation. We do find a tentative peak of the NaD2 line at $P=1940\,\mathrm{d}$, which, however, closely coincides with a peak in the window function of the CARMENES observations (bottom panel). Apart from that, there is (so far) no indication of any significant periodicity at periods similar to the RV period in either plot. While we cannot rule out a long-period activity variation with certainty, we conclude that the NaD2 variation is likely a consequence of the sampling of the observations. 

The linear correlation analysis reveals two tentative correlations between the RVs and the contrast of the CCF in both the VIS and NIR channels. The $p$-values of the correlations are $3.3\%$, and $3.2\%$ in the VIS and the NIR, respectively. However, we note that the correlation between the contrast of the CCF and the RVs is of opposite sign in both channels. That is, it is anti-correlated in the VIS, while it is positively correlated in the NIR. Combined with our general finding that the correlation analysis seems to be relatively ineffective for the giant stars in this work, we find it likely that these correlations are mostly a consequence of large uncertainties in the determination of the CCF parameters coupled with a relatively small number of measurements, as well as short-term jitter or the potential presence of intrinsic variations at shorter periods (see \autoref{sec:hip64823}). 

Analyzing the residuals in \autoref{fig:residuals}, we find the closest significant periodicity at $P=410\,\mathrm{d}$, with a slight secondary peak at $P=436\,\mathrm{d}$. In contrast to the previous stars, this additional periodicity does not seem to be a sign of a changing periodicity of the dominating period. Overall, we conclude that the CARMENES data available so far do not argue against the planetary nature of the RV periodicity. However, due to the insufficient baseline of the CARMENES observations, we cannot conclusively exclude any connection to intrinsic processes. We thus conclude hat HIP~64823 remains a good planet candidate and examine the Keplerian model in more detail in \autoref{sec:hip64823}. 

\subsection{HIP 73620}
The second least luminous star in the sample, \object{HIP~73620} (110~Vir, HD~133165), has the smallest RV variations of the sample with semi-amplitude $K\sim20.3\,\mathrm{m\,s^{-1}}$, only marginally larger than the estimated and expected jitter $\sigma_\mathrm{jit} \sim 13.3\,\mathrm{m\,s^{-1}}$. The star is located in the red clump \citep{Alves2000, Laney2012}. Examining \autoref{fig:mlp_part2}, one can observe a rather complex combined periodogram structure with the period $P=512.5\,\mathrm{d}$ being picked up by our nested sampling approach. This period coincides with the dominant periodicity in the Lick data set. The CARMENES RVs show a slightly offset peak at $P=554\,\mathrm{d}$ in the VIS channel, and at $P=547\,\mathrm{d}$ in the NIR. In the vicinity, we further observe several peaks present in the CRX and dLW of the VIS channel with varying levels of FAPs.

When analyzed using the \texttt{DBSCAN} algorithm in \autoref{fig:cluster}, we find a cluster of peaks centered at $P=539\,\mathrm{d}$, very close to the RV period. We note, however, that two of the peaks each stem from the CRX and dLW indicators. We further observe four tentative correlations for dLW, FWHM, and NaD1 in the VIS, as well as between the BIS and the RVs in the NIR. Using the period information from the RV residuals in \autoref{fig:residuals}, we find that subtracting the best fit reveals periodicity close-by at $P=597\,\mathrm{d}$, $P=544\,\mathrm{d}$, and $P=421\,\mathrm{d}$, with the central period closely coinciding with the CARMENES periodicity. The latter finding indicates that the shift of the periodicity in the CARMENES RVs is likely real and not a consequence of insufficient sampling of the phase curve. Overall, we conclude that the relatively small RV variations (within the context of this sample) of HIP~73620 are most likely caused by intrinsic variations instead of a planetary companion. Nevertheless, the star would benefit from continued observations to increase the temporal baseline and acquire additional measurements of activity indicators before a firm conclusion can be drawn.  

\subsection{HIP 75458}
Next, we examine \object{HIP~75458} ($\iota$~Dra, HD~137759), one of two published planet systems hosting a super-Jovian exoplanet in a $511\,\mathrm{d}$, very eccentric orbit \citep{Frink2002}. The RVs further show a long-period trend \citep{Zechmeister2008, Kane2010} attributed to a further-out brown-dwarf companion \citep{Hill2021}. The star was observed using SONG and added to the CARMENES sample in 2021. 

Examining the periodograms in \autoref{fig:mlp_part2}, one can observe that the strong RV periodicity at $P=510.9\,\mathrm{d}$ is consistently present in all data sets. Due to the very eccentric orbit of $\iota$~Dra~b (named Hypatia by the IAU) and the sinusoidal nature of the periodogram analysis, one can identify clear and significant periodicity also at the harmonics and aliases of the RV period. Examining the activity indicators, there is no strong peak at the planet period. We observe tentative periodicity of the NaD1 and NaD2 lines at $P=402\,\mathrm{d}$ and $P=349\,\mathrm{d}$, respectively. However, we note that these periods are also in proximity to one year, which suggests a connection to variable observing conditions or telluric contamination. This suspicion is reinforced by the \texttt{DBSCAN} clustering algorithm (\autoref{fig:cluster}), which detects a cluster of peaks (two of which are insignificant) centered at $P=368\,\mathrm{d}$. We thus find no evidence against the planetary hypothesis from the activity time series analysis. 

As mentioned in \autoref{sec:results:corr}, we find tentative correlations (see \autoref{fig:corr}) between the RVs and the CaIRT1 line in the VIS, as well as the dLW, FWHM, and contrast of the CCF in the NIR, with the latter having a $p$-value $p=0.3\%$ close to the significance threshold. It is noteworthy that all three indicators that show these tentative correlations in the NIR are based on similar variations of the spectral line shape (see, for instance, \citealt{Zechmeister2018}), and are thus not entirely independent. As discussed in \autoref{sec:results:coherence}, we find two significant peaks in the residual periodogram, excluding the SONG RVs, at $P=407\,\mathrm{d}$ and $P=634\,\mathrm{d}$. Both are much less significant than the initial peak caused by the orbital companion and lie in proximity to alias periods of the Keplerian period and the structured, combined window function. We thus suspect that these are a consequence of the large eccentricity of the orbit of $\iota$~Dra~b, which might be imperfectly constrained by the orbital fit.  

Overall, we observe some noteworthy features in the activity indicator and residual analysis of $\iota$~Dra. However, individually we regard none of these as strong arguments against a planetary nature of the RV signal. Moreover, the Keplerian orbit has a very high eccentricity ($e \sim 0.71$), which can hardly be mimicked by, for instance, non-radial oscillations. It is further extremely well determined and fulfills all stability constraints \citep{Hill2021}. The star is furthermore the least luminous star in the sample ($L\sim57\,L_\sun$) with radius $R\sim12\,R_\sun$, placing it relatively far away from the region of suspicious planet detections identified by \citet{Reichert2019} and \citet{Dollinger2021}. We therefore have no reason to question the existence of $\iota$~Dra~b. The results in this section rather emphasize that none of the metrics, particularly the correlation analysis, provide definitive conclusions regarding the existence of a planet responsible for the RV signal. 

\subsection{HIP 84671}
\object{HIP~84671} (66~Her, HD~156681) belongs to the most luminous stars in the sample with $L\sim472\,L_\sun$. It shows similar behavior to the two even more luminous sample stars, HIP~38253 and HIP~46390. While the RV periodicity at $P=447.2\,\mathrm{d}$ is strongly present in the Lick data set (see \autoref{fig:mlp_part2}), it is not properly picked up in the CARMENES RV MLPs (black lines). The strongest peak in the vicinity of the fitted period is instead caused by the CRX in the VIS at $P=423\,\mathrm{d}$, with further tentative peaks close to the alias periods from the CRX and the H$\alpha$ line. We further find a long-period signal at $P=1105\,\mathrm{d}$ in the VIS (and $P=1094\,\mathrm{d}$ in the NIR) to be the most significant peaks in the CARMENES RVs. This peak is accompanied, however, by an even more significant peak in the H$\alpha$ indicator. These are not present in the longer Lick time series. Finally, we find a significant peak in the Lick H$\alpha$ time series, within the peak width of the peak in the Lick RV MLP. Combining the peak information into the cluster analysis in \autoref{fig:cluster}, one can observe a cluster centered at $432\,\mathrm{d}$ (very close to the fitted period), with another longer-period cluster potentially linked to the long-period RV variation present in the CARMENES data. 

The H$\alpha$ variations in both Lick and CARMENES are also significantly correlated with the RVs, albeit likely linked to different periodicity in the different instruments. These correlations are all with a positive sign, strengthening the significance of the finding. HIP~84671 thus appears as one of the few stars for which the correlation analysis can provide useful additional information.
Subtracting the Keplerian model leads to residuals with additional significant periodicity at $P=470\,\mathrm{d}$ and $P=506\,\mathrm{d}$. Thus, we conclude that the RV variations with a large semi-amplitude $K\sim220\,\mathrm{m\,s^{-1}}$ are of intrinsic origin.

\subsection{HIP 88048}
The other published planetary system in our sample orbits \object{HIP~88048} ($\nu$~Oph, HD~163917) and consists of two brown-dwarf companions identified by \citet{Quirrenbach2011, Quirrenbach2019}.  The MLP analysis in \autoref{fig:mlp_part2} reveals no tentative or significant peaks in the vicinity of the RV period of $\nu$~Oph~b ($P=529.9\,\mathrm{d}$), and likewise, no significant clusters are detected in \autoref{fig:cluster}. We do observe a number of insignificant peaks by the Lick H$\alpha$, and CARMENES H$\alpha$, NaD1, and CaIRT2 indicators close to the one-year period. 

The period of the outer brown-dwarf companion at $P=3173.4\,\mathrm{d}$ is much longer than the baseline of the CARMENES observations. Therefore, no reliable conclusions can be drawn from the CARMENES activity indicators for that period. We do find, however, a rise of the likelihood for the Lick H$\alpha$ indicator at even longer periods, which does not seem to be directly connected to the RV period of $\nu$~Oph~c.
This long period is at the limit of the baseline of the Lick observations ($4009\,\mathrm{d}$) and could also be linked to aging effects of the Lick CCDs. This trend of the H$\alpha$ indicator is likely also causing the two tentative correlations between the Lick H$\alpha$ indicators and the RVs (see \autoref{fig:corr}), with $p$-values of $p=1.8\%$ and $p=2.7\%$ for the first and second Lick CCD, respectively. No correlations are detected between the CARMENES activity indicators and the RVs.

After removing the two-planet model from the RVs, we do not find any significant power close to the orbital periods in \autoref{fig:residuals}. Overall, we thus conclude that there are no indications in the CARMENES data that argue against the real nature of the inner brown-dwarf companion, while the outer is still outside the baseline of the CARMENES observations. 

\subsection{HIP 89826}
\object{HIP~89826} ($\kappa$~Lyr, HD~168775) is the second star among the planet candidates that stands out due to the relatively long-period nature of the RV variations which have a fitted period of $P=1642.6\,\mathrm{d}$. Looking at \autoref{fig:mlp_part2}, these are clearly apparent in the Lick data set, while they are (still) insignificant in the CARMENES--VIS RVs and only tentatively present in CARMENES--NIR. However, the baseline of the CARMENES RVs is only a few hundred days longer than the fitted period, which is a potential reason for the lack of significance in the CARMENES measurements. The semi-amplitude of the RVs, $K\sim32.5\,\mathrm{m\,s^{-1}}$, is relatively small compared to the average of the sample and compared to the jitter $\sigma_\mathrm{jit}=12.8\,\mathrm{m\,s^{-1}}$ likely caused by short-term pulsations and expected for this star. This can provide an additional explanation for the lack of periodicity in the VIS data set. Finally, the measurements (see \autoref{fig:multi-rv}) do not ideally sample the phase curve of the proposed planet due to a gap in the CARMENES data in 2019 and 2020.

Analyzing the activity indicators in \autoref{fig:mlp_part2}, we find no significant periodicity at the Keplerian period. We find two tentative peaks by the CARMENES--NIR CRX and BIS close to the second harmonic of the RVs, which also form a small cluster with another tentative peak of the CARMENES--VIS CRX indicator centered at $P=781\,\mathrm{d}$ (see \autoref{fig:cluster}). Another tentative cluster of peaks can be found close to the fourth harmonic of the Keplerian period. 

We further find a significant correlation between the RVs and the CRX in the NIR channel (see \autoref{fig:corr}), likely linked to the tentative periodicity of the CRX at the second harmonic of the RV period. We also find a tentative correlation (with $p$-value close to being significant) between H$\alpha$ and the RVs using the second Lick CCD. Removing the Keplerian fit from the RVs and analyzing the residual MLP in \autoref{fig:residuals} does not reveal any residuals peaks in the vicinity. 

Altogether, we conclude that the CARMENES activity indicators so far do not decisively rule out a planet causing the RV variations. Nevertheless, since the long-period signal of the RV variations is not yet properly picked up by the CARMENES data given the sizable stellar jitter, we conclude that continued observations would be necessary to reveal its origin.

\subsection{Summary of the results}
Overall, we can find strong reasons to suspect that the RV variations of seven out of the ten planet candidates are caused by intrinsic processes. These are the stars HIP~7607, HIP~7884, HIP~38253, HIP~46390, HIP~47959, HIP~84671, and, notably, the published planet host HIP~16335 \citep{Lee2014a}. We further lean toward an intrinsic origin of the RV variations of HIP~73620. The CARMENES activity indicators so far do not rule out a planetary origin of the long-period RV variations of HIP~89826, but continued monitoring is necessary to draw firm conclusions. For HIP~64823, we favor the planet hypothesis, which is examined in more detail in \autoref{sec:hip64823}. 

For the two comparison stars, HIP~75458 and HIP~88048, we find some tentative correlations between activity indicators and the RVs, which likely highlights the limitations of this analysis. We do not find any significant periodic signals in the activity indicators which would argue against the existence of the orbital companions. For HIP~75458, we note some residual periodicity, which is likely due to the large eccentricity of the orbit. 

We generally conclude that testing for linear correlations does not provide strong and reliable insights for distinguishing between planets and intrinsic processes. We further find that the infrared test is not sensitive enough in the presence of large RV jitter and relatively little data. Overall, we find that a very long baseline of RV measurements is necessary to detect the subtle differences between the intrinsically induced variations and real planets. A detection of different periods in different time windows, as well as an analysis of the residual RVs can provide the strongest clues toward the origin of the RVs. We note that the most luminous stars in the sample (HIP~38253, HIP~46390, and HIP~84671) seem to show the strongest signs of changing periodicities in data sets taken at different epochs and with different instruments. A complementary periodogram analysis of activity indicators can often confirm these signals to be of intrinsic origin. We cannot identify any indicator that would consistently be more sensitive than the others in our analysis. 

\section{The long-period planet candidate HIP 64823}
\label{sec:hip64823}
\subsection{Radial velocity modeling}
In \autoref{sec:results:individual_stars}, we find that the long-period planet candidate HIP~64823 stands out from the rest of the sample as there are currently no indications from the activity analysis or the RV residuals that argue against a planet causing the RV periodicity. Furthermore, the RV variation with a period $P=2878\,\mathrm{d}$ is, by far, the longest among the planet candidates, and far outside the typical range of periods detected to be false positives, which is several hundred days.

The best one-planet model presented in \autoref{fig:multi-rv} has stellar jitter $\sigma_\mathrm{jit}=36.7\,\mathrm{m\,s^{-1}}$, which lies between the jitter estimates $\sigma_\mathrm{jit}=18.0\pm2.3\mathrm{m\,s^{-1}}$ \citep{Kjeldsen1995} and $\sigma_\mathrm{jit}=50.8\pm8.8\mathrm{m\,s^{-1}}$ \citep{Kjeldsen2011}. Thus, we tested whether some residual periodic signal was absorbed into the jitter estimate in the one-planet model. 

Examining the RV residuals in \autoref{fig:residuals}, we find significant periodicity at several periods, with the most prominent peaks consisting of two blended peaks with maxima at $P=410\,\mathrm{d}$ and $P=436\,\mathrm{d}$, as well as other peaks at $P=85.6\,\mathrm{d}$ and $P=15.7\,\mathrm{d}$. We tested a 2-planet model (2P) with a uniform period prior $\mathcal{U}(300\,\mathrm{d}, 500\,\mathrm{d})$ for the inner planet, trying to capture the most significant periodicity. We give the full set of priors in \autoref{tab:HIP64823_priors}. The resulting model attributes a period $P=438\,\mathrm{d}$ to the inner companion. We compare the Bayesian evidence resulting from the nested sampling run and find a moderate improvement of $\Delta\mathcal{Z}=3.41$ \citep[$\Delta\mathrm{BIC}=-2.8$;][]{Trotta2008} compared to the one-planet model. We further find that the estimated RV jitter is only reduced to $\sigma_\mathrm{jit}=31.5\pm3.6\,\mathrm{m\,s^{-1}}$, still exceeding the lower estimate by \citet{Kjeldsen1995}.

Examining the activity MLPs in \autoref{fig:mlp_part2}, however, we find several peaks of the activity indicators in the period region between $100\,\mathrm{d}$ and $500\,\mathrm{d}$, with a tentative peak of the CaIRT2 line at $P=397\,\mathrm{d}$, not far off the residual period of the RVs or the period of the inner planet in the two-planet model. We thus suspect that the periodicity at shorter periods might be linked to stellar activity or another form of intrinsic variability and attempt to model it using a Gaussian process (GP) regression model as implemented via the \texttt{celerite} \citep{Foreman-Mackey2017} package included in the \texttt{Exo-Striker}. While cool spots such as those commonly found for low-mass dwarfs are practically ruled out by the low level of photometric variability ($\mathrm{Hpscat\sim0.005\,\mathrm{mag}}$), other sources of intrinsic variations, such as variable levels of convection or long-period stellar oscillations, are not well understood for giant stars. We thus attempted to model these variations with a commonly used but flexible rotational kernel defined by \citet{Foreman-Mackey2017}. The kernel is characterized by the amplitude of the covariance $B$, the kernel period $P$, the coherence timescale $L$ representing the decay of correlations over time, and a dimensionless factor $C$ that adjusts the relative contributions of the periodic and constant terms. 

In initial tests, using a joint model combining the contributions of the outer planetary candidate and the GP (1P+GP) and broad priors, we found that the GP mostly preferred a kernel period $P\sim460\,\mathrm{d}$ but with small coherence timescales, which tended to overfit the short-term noise inherent to the RVs of the giant due to p-mode pulsations. Similar to \citet[their Fig.~2]{Stock2023}, we observed a triangular shape in the $L$-$P$ plane of the posterior samples. That is, we detected many samples having small $L$ but spanning a large range of periods $P$, as well as a distinct branch of solutions close to $P\sim460\,\mathrm{d}$ extending to larger values of $L$. We thus restricted the lower limit of the prior on $L$ to $450\,\mathrm{d}$ and clipped this plateau.

\begin{figure*}
    \centering
    \includegraphics{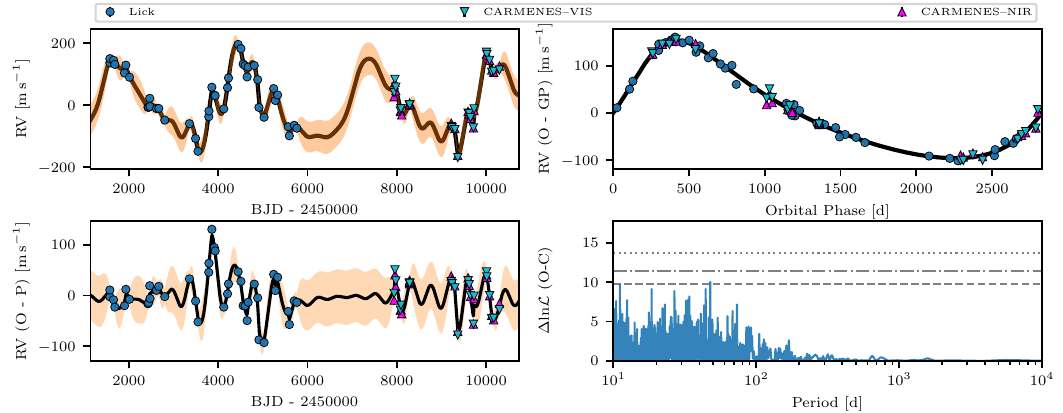}
    \caption{Model combining a long-period planet with a GP model at shorter periods for HIP~64823. Top left: The combined 1P+GP model fitted to the data. Top right: RV residuals after subtracting the GP model, phase-folded to highlight the contribution of the planet. Bottom left: RV residuals after removing the contribution of the planet to highlight the GP model. Bottom right: Residual RV periodogram showing only tentative periodicity at shorter periods. The FAPs of $5\%$ (dashed), $1\%$ (dash-dotted), and $0.1\%$ (dotted) are plotted as gray horizontal lines.}
    \label{fig:hip64823_rv}
\end{figure*}

 We selected a rather constrained prior on the factor $C$ ($\mathcal{U}(0, 1)$), expecting the oscillating component to have a significant contribution. The final 1P+GP model is presented in \autoref{fig:hip64823_rv} (top left), decomposing the contributions of the planet (top right), and the contribution of the GP (bottom left). We find that the Bayesian evidence $\Delta \mathcal{Z} = 33.3$ and $\Delta \mathrm{BIC} = -72.7$ of the 1P+GP model are strongly favored compared to the one-planet model, as well as the two-planet model. We tested that the period prior has an insignificant effect on the Bayesian evidence computed for model comparison with the two-planet model. The residual RV scatter modeled as jitter results in $\sigma_\mathrm{jit}=13.7_{-4.8}^{+4.7}\,\mathrm{m\,s^{-1}}$, which is consistent with the jitter estimate by \citet{Kjeldsen1995}. The determined kernel period is $P_\mathrm{GP}=459_{-38}^{+72}\,\mathrm{d}$. Given the strongly favored 1P+GP model compared to the two-planet model, we find the suspicion justified that the variability at $P \sim400\,\mathrm{d}$ is caused by intrinsic variations rather than another planetary candidate. We give the mode of the posterior distributions in \autoref{tab:64283_params} and provide the corner plot of the posterior samples in \autoref{fig:64823_corner}. The GP parameters are not very well constrained. We specifically find a broad distribution for the hyperparameter $C_\mathrm{GP}$. We stress that we regard HIP~64823 as a planet host candidate due to the insufficient time baseline of the CARMENES observations that cannot entirely rule out that the long-period variations could be caused by intrinsic processes. 

\begin{table}[]
    \centering
    \caption{Results of the 1P+GP nested sampling modeling.}
    \renewcommand{\arraystretch}{1.15}
    \begin{tabular}{llr}

    \hline  \hline \noalign{\vskip 0.15mm}
    Parameter \hspace{0.0 mm}& Unit & Value \\
    \hline \noalign{\vskip 0.15mm}

        $K$ & (m\,s$^{-1}$)             &     $125_{-17}^{+16}$ \\ \noalign{\vskip 0.15mm}
        $P$ & (d)                     &    $2827_{-48}^{+61}$ \\ \noalign{\vskip 0.15mm}
        $e$              &             &       $0.33_{-0.11}^{+0.12}$ \\ \noalign{\vskip 0.15mm}
        $\omega$ & (deg)                &     $315_{-21}^{+20}$ \\ \noalign{\vskip 0.15mm}
        $M_{\rm 0}$ & (deg)             &       $8_{-10}^{+27}$ \\ \noalign{\vskip 0.15mm}
        $\sigma_{\rm jit}$ & (m\,s$^{-1}$)&      $13.7_{-4.8}^{+4.7}$ \\ \noalign{\vskip 0.15mm}
        RV$_{\rm off, Lick}$ & (m\,s$^{-1}$)&     $-49_{-27}^{+26}$ \\ \noalign{\vskip 0.15mm}
        RV$_{\rm off, NIR}$ & (m\,s$^{-1}$)&     $133_{-32}^{+32}$ \\ \noalign{\vskip 0.15mm}
        RV$_{\rm off, VIS}$ & (m\,s$^{-1}$)&    $-124_{-32}^{+32}$ \\ \noalign{\vskip 0.15mm}
        $B_\mathrm{GP}$ & ($\mathrm{(m\,s^{-1})^2}$)       &    $2902_{-1015}^{+1871}$ \\ \noalign{\vskip 0.15mm}
        $P_\mathrm{GP}$ & (d)    &     $459_{-38}^{+72}$ \\ \noalign{\vskip 0.15mm}
        $L_\mathrm{GP}$ & (d) &     $484_{-34}^{+433}$ \\ \noalign{\vskip 0.15mm}
        $C_\mathrm{GP}$      &    &   $0.04_{-0.04}^{+0.74}$ \\ \noalign{\vskip 0.15mm}
        \hline 
        $a_\mathrm{p}$ & (au)                      &       4.40$_{-0.13}^{+0.19}$ \\ \noalign{\vskip 0.15mm}
        $m_\mathrm{p}\sin(i)$ &  ($M_{\rm Jup}$)    &      10.4$_{-1.9}^{+1.4}$ \\ \noalign{\vskip 0.15mm}
        $T_0$ &  (BJD) & $2451510_{-202}^{+92}$ \\ \noalign{\vskip 0.15mm}
    \hline \noalign{\vskip 0.15mm}
\end{tabular}
    \tablefoot{$M_0$ represents the mean anomaly at the epoch of the first data point ($\mathrm{BJD} = 2451572.029297$). $T_0$ denotes the time of periastron passage and was derived from the best fit parameters. For each parameter, we give the mode of the posterior distribution. As uncertainties, we adopt the interval around the mode that encompasses 68\% of the posterior samples. For the GP timescale $L_\mathrm{GP}$, due to the large plateau of low timescale solutions, the mode of the very asymmetric posterior is close to the prior boundary at $450\,\mathrm{d}$. The GP hyperparameter $C_\mathrm{GP}$ is poorly constrained. The full posteriors can be seen in \autoref{fig:64823_corner}. $a_\mathrm{p}$ and $m_\mathrm{p} \sin i$ were derived from the best parameters above. The uncertainties include the uncertainties of the stellar mass. If excluded, the uncertainties reduce to $a_\mathrm{p}=4.40_{-0.04}^{+0.07}\,\mathrm{au}$ and $m_\mathrm{p} \sin(i)=10.4_{-1.7}^{+1.3}\,M_\mathrm{Jup}$.}
    \label{tab:64283_params}
\end{table}

\subsection{Stacked Bayesian GLS analysis}
Another test to distinguish RV periodicity caused by planets from those caused by intrinsic processes is proposed by \citet{Mortier2017}. The idea is to test whether the periodogram power of an RV signal increases by adding new data points as expected from a long-lived, coherent periodic signal. \citet{Mortier2017} use stacked Bayesian GLS (S-BGLS) periodograms based on the formalism by \citet{Mortier2015}. The S-BGLS periodograms have been used successfully to distinguish stellar activity from planets around main-sequence dwarfs or T Tauri stars (see, for instance, \citealt{Zaire2024, VonStauffenberg2024, Burt2024, Dalal2024, Dreizler2024}). The S-BGLS and similar tests have also been successfully applied to rule out the orbital companions around the giant stars Aldebaran \citep{Reichert2019} and Sanders 364 \citep{Zhou2023}.

\begin{figure*}
    \centering
    \includegraphics{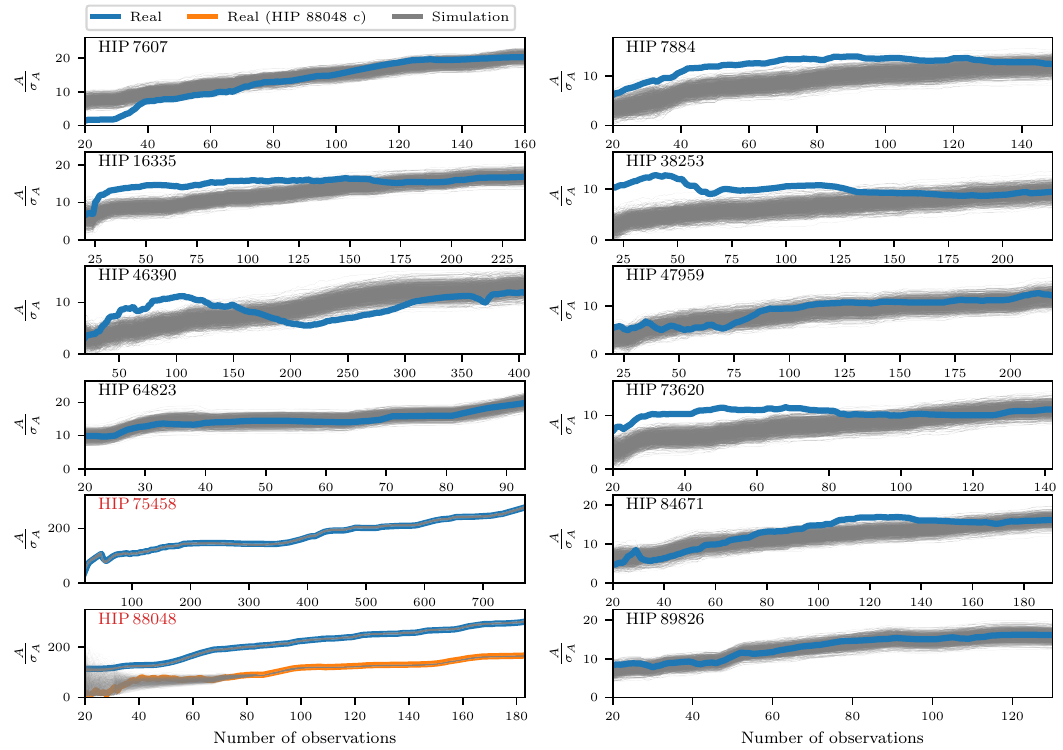}
    \caption{Significance of the amplitude $A$ as a function of the number of observations. The real data are portrayed in blue, while the 1000 realizations of the simulated data are portrayed as gray lines. We evaluated the S-BGLS at the period of the best Keplerian models summarized in \autoref{tab:nest_samp_params}. For the two secure planet hosts, HIP~75458 and HIP~88048 (identifiers marked in red), the simulated relations are plotted on top of the real ones for increased visibility. For HIP~88048, we further plot the significance of the long-period signal in orange. The contributions of the respective other companions were subtracted for the two secure systems.}
    \label{fig:sbgls}
\end{figure*}

However, similar to \citet{Zhou2023}, we find the color map representation of the S-BGLS (see Fig.~1 in \citealt{Mortier2017}) somewhat subjective to interpret. Instead, we opt for an approach similar to that by \citet{Hatzes2013}, \citet{Reichert2019}, and \citet{Zhou2023}, as also proposed by \citet{Mortier2017}. That is, we plot the growth of the significance of the periodic signal, as expressed by the ratio between the amplitude of the sinusoidal $A$, derived by the circular S-BGLS periodogram, and its uncertainty $\sigma_A$, as a function of the number of data points $N$ that have been added in in sequence. The S-BGLS power was evaluated at the period of the best Keplerian model. These are portrayed in \autoref{fig:sbgls}, as the blue lines. Due to the presence of secure long-period orbital companions in the HIP~75458 and HIP~88048 systems, we prewhitened the data sets for these stars by subtracting the contributions of the respective other orbital companions in the analysis. For HIP~88048, we also plot the growth of the signal of HIP~88048~c in orange. For HIP~75458, we omit plotting the change of the significance of the outer long-period companion, as its phase curve is only fractionally sampled.

As can be seen for both published planet hosts, HIP~75458 and HIP~88048, the significance of $A$ can drop significantly in parts of the data set. However, these can be the results of an inadequate sampling of the RV phase curve in the presence of stellar jitter. For HIP~75458, for instance, the SONG spectrograph took numerous spectra in subsequent nights. Coupled with the eccentric orbit of HIP~75458~b and the sizable stellar jitter, these spectra in practice mostly contribute stellar ``noise'' and reduce the significance of the determined RV signal. This also helps one understand why the color maps of the S-BGLS analysis can sometimes be misleading for giant stars. 

To study how the significance of $A$ would evolve if the Keplerian model can adequately describe the data, we simulated 1000 realizations of the Keplerian model evaluated at the epochs of the real spectra and adding a Gaussian noise component with the standard deviation set to the RV jitter of the star. The latter was also used as a white noise component for the weights to compute the S-BGLS (see \citealt{Mortier2017}). These simulations are plotted as the gray lines. As one can see for both HIP~75458 and HIP~88048, the simulated significances follow the real data very closely, including the drop of the significance in certain parts of the relation. We note the change of the significance of $A$ for HIP~88048~c (orange) is initially very noisy and behaves slightly differently than the simulated relations. However, this can easily be explained by the long-period nature of the orbit, which has the longest period among all systems apart from HIP~75458~c, as well as the relative dense sampling of the early Lick observations (see \autoref{fig:multi-rv}) compared to other long-period planet candidates such as HIP~64823.  

To allow for a fair comparison of the model of HIP~64823 against the other planet candidates, we used the one-planet model with its respective jitter as given in \autoref{tab:nest_samp_params}. As can be seen from \autoref{fig:sbgls}, HIP~64823, along with HIP~89826, are the only planet candidates for which there are no striking differences between the real data and the simulated relations. For all other stars, there are parts of the relation in which the real data behave differently than the model. The relation for HIP~47959 does not deviate as strongly from the simulations as for most other planet candidates. Nevertheless, especially in the early part of the relation, the significance of the real data increases less quickly than the simulations predict. We note that, by definition, the simulated and real significances converge at large $N$, as the Keplerian models were derived using the complete data sets. 

It has to be noted that the test, in essence, can only examine whether a specific Keplerian model including its RV jitter is consistent with the whole data set. Thus, altering the Keplerian models, for instance by including additional orbital components, which are however challenged based on dynamical assumptions, could alter the interpretation of this specific S-BGLS test. Nevertheless, having tested different models for the planet candidates we could not find any differences in the interpretation. We thus conclude that HIP~64823 and HIP~89826 are the only planet candidates that can be consistently described by the Keplerian models. For the latter, the CARMENES data are not yet sufficient to draw reliable conclusions. For HIP~64823, we conclude that the exoplanet of minimum mass $m_\mathrm{p}\sin(i)=10.4_{-1.9}^{+1.4}\,M_\mathrm{Jup}$ with orbital period $P = 2827_{-48}^{+61}\,\mathrm{d} \sim 7.75\,\mathrm{yr}$ is likely real. Nevertheless, as the baseline of the CARMENES observations is still too short to properly test the activity indicators and given the general poor understanding of the physical nature of long-period intrinsic variations of giant stars, we regard HIP~64823~b as a planet candidate until confirmed by further RV monitoring. 

\section{Discussion}
\label{sec:discussion}
In recent years, a number of intrinsically variable giant stars, some of which had previously been reported to host planets using RV data, have been identified \citep{Hatzes2018, DelgadoMena2018, DelgadoMena2023, Reichert2019, Heeren2021, Dollinger2021, Teng2023a, Zhou2023, Lee2023a}. Given the varying temporal baselines of RV data sets and availability of activity indicators, there could be more false positives among luminous evolved planet hosts. These could have an impact on occurrence rate studies if not treated carefully. \citet{Wolthoff2022}, for instance, perform several ``homogenizing cuts'' to their underlying sample to remove very luminous ($\log_{10}(L/L_\sun) > 3$) giants, as well as stars with complex RV signatures. Thus, they reduce the impact false-positive planet detections could have on their conclusions. Despite this careful treatment, the published ``planet host'' HIP~16335 \citep{Lee2014a} is included as such in their analysis, although we now conclude that its variations are intrinsic.

Within the literature, there is no consensus which process causes these false-positive detections. One candidate could be the rotational modulation of stellar surface features such as hot plages or cool spots. Unfortunately, for most of the stars there are no reliable estimates of the rotation periods due to a lack of photometric variations (see also \autoref{tab:params}). This lack of variations, however, also generally argues against large stellar spots \citep{Reffert2015, Reichert2019}.

On the other hand, the variations of surface features linked to magnetic activity are yet largely unexplored. As suggested by \citet[and references therein]{DelgadoMena2023}, magnetic structures on giant stars could locally suppress convection, potentially leading to RV variations without associated photometric variations.
 This idea seems plausible, given the detection of a weak magnetic surface field of \object{Pollux} \citep{Auriere2009, Auriere2014, Auriere2021}, which varies on similar timescales as the RVs. These RV variations have been attributed to an exoplanet companion \citep{Hatzes2006, Reffert2006}. Recent simulations show that this magnetic field could be dynamo-driven and that the strength of the magnetic field is coupled to the scale of convection cells \citep{Amard2024}. \citet{Rolo2024} discuss a connection between the amplitudes of the activity indicator variations and the Li abundance of the stars. Their work includes the intrinsically variable giants discussed in \citet{DelgadoMena2023}, for several of which the H$\alpha$ index is shown to be periodic close to the RV period. All of these stars are Li-rich, potentially providing another link toward magnetic activity, as high Li abundance is associated with enhanced chromospheric activity \citep{Sneden2022}. Interestingly, strong magnetic fields have also been detected in the cores of some giant stars using \textit{Kepler} data \citep{Li2022, Deheuvels2023, Li2023}. Of the 12 stars in this sample, literature Li abundance measurements are available for eleven. None of them were found to be Li-rich. \citep{Brown1989, Lebre2006, Liu2014, Charbonnel2020}.

In this context, we find it noteworthy that the chromospheric activity indicators, namely those targeting the H$\alpha$, NaD, and CaIRT lines, which are sensitive to magnetic activity \citep{Sarkis2018}, show at least tentative periodic variations at periods similar to the RVs for several of the stars examined in this work. Thus, a connection between magnetic activity and RV variations resembling planets seems plausible. However, further modeling of the RV curves remains challenging lacking a better physical understanding of the magnetic surface fields and convection patterns of giants. It also remains uncertain whether such variations in the chromospheric indicators might also be attributed to other intrinsic mechanisms.

Another promising alternative, long-period non-radial oscillations, was already discussed for giant stars by \citet{Hatzes1993}. More recently, \citet{Hatzes2018} propose that the RV variations of the luminous giant $\gamma$~Dra show a beating of different periods that seem to be most elegantly explained by non-radial oscillations. \citet{Hatzes2018} also propose that this new type of variability might be linked to oscillatory convective modes which have been developed by \citet{Saio2015} to explain the so-called sequence D in the period-luminosity plot of bright, variable giants showing long-secondary periods (LSPs) \citep{Wood1999}. Although contested in the literature as the origin of these LSPs (see, for instance, \citealt{Soszynski2014, Soszynski2021, Goldberg2024}), \citet{Hatzes2018} and \citet{Reichert2019} argue that $\gamma$~Dra and Aldebaran occupy a region in a period--luminosity plot that could be crossed by an extrapolation of the models by \citet{Saio2015}. This could make oscillatory convective modes a plausible explanation for the RV variations. In \citet{Spaeth2024}, we use this idea to show that a simple model of an $l=1, m=1$ non-radial oscillation could explain the RV variations, as well as the correlations with the available activity indicator, for the known false-positive evolved planet host NGC~4349~No.~127 \citep{DelgadoMena2018, DelgadoMena2023}. Although more massive, NGC 4349 No.\ 127 has a similar luminosity to that of $\gamma$~Dra and Aldebaran, and shows RV variations of similar period.

\begin{figure*}
    \centering
    \includegraphics{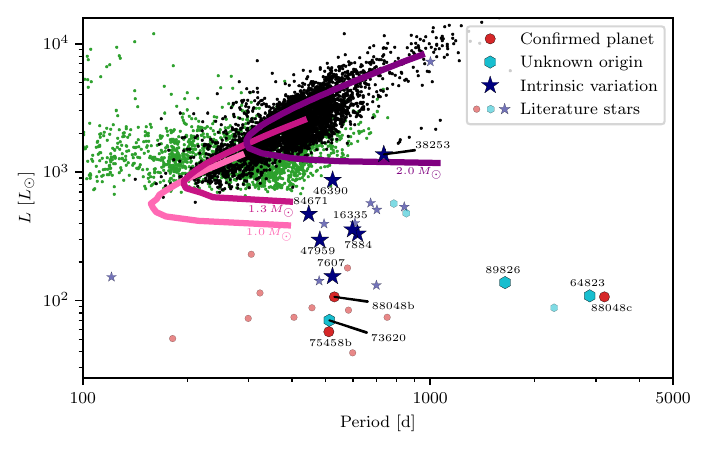}
    \caption{Luminosity of the stars plotted against the RV period. Variable stars from the OGLE survey are plotted in green, with the subset of those showing LSPs on sequence D plotted in black. Overplotted in pink are models from the expected relations of oscillatory convective modes \citep{Saio2015} for different stellar masses. We plot confirmed planets within the Lick sample as red circles, and confirmed intrinsically RV-variable stars that mimic planets as blue star markers. Stars with as yet undecided origins of their RV variations are shown as cyan hexagons. Stars taken from the literature are shown with reduced marker sizes and lighter shades. We mark the \textsc{Hipparcos} identifiers of the stars discussed in this work. We omit plotting the short-period planet hosts \object{HIP~89587} \citep{Teng2023} and \object{HIP~70791} \citep{Takarada2018} for visual clarity. The plot is adapted from Fig.~8 of \citet{Reichert2019} and Fig.~9 of \citet{Spaeth2024}.}
    \label{fig:LP}
\end{figure*}

Having identified seven additional giant stars with RV variations that resemble planets (at least during part of their temporal baseline) that we regard to be securely of intrinsic origin, we can update Fig.~8 of \citet{Reichert2019} (and Fig.~9 in \citealt{Spaeth2024}). In \autoref{fig:LP}, we show luminous and variable giant stars identified in the OGLE-III survey \citep{Soszynski2009} as small green dots, with the subset of those showing LSPs on sequence D in black. We further plot models of oscillatory convective modes for different stellar masses taken from \citet{Saio2015} in different shades of pink. On top, we present published planet hosts from the Lick survey as red circles, stars that show RV curves resembling planets but have been shown to be of intrinsic origin as blue star markers, and remaining planet candidates, or stars for which we regard the analysis to not yet provide clear conclusions, as cyan hexagons. We highlight the stars examined within this work, marking these with their \textsc{Hipparcos} identifiers, while we show the literature stars with somewhat reduced sizes and lighter shades. We exclude the lower-luminosity false-positive hosts \object{41~Lyn} and \object{14~And} \citep{Teng2023a} from this analysis. The RV periodicity of both stars closely aligns with a half-yearly cycle, accompanied by a strong yearly signal in the window function. These characteristics suggest that the observed periodicities are more likely related to the observational sampling than to a physical process.

As can be seen, the clear distinction between the secure planet hosts and the false-positive detections highlighted by \citet{Reichert2019} becomes somewhat less striking as more stars are included. While we can still observe a clear trend that the false-positive detections mostly occur for larger luminosities, we can see several such false positives also in the transition region, among these HIP~7607. We note that one of the supposedly confirmed planet hosts in the transition region, HIP~60202 \citep{Liu2008}, was recently challenged by \citet{Teng2023a} due to additional periodicity in the residuals RVs that would be challenging to attribute to an additional companion due to dynamical constraints, akin to the results of the residual analysis in \autoref{sec:results:coherence} in this work. We also find HIP~16335, whose published planet is likely a misidentified intrinsic signal according to our analysis, to be in close vicinity to other intrinsically variable giants, including $\gamma$~Dra.  The two stars with RV variations of unknown origin at high luminosities, \object{3 Cnc} and \object{44 UMa} \citep{TalaPinto2020}, have convincing Keplerian models but lack a set of reliable activity indicators other than Lick H$\alpha$ measurements, which were inconspicuous for these stars. Noteworthy is also the location of HIP~73620, whose RV variations fall within the region occupied by stars who are considered to be secure planet hosts. Although we favor an intrinsic origin of the RV variations, we cannot rule out a planetary companion with certainty. However, assuming our assessment of an intrinsic origin to be correct, this would certainly make the distinction between the groups of the secure hosts and the intrinsically RV-variable stars less clear. However, we also note that HIP~73620 shows the smallest amplitude among the stars in this study. 

Finally, we note the locations of HIP~64823 and HIP~89826, which fall far outside the period regime in which the other false positives have been found. In terms of their luminosity they fall into the transition region between the secure planets and the RV-variable stars. At least for HIP~64823, this further supports our interpretation that the RV variations are best explained by a planetary companion. HIP~89826 remains uncertain.

\begin{figure}
    \centering
    \includegraphics{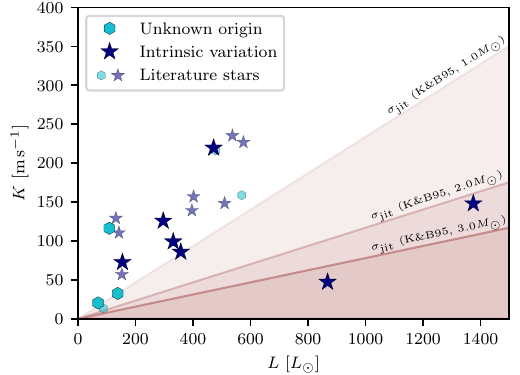}
    \caption{Semi-amplitude $K$ of the intrinsically RV-variable stars (blue star markers) and the stars with as yet unknown origin of their RV variations (cyan hexagons) plotted against the luminosity. The literature stars are plotted with reduced sizes. We further plot the estimated p-mode jitter following the scaling relations by \citet{Kjeldsen1995}, labeled K\&B95, for three different stellar masses. We observe that many of the stars follow a roughly linear relation between the amplitude of their RV variation and the luminosity. Two outliers at large $L$ but small $K$ are noted. These are HIP~46390 and HIP~38253. We exclude the very luminous star NGC 2345 No.\ 50 from the plot for visual clarity.}
    \label{fig:k_vs_L}
\end{figure}

In \autoref{fig:k_vs_L}, we select all stars that are intrinsically RV variable and those of so-far unknown origin from \autoref{fig:LP}, and plot their semi-amplitude $K$ against their luminosity. We observe that the majority of the stars follows a roughly linear trend with increased $K$ for higher luminosities. We note two strong outliers at high luminosities, which are HIP~46390 and HIP~38253. Especially for the latter, the amplitude in the Lick RVs alone would be somewhat larger than the amplitude derived combining the Lick data with the SONG and CARMENES measurements (see \autoref{fig:multi-rv}), such that it might also lie at higher $K$ depending on the part of the RV curve that was observed. We emphasize that we also include more likely planet candidates, such as HIP~64823, among the stars of unknown origin.

We further mark the estimated p-mode jitter derived using the scaling relations by \citet{Kjeldsen1995} for three different stellar masses in shades of red. These give a rough lower limit of semi-amplitude $K$ that could have been detected. That is, it is possible that the linear trend is merely a consequence of selection effects due to the higher short-term jitter that would likely prevent the detection of longer-period RV variability for more luminous stars. Nevertheless, we find the observed trend noteworthy as it might contribute toward a physical interpretation. A similar trend (expressed in terms of $\log\,g$, which, however, correlates with $L$ and that can also be observed for our stars) has already been discussed by \citet{Hatzes1998a}.
 
Overall, we have to conclude that the full picture remains elusive. We can certainly strengthen the finding that planet detections around giant stars with luminosities $L \gtrsim 125\,L_\sun$ should be regarded with caution. We, however, also note the suspicion that HIP~73620, which has much lower luminosity, has intrinsically induced RV variations. While we can somewhat strengthen the suggestions by \citet{Hatzes2018} and \citet{Reichert2019}, that an extrapolation of the models of oscillatory convective modes could explain the majority of the false-positive sample, we also find hints of variations in the chromospheric activity indicators for several of the stars, which might suggest a link toward magnetic activity of, perhaps, rather exotic nature. Unlike the case of NGC 4349 No.\ 127 \citep{DelgadoMena2018, DelgadoMena2023, Spaeth2024}, we do not find distinctive correlations between the activity indicators and the RVs for any of our stars, which could be easily interpreted as the fingerprints of a single dominating oscillation mode. However, we discuss in \citet{Spaeth2024} that a superposition of different modes along with sizable stellar jitter can hide such correlations. We also note that NGC 4349 No.\ 127 was also shown to have periodic variations in the H$\alpha$ indicator \citep{DelgadoMena2018, DelgadoMena2023}, which might suggest that such chromospheric activity could also be linked to oscillations. More detailed investigations, both from the observational and the theoretical sides, are certainly necessary to resolve the mysterious RV variations of luminous evolved stars. 

\section{Summary}

\label{sec:summary}
The goal of this study was to test ten bright evolved stars identified as planet candidates after the conclusion of the 12-year Lick RV survey for the presence of planets using the SONG and CARMENES spectrographs. We further aimed to assess the effectiveness of different standard metrics used in the literature to differentiate between planets and intrinsic variations when applied to giant stars. Finally, we were motivated to also highlight the false-positive detections in order to provide new evidence of the observational fingerprints of the mysterious RV variations of luminous giant stars.

We find that seven out of the ten former planet candidate stars have an intrinsic origin of their RV variations. These are HIP~7607, HIP~7884, HIP~38253, HIP~46390, HIP~47959, HIP~84671, and (most notably) HIP~16335. Significantly, HIP~16335 was previously reported by \citet{Lee2014a} to host a $6.5\,M_\mathrm{Jup}$ minimum mass giant exoplanet in a $579.8\,\mathrm{d}$ orbit, which we reveal to be a misidentified intrinsic RV signal. Furthermore, we lean toward an intrinsic origin for the (comparably) low-amplitude RV variations of HIP~73620, whereas we cannot draw reliable conclusions for HIP~89826. Finally, we find that HIP~64823 remains a promising candidate to host a long-period planet ($P\sim2827_{-48}^{+61}\,\mathrm{d}$), and we showed that a GP model can be used to explain residual variability at shorter periodicity. However, as the baseline of the CARMENES observations is not yet long enough, we cannot confirm the presence of the exoplanet candidate with minimum mass $m_\mathrm{p}\sin(i)=10.4_{-1.9}^{+1.4}\,M_\mathrm{Jup}$ with certainty. 

Although the RV variations of most of the intrinsically variable giants are convincingly not caused by exoplanet companions, we find that testing the consistency of the RV amplitude between the optical and the infrared is not effective enough to rule out the possibility that the giants are being orbited by planets given the large stellar jitter. We further find that testing for linear correlations between the activity indicators and the RVs does not detect the majority of the false-positive planet hosts and thus caution relying solely on such metrics. We instead find that for the most luminous stars, clear changes of the dominating periodicity on timescales of tens of years can be observed. A similar effect is also apparent from analyzing the residual RVs, which often show significant remaining periodicity close to the proposed orbital period, which we interpret as the consequences of changing periods or amplitudes. We also observed that testing the consistency of the best Keplerian model with the real data, based on the significance of the amplitude derived from stacked Bayesian GLS periodograms, reliably picks out the false-positive hosts. For such tests, long-term RV monitoring is critical.

Generally, we find that analyzing the CARMENES activity indicators using maximum likelihood periodograms as well as a \texttt{DBSCAN} cluster analysis can, in most cases, detect the intrinsic variability. However, a detailed analysis combining the findings from different indicators is necessary to rule out planetary companions with certainty, as the signals in the activity indicators are typically not very strong, meaning that they can be easily missed. We also observed that even for the well-studied exoplanet host HIP~75458, there are hints in the activity indicators and RV residuals that suggest the current metrics may not be entirely robust at differentiating between planetary and stellar activity signals. 

Finally, we discuss the additional detection of intrinsically RV-variable giant stars mimicking exoplanets in the context of other false-positive evolved planet hosts. We find that the general trend that most false positives have high luminosities holds up, but we also found one suspicious case at lower luminosity. We observed a slight trend of increasing RV amplitudes with increasing luminosity. Given the high luminosities of the majority of the stars, a connection to oscillatory convective modes \citep{Saio2015} seems plausible. However, we do not find any clear correlations between the activity indicators and the RVs in this sample, which would allow for a more detailed assessment of potential oscillation modes \citep{Spaeth2024}. We also note that for several of the stars in the sample and in the literature, chromospheric activity indicators contribute to refuting the planetary companions. These might suggest a connection to magnetic or convective processes in these giants. The phenomenon as a whole remains poorly understood. 

With this work, we hope to contribute to a better physical understanding of RV variations of luminous giant stars, which convincingly mimic exoplanet companions. Future research both from the observational and theoretical side are certainly necessary to ultimately resolve the mystery.

\begin{acknowledgements}
      We thank the anonymous referee for their valuable feedback, which helped improve the quality of this work. This work was supported by the Deutsche Forschungsgemeinschaft (DFG) within the Priority Program SPP 1992 “The Diversity of Exoplanets” (RE 2694/7-1). T.T. acknowledges support by the BNSF program "VIHREN-2021" project No. KP-06-DV/5. We would like to thank the staff at Lick observatory for their support and thank the CAT observers who assisted with this project, including David Mitchell, Saskia Hekker, Christian Schwab, Christoph Bergmann, Kirsten Vincke, Julian Stürmer, Kelsey Clubb, David Bauer, Dominika Wylezalek, Dennis Kügler, and Debra Fischer. This work is based on observations made with the Hertzsprung SONG telescope operated on the Spanish Observatorio del Teide on the island of Tenerife by the Aarhus and Copenhagen Universities and by the Instituto de Astrofísica de Canarias. It is further based on observations collected at the Centro Astronómico Hispano en Andalucía (CAHA) at Calar Alto, proposals F16-3.5-024, H17-3.5-014, F18-3.5-016, H18-3.5-015, 21A-3.5-051, 21B-3.5-052, 22A-3.5-051,  22B-3.5-051, 23A-3.5-051, 23B-3.5-052, operated jointly by Junta de Andalucía and Consejo Superior de Investigaciones Científicas (IAA-CSIC). Based on data from the CAHA Archive at CAB (INTA-CSIC).\footnote{\url{http://cab.inta.es/}} The CAHA Archive is part of the Spanish Virtual Observatory\footnote{\url{http://svo.cab.inta-csic.es/}} project funded by MCIN/AEI/10.13039/501100011033 through grant PID2020-112949GB-I00. This project has received funding from the European Union’s Horizon 2020 research and innovation programme under grant agreement No 101004719. This research has made use of the VizieR catalog access tool and the SIMBAD database, operated at CDS, Strasbourg, France. This work has made use of data from the European Space Agency (ESA) mission {\it Gaia} (\url{https://www.cosmos.esa.int/gaia}), processed by the {\it Gaia} Data Processing and Analysis Consortium (DPAC, \url{https://www.cosmos.esa.int/web/gaia/dpac/consortium}). Funding for the DPAC has been provided by national institutions, in particular the institutions participating in the {\it Gaia} Multilateral Agreement. This research made use of Astropy,\footnote{\url{http://www.astropy.org}} a community-developed core Python package for Astronomy . We further acknowledge use of the python packages NumPy \citep{Harris2020}, SciPy \citep{Virtanen2020}, Matplotlib \citep{Hunter2007}, seaborn \citep{Waskom2021}, pandas \citep{pandas2024, McKinney2010}, PyAstronomy\footnote{\url{https://github.com/sczesla/PyAstronomy}} \citep{Czesla2019}, lmfit \citep{Newville2021}, {\sc dynesty} \citep{Speagle2020}, \texttt{celerite} \citep{Foreman-Mackey2017}, corner \citep{corner}, \texttt{scikit-learn} \citep{Pedregosa2011}, bgls \citep{Mortier2015}, astroquery \citep{Ginsburg2019}, dill \citep{McKerns2012}, and the RV fitting tool \texttt{Exo-Striker} \citep{Trifonov2019a}.
\end{acknowledgements}

%
%
\bibpunct{(}{)}{;}{a}{}{,} 
\bibliographystyle{aa} 
\bibliography{Literature.bib}

\begin{thebibliography}{153}
\expandafter\ifx\csname natexlab\endcsname\relax\def\natexlab#1{#1}\fi

\bibitem[{Alves(2000)}]{Alves2000}
Alves, D.~R. 2000, ApJ, 539, 732

\bibitem[{Amard {et~al.}(2024)Amard, Brun, \& Palacios}]{Amard2024}
Amard, L., Brun, A.~S., \& Palacios, A. 2024, ApJ, 974, 311

\bibitem[{Andersen {et~al.}(2014)Andersen, Grundahl, {Christensen-Dalsgaard}, Frandsen, J{\o}rgensen, Kjeldsen, Pall{\'e}, Skottfelt, S{\o}rensen, \& Weiss}]{Andersen2014}
Andersen, M.~F., Grundahl, F., {Christensen-Dalsgaard}, J., {et~al.} 2014, in Rev. Mex. Astron. Astrofis. Conf. Ser., Vol.~45, 83

\bibitem[{Assef {et~al.}(2009)Assef, Gaudi, \& Stanek}]{Assef2009}
Assef, R.~J., Gaudi, B.~S., \& Stanek, K.~Z. 2009, ApJ, 701, 1616

\bibitem[{{Auri{\`e}re} {et~al.}(2014){Auri{\`e}re}, {Konstantinova-Antova}, {Espagnet}, {Petit}, {Roudier}, {Charbonnel}, {Donati}, \& {Wade}}]{Auriere2014}
{Auri{\`e}re}, M., {Konstantinova-Antova}, R., {Espagnet}, O., {et~al.} 2014, in IAU Symposium, Vol. 302, Magnetic Fields throughout Stellar Evolution, ed. P.~{Petit}, M.~{Jardine}, \& H.~C. {Spruit}, 359--362

\bibitem[{Auri{\`e}re {et~al.}(2021)Auri{\`e}re, Petit, Mathias, {Konstantinova-Antova}, Charbonnel, Donati, Espagnet, Folsom, Roudier, \& Wade}]{Auriere2021}
Auri{\`e}re, M., Petit, P., Mathias, P., {et~al.} 2021, A\&A, 646, A130

\bibitem[{Auri{\`e}re {et~al.}(2009)Auri{\`e}re, Wade, {Konstantinova-Antova}, Charbonnel, Catala, Weiss, Roudier, Petit, Donati, Alecian, Cabanac, {van Eck}, Folsom, \& Power}]{Auriere2009}
Auri{\`e}re, M., Wade, G.~A., {Konstantinova-Antova}, R., {et~al.} 2009, A\&A, 504, 231

\bibitem[{Boisse {et~al.}(2009)Boisse, Moutou, {Vidal-Madjar}, Bouchy, Pont, H{\'e}brard, Bonfils, Croll, Delfosse, Desort, Forveille, Lagrange, Loeillet, Lovis, Matthews, Mayor, Pepe, Perrier, Queloz, Rowe, Santos, S{\'e}gransan, \& Udry}]{Boisse2009}
Boisse, I., Moutou, C., {Vidal-Madjar}, A., {et~al.} 2009, A\&A, 495, 959

\bibitem[{Bressan {et~al.}(2012)Bressan, Marigo, Girardi, Salasnich, Dal~Cero, Rubele, \& Nanni}]{Bressan2012}
Bressan, A., Marigo, P., Girardi, {\relax L{\'e}o}., {et~al.} 2012, MNRAS, 427, 127

\bibitem[{Brown {et~al.}(1989)Brown, Sneden, Lambert, \& Dutchover}]{Brown1989}
Brown, J.~A., Sneden, C., Lambert, D.~L., \& Dutchover, Jr., E. 1989, ApJS, 71, 293

\bibitem[{Burt {et~al.}(2024)Burt, Hooton, Mamajek, Barrag{\'a}n, Millholland, Fairnington, Fisher, Halverson, Huang, Brady, Seifahrt, Gaidos, Luque, Kasper, \& Bean}]{Burt2024}
Burt, J.~A., Hooton, M.~J., Mamajek, E.~E., {et~al.} 2024, ApJ, 971, L12

\bibitem[{Butler {et~al.}(1996)Butler, Marcy, Williams, McCarthy, Dosanjh, \& Vogt}]{Butler1996}
Butler, R.~P., Marcy, G.~W., Williams, E., {et~al.} 1996, PASP, 108, 500

\bibitem[{Campante {et~al.}(2023)Campante, Li, Ong, Corsaro, Cunha, Bedding, Bossini, Breton, Buzasi, Chaplin, Deal, Garc{\'i}a, Hill, Hon, Huber, Jiang, Kane, Kayhan, Kuszlewicz, {Lillo-Box}, Mathur, Monteiro, Pereira, Santos, Serenelli, \& Stello}]{Campante2023}
Campante, T.~L., Li, T., Ong, J. M.~J., {et~al.} 2023, AJ, 165, 214

\bibitem[{Charbonnel {et~al.}(2020)Charbonnel, Lagarde, Jasniewicz, North, Shetrone, Krugler~Hollek, Smith, Smiljanic, Palacios, \& Ottoni}]{Charbonnel2020}
Charbonnel, C., Lagarde, N., Jasniewicz, G., {et~al.} 2020, A\&A, 633, A34

\bibitem[{Cox {et~al.}(1972)Cox, King, \& Stellingwerf}]{Cox1972}
Cox, J.~P., King, D.~S., \& Stellingwerf, R.~F. 1972, ApJ, 171, 93

\bibitem[{Czesla {et~al.}(2018)Czesla, Molle, \& Schmitt}]{Czesla2018}
Czesla, S., Molle, T., \& Schmitt, J. H. M.~M. 2018, A\&A, 609, A39

\bibitem[{Czesla {et~al.}(2019)Czesla, Schr{\"o}ter, Schneider, Huber, Pfeifer, Andreasen, \& Zechmeister}]{Czesla2019}
Czesla, S., Schr{\"o}ter, S., Schneider, C.~P., {et~al.} 2019, {{PyA}}: {{Python}} Astronomy-Related Packages

\bibitem[{Dalal {et~al.}(2024)Dalal, Rescigno, Cretignier, Anna~John, Majidi, Malavolta, Mortier, Pinamonti, Buchhave, Haywood, Sozzetti, Dumusque, Lienhard, Rice, Vanderburg, Lakeland, Bonomo, Cameron, Damasso, Affer, Boschin, Cooke, Cosentino, Fabrizio, Ghedina, Harutyunyan, Latham, {L{\'o}pez-Morales}, Lovis, Fiorenzano, Mayor, Nicholson, Pepe, Stalport, Udry, Watson, \& Wilson}]{Dalal2024}
Dalal, S., Rescigno, F., Cretignier, M., {et~al.} 2024, MNRAS, 531, 4464

\bibitem[{Deheuvels {et~al.}(2023)Deheuvels, Li, Ballot, \& Ligni{\`e}res}]{Deheuvels2023}
Deheuvels, S., Li, G., Ballot, J., \& Ligni{\`e}res, F. 2023, A\&A, 670, L16

\bibitem[{Delgado~Mena {et~al.}(2023)Delgado~Mena, {Gomes da Silva}, Faria, Santos, Martins, Tsantaki, Mortier, Sousa, \& Lovis}]{DelgadoMena2023}
Delgado~Mena, E., {Gomes da Silva}, J., Faria, J.~P., {et~al.} 2023, A\&A, 679, A94

\bibitem[{Delgado~Mena {et~al.}(2018)Delgado~Mena, Lovis, Santos, {Gomes da Silva}, Mortier, Tsantaki, Sousa, Figueira, Cunha, Campante, Adibekyan, Faria, \& Montalto}]{DelgadoMena2018}
Delgado~Mena, E., Lovis, C., Santos, N.~C., {et~al.} 2018, A\&A, 619, A2

\bibitem[{Desort {et~al.}(2007)Desort, Lagrange, Galland, Udry, \& Mayor}]{Desort2007}
Desort, M., Lagrange, A.~M., Galland, F., Udry, S., \& Mayor, M. 2007, A\&A, 473, 983

\bibitem[{D{\'i}az {et~al.}(2007)D{\'i}az, Cincunegui, \& Mauas}]{Diaz2007}
D{\'i}az, R.~F., Cincunegui, C., \& Mauas, P. J.~D. 2007, MNRAS, 378, 1007

\bibitem[{D{\"o}llinger \& Hartmann(2021)}]{Dollinger2021}
D{\"o}llinger, M.~P. \& Hartmann, M. 2021, ApJS, 256, 10

\bibitem[{Dreizler {et~al.}(2024)Dreizler, Luque, Ribas, Koseleva, Ruh, Nagel, Pozuelos, Zechmeister, Reiners, Caballero, Amado, B{\'e}jar, Bean, Brady, Cifuentes, Gillon, Hatzes, Henning, Kasper, Montes, Morales, Murray, Pall{\'e}, Quirrenbach, Seifahrt, Schweitzer, St{\"u}rmer, Stef{\'a}nsson, \& Linares}]{Dreizler2024}
Dreizler, S., Luque, R., Ribas, I., {et~al.} 2024, A\&A, 684, A117

\bibitem[{Ernst(2004)}]{Ernst2004}
Ernst, M.~D. 2004, Stat. Sci., 19, 676

\bibitem[{{ESA}(1997)}]{ESA1997}
{ESA}. 1997, ESA Spec. Publ., 1200

\bibitem[{Ester {et~al.}(1996)Ester, Kriegel, Sander, \& Xu}]{Ester1996}
Ester, M., Kriegel, H.-P., Sander, J., \& Xu, X. 1996, in Proc 2nd {{Int}}. {{Conf}}. {{Knowl}}. {{Discov}}., 226--231

\bibitem[{Feroz {et~al.}(2009)Feroz, Hobson, \& Bridges}]{Feroz2009}
Feroz, F., Hobson, M.~P., \& Bridges, M. 2009, MNRAS, 398, 1601

\bibitem[{Fischer {et~al.}(2013)Fischer, Marcy, \& Spronck}]{Fischer2013}
Fischer, D.~A., Marcy, G.~W., \& Spronck, J. F.~P. 2013, ApJS, 210, 5

\bibitem[{Fischer \& Valenti(2005)}]{Fischer2005}
Fischer, D.~A. \& Valenti, J. 2005, ApJ, 622, 1102

\bibitem[{{Foreman-Mackey}(2016)}]{corner}
{Foreman-Mackey}, D. 2016, J. Open Source Softw., 1, 24

\bibitem[{{Foreman-Mackey} {et~al.}(2017){Foreman-Mackey}, Agol, Ambikasaran, \& Angus}]{Foreman-Mackey2017}
{Foreman-Mackey}, D., Agol, E., Ambikasaran, S., \& Angus, R. 2017, AJ, 154, 220

\bibitem[{Fredslund~Andersen {et~al.}(2019)Fredslund~Andersen, Handberg, Weiss, Frandsen, {Sim{\'o}n-D{\'i}az}, Grundahl, \& Pall{\'e}}]{FredslundAndersen2019}
Fredslund~Andersen, M., Handberg, R., Weiss, E., {et~al.} 2019, PASP, 131, 045003

\bibitem[{Frink {et~al.}(2002)Frink, Mitchell, Quirrenbach, Fischer, \& Marcy}]{Frink2002}
Frink, S., Mitchell, D.~S., Quirrenbach, A., Fischer, D.~A., \& Marcy, G.~W. 2002, ApJ, 576, 7

\bibitem[{Frink {et~al.}(2001)Frink, Quirrenbach, Fischer, R{\"o}ser, \& Schilbach}]{Frink2001}
Frink, S., Quirrenbach, A., Fischer, D., R{\"o}ser, S., \& Schilbach, E. 2001, PASP, 113, 173

\bibitem[{{Gaia Collaboration} {et~al.}(2016){Gaia Collaboration}, Prusti, {de Bruijne}, Brown, Vallenari, Babusiaux, {Bailer-Jones}, Bastian, Biermann, Evans, Eyer, Jansen, Jordi, Klioner, Lammers, Lindegren, Luri, Mignard, Milligan, Panem, Poinsignon, Pourbaix, Randich, Sarri, Sartoretti, Siddiqui, Soubiran, Valette, {van Leeuwen}, Walton, Aerts, Arenou, Cropper, Drimmel, H{\o}g, Katz, Lattanzi, O'Mullane, Grebel, Holland, Huc, Passot, Bramante, Cacciari, Casta{\~n}eda, Chaoul, Cheek, De~Angeli, Fabricius, Guerra, Hern{\'a}ndez, {Jean-Antoine-Piccolo}, Masana, Messineo, Mowlavi, Nienartowicz, {Ord{\'o}{\~n}ez-Blanco}, Panuzzo, Portell, Richards, Riello, Seabroke, Tanga, Th{\'e}venin, Torra, Els, {Gracia-Abril}, Comoretto, {Garcia-Reinaldos}, Lock, Mercier, Altmann, Andrae, Astraatmadja, {Bellas-Velidis}, Benson, Berthier, Blomme, Busso, Carry, Cellino, Clementini, Cowell, Creevey, Cuypers, Davidson, De~Ridder, {de Torres}, Delchambre, Dell'Oro, Ducourant, Fr{\'e}mat, {Garc{\'i}a-Torres}, Gosset, Halbwachs, Hambly, Harrison, Hauser, Hestroffer, Hodgkin, Huckle, Hutton, Jasniewicz, Jordan, Kontizas, Korn, Lanzafame, Manteiga, Moitinho, Muinonen, Osinde, Pancino, Pauwels, Petit, {Recio-Blanco}, Robin, Sarro, Siopis, Smith, Smith, Sozzetti, Thuillot, {van Reeven}, Viala, Abbas, Abreu~Aramburu, Accart, Aguado, Allan, Allasia, Altavilla, {\'A}lvarez, Alves, Anderson, Andrei, Anglada~Varela, Antiche, Antoja, Ant{\'o}n, Arcay, Atzei, Ayache, Bach, Baker, {Balaguer-N{\'u}{\~n}ez}, Barache, Barata, Barbier, Barblan, Baroni, {Barrado y Navascu{\'e}s}, Barros, Barstow, Becciani, Bellazzini, Bellei, Bello~Garc{\'i}a, Belokurov, Bendjoya, Berihuete, Bianchi, Bienaym{\'e}, Billebaud, Blagorodnova, {Blanco-Cuaresma}, Boch, Bombrun, Borrachero, Bouquillon, Bourda, Bouy, Bragaglia, Breddels, Brouillet, Br{\"u}semeister, Bucciarelli, Budnik, Burgess, Burgon, Burlacu, Busonero, Buzzi, Caffau, Cambras, Campbell, Cancelliere, {Cantat-Gaudin}, Carlucci, Carrasco, Castellani, Charlot, Charnas, Charvet, Chassat, Chiavassa, Clotet, Cocozza, Collins, Collins, Costigan, Crifo, Cross, Crosta, Crowley, Dafonte, Damerdji, Dapergolas, David, David, De~Cat, {de Felice}, {de Laverny}, De~Luise, De~March, {de Martino}, {de Souza}, Debosscher, {del Pozo}, Delbo, Delgado, Delgado, {di Marco}, Di~Matteo, Diakite, Distefano, Dolding, Dos~Anjos, Drazinos, Dur{\'a}n, Dzigan, Ecale, Edvardsson, Enke, Erdmann, Escolar, Espina, Evans, Eynard~Bontemps, Fabre, Fabrizio, Faigler, Falc{\~a}o, Farr{\`a}s~Casas, Faye, Federici, Fedorets, {Fern{\'a}ndez-Hern{\'a}ndez}, Fernique, Fienga, Figueras, Filippi, Findeisen, Fonti, Fouesneau, Fraile, Fraser, Fuchs, Furnell, Gai, Galleti, Galluccio, Garabato, {Garc{\'i}a-Sedano}, Gar{\'e}, Garofalo, Garralda, Gavras, Gerssen, Geyer, Gilmore, Girona, Giuffrida, Gomes, {Gonz{\'a}lez-Marcos}, {Gonz{\'a}lez-N{\'u}{\~n}ez}, {Gonz{\'a}lez-Vidal}, Granvik, Guerrier, Guillout, Guiraud, G{\'u}rpide, {Guti{\'e}rrez-S{\'a}nchez}, Guy, Haigron, Hatzidimitriou, Haywood, Heiter, Helmi, Hobbs, Hofmann, Holl, Holland, Hunt, Hypki, Icardi, Irwin, {Jevardat de Fombelle}, Jofr{\'e}, Jonker, Jorissen, Julbe, Karampelas, Kochoska, Kohley, Kolenberg, Kontizas, Koposov, Kordopatis, Koubsky, Kowalczyk, {Krone-Martins}, Kudryashova, Kull, Bachchan, {Lacoste-Seris}, Lanza, Lavigne, {Le Poncin-Lafitte}, Lebreton, Lebzelter, Leccia, Leclerc, {Lecoeur-Taibi}, Lemaitre, Lenhardt, Leroux, Liao, Licata, Lindstr{\o}m, Lister, Livanou, Lobel, L{\"o}ffler, L{\'o}pez, {Lopez-Lozano}, Lorenz, Loureiro, MacDonald, Magalh{\~a}es~Fernandes, Managau, Mann, Mantelet, Marchal, Marchant, Marconi, Marie, Marinoni, Marrese, Marschalk{\'o}, Marshall, {Mart{\'i}n-Fleitas}, Martino, Mary, Matijevi{\v c}, Mazeh, McMillan, Messina, Mestre, Michalik, Millar, Miranda, Molina, Molinaro, Molinaro, Moln{\'a}r, Moniez, Montegriffo, Monteiro, Mor, Mora, Morbidelli, Morel, Morgenthaler, Morley, Morris, Mulone, Muraveva, Musella, Narbonne, Nelemans, Nicastro, Noval, Ord{\'e}novic, {Ordieres-Mer{\'e}}, Osborne, Pagani, Pagano, Pailler, Palacin, Palaversa, Parsons, Paulsen, Pecoraro, Pedrosa, Pentik{\"a}inen, Pereira, Pichon, Piersimoni, Pineau, Plachy, Plum, Poujoulet, Pr{\v s}a, Pulone, Ragaini, Rago, Rambaux, {Ramos-Lerate}, Ranalli, Rauw, Read, Regibo, Renk, Reyl{\'e}, Ribeiro, Rimoldini, Ripepi, Riva, Rixon, Roelens, {Romero-G{\'o}mez}, Rowell, Royer, Rudolph, {Ruiz-Dern}, Sadowski, Sagrist{\`a}~Sell{\'e}s, Sahlmann, Salgado, Salguero, Sarasso, Savietto, Schnorhk, Schultheis, Sciacca, Segol, Segovia, Segransan, Serpell, Shih, Smareglia, Smart, Smith, Solano, Solitro, Sordo, Soria~Nieto, Souchay, Spagna, Spoto, Stampa, Steele, Steidelm{\"u}ller, Stephenson, Stoev, Suess, S{\"u}veges, Surdej, Szabados, {Szegedi-Elek}, Tapiador, Taris, Tauran, Taylor, Teixeira, Terrett, Tingley, Trager, Turon, Ulla, Utrilla, Valentini, {van Elteren}, Van~Hemelryck, {van Leeuwen}, Varadi, Vecchiato, Veljanoski, Via, Vicente, Vogt, Voss, Votruba, Voutsinas, Walmsley, Weiler, Weingrill, Werner, Wevers, Whitehead, Wyrzykowski, Yoldas, {\v Z}erjal, Zucker, Zurbach, Zwitter, Alecu, Allen, Allende~Prieto, Amorim, {Anglada-Escud{\'e}}, Arsenijevic, Azaz, Balm, Beck, Bernstein, Bigot, Bijaoui, Blasco, Bonfigli, Bono, Boudreault, Bressan, Brown, Brunet, Bunclark, Buonanno, Butkevich, Carret, Carrion, Chemin, Ch{\'e}reau, Corcione, Darmigny, {de Boer}, {de Teodoro}, {de Zeeuw}, Delle~Luche, Domingues, Dubath, Fodor, Fr{\'e}zouls, Fries, Fustes, Fyfe, Gallardo, Gallegos, Gardiol, Gebran, Gomboc, G{\'o}mez, Grux, Gueguen, Heyrovsky, Hoar, Iannicola, Isasi~Parache, Janotto, Joliet, Jonckheere, Keil, Kim, Klagyivik, Klar, Knude, Kochukhov, Kolka, Kos, Kutka, Lainey, LeBouquin, Liu, Loreggia, Makarov, Marseille, Martayan, {Martinez-Rubi}, Massart, Meynadier, Mignot, Munari, Nguyen, Nordlander, Ocvirk, O'Flaherty, Olias~Sanz, Ortiz, Osorio, Oszkiewicz, Ouzounis, Palmer, Park, Pasquato, Peltzer, Peralta, P{\'e}turaud, Pieniluoma, Pigozzi, Poels, Prat, Prod'homme, Raison, Rebordao, Risquez, {Rocca-Volmerange}, Rosen, {Ruiz-Fuertes}, Russo, Sembay, Serraller~Vizcaino, Short, Siebert, Silva, Sinachopoulos, Slezak, Soffel, Sosnowska, Strai{\v z}ys, {ter Linden}, Terrell, Theil, Tiede, Troisi, Tsalmantza, Tur, Vaccari, Vachier, Valles, Van~Hamme, Veltz, Virtanen, Wallut, Wichmann, Wilkinson, Ziaeepour, \& Zschocke}]{GaiaCollaboration2016}
{Gaia Collaboration}, Prusti, T., {de Bruijne}, J. H.~J., {et~al.} 2016, A\&A, 595, A1

\bibitem[{{Gaia Collaboration} {et~al.}(2023){Gaia Collaboration}, Vallenari, Brown, Prusti, {de Bruijne}, Arenou, Babusiaux, Biermann, Creevey, Ducourant, Evans, Eyer, Guerra, Hutton, Jordi, Klioner, Lammers, Lindegren, Luri, Mignard, Panem, Pourbaix, Randich, Sartoretti, Soubiran, Tanga, Walton, {Bailer-Jones}, Bastian, Drimmel, Jansen, Katz, Lattanzi, {van Leeuwen}, Bakker, Cacciari, Casta{\~n}eda, De~Angeli, Fabricius, Fouesneau, Fr{\'e}mat, Galluccio, Guerrier, Heiter, Masana, Messineo, Mowlavi, Nicolas, Nienartowicz, Pailler, Panuzzo, Riclet, Roux, Seabroke, Sordo, Th{\'e}venin, {Gracia-Abril}, Portell, Teyssier, Altmann, Andrae, Audard, {Bellas-Velidis}, Benson, Berthier, Blomme, Burgess, Busonero, Busso, C{\'a}novas, Carry, Cellino, Cheek, Clementini, Damerdji, Davidson, {de Teodoro}, Nu{\~n}ez~Campos, Delchambre, Dell'Oro, Esquej, {Fern{\'a}ndez-Hern{\'a}ndez}, Fraile, Garabato, {Garc{\'i}a-Lario}, Gosset, Haigron, Halbwachs, Hambly, Harrison, Hern{\'a}ndez, Hestroffer, Hodgkin, Holl, Jan{\ss}en, {Jevardat de Fombelle}, Jordan, {Krone-Martins}, Lanzafame, L{\"o}ffler, Marchal, Marrese, Moitinho, Muinonen, Osborne, Pancino, Pauwels, {Recio-Blanco}, Reyl{\'e}, Riello, Rimoldini, Roegiers, Rybizki, Sarro, Siopis, Smith, Sozzetti, Utrilla, {van Leeuwen}, Abbas, {\'A}brah{\'a}m, Abreu~Aramburu, Aerts, Aguado, Ajaj, {Aldea-Montero}, Altavilla, {\'A}lvarez, Alves, Anders, Anderson, Anglada~Varela, Antoja, Baines, Baker, {Balaguer-N{\'u}{\~n}ez}, Balbinot, Balog, Barache, Barbato, Barros, Barstow, Bartolom{\'e}, Bassilana, Bauchet, Becciani, Bellazzini, Berihuete, Bernet, Bertone, Bianchi, Binnenfeld, {Blanco-Cuaresma}, Blazere, Boch, Bombrun, Bossini, Bouquillon, Bragaglia, Bramante, Breedt, Bressan, Brouillet, Brugaletta, Bucciarelli, Burlacu, Butkevich, Buzzi, Caffau, Cancelliere, {Cantat-Gaudin}, Carballo, Carlucci, Carnerero, Carrasco, Casamiquela, Castellani, {Castro-Ginard}, Chaoul, Charlot, Chemin, Chiaramida, Chiavassa, Chornay, Comoretto, Contursi, Cooper, Cornez, Cowell, Crifo, Cropper, Crosta, Crowley, Dafonte, Dapergolas, David, David, {de Laverny}, De~Luise, De~March, De~Ridder, {de Souza}, {de Torres}, {del Peloso}, {del Pozo}, Delbo, Delgado, Delisle, Demouchy, Dharmawardena, Di~Matteo, Diakite, Diener, Distefano, Dolding, Edvardsson, Enke, Fabre, Fabrizio, Faigler, Fedorets, Fernique, Fienga, Figueras, Fournier, Fouron, Fragkoudi, Gai, {Garcia-Gutierrez}, {Garcia-Reinaldos}, {Garc{\'i}a-Torres}, Garofalo, Gavel, Gavras, Gerlach, Geyer, Giacobbe, Gilmore, Girona, Giuffrida, Gomel, Gomez, {Gonz{\'a}lez-N{\'u}{\~n}ez}, {Gonz{\'a}lez-Santamar{\'i}a}, {Gonz{\'a}lez-Vidal}, Granvik, Guillout, Guiraud, {Guti{\'e}rrez-S{\'a}nchez}, Guy, Hatzidimitriou, Hauser, Haywood, Helmer, Helmi, Sarmiento, Hidalgo, Hilger, H{\l}adczuk, Hobbs, Holland, Huckle, Jardine, Jasniewicz, {Jean-Antoine Piccolo}, {Jim{\'e}nez-Arranz}, Jorissen, Juaristi~Campillo, Julbe, Karbevska, Kervella, Khanna, Kontizas, Kordopatis, Korn, K{\'o}sp{\'a}l, {Kostrzewa-Rutkowska}, Kruszy{\'n}ska, Kun, Laizeau, Lambert, Lanza, Lasne, Le~Campion, Lebreton, Lebzelter, Leccia, Leclerc, {Lecoeur-Taibi}, Liao, Licata, Lindstr{\o}m, Lister, Livanou, Lobel, Lorca, Loup, Madrero~Pardo, Magdaleno~Romeo, Managau, Mann, Manteiga, Marchant, Marconi, Marcos, Marcos~Santos, Mar{\'i}n~Pina, Marinoni, Marocco, Marshall, Martin~Polo, {Mart{\'i}n-Fleitas}, Marton, Mary, Masip, Massari, {Mastrobuono-Battisti}, Mazeh, McMillan, Messina, Michalik, Millar, Mints, Molina, Molinaro, Moln{\'a}r, Monari, Mongui{\'o}, Montegriffo, Montero, Mor, Mora, Morbidelli, Morel, Morris, Muraveva, Murphy, Musella, Nagy, Noval, Oca{\~n}a, Ogden, Ordenovic, Osinde, Pagani, Pagano, Palaversa, Palicio, {Pallas-Quintela}, Panahi, {Payne-Wardenaar}, Pe{\~n}alosa~Esteller, Penttil{\"a}, Pichon, Piersimoni, Pineau, Plachy, Plum, Poggio, Pr{\v s}a, Pulone, Racero, Ragaini, Rainer, Raiteri, Rambaux, Ramos, {Ramos-Lerate}, Re~Fiorentin, Regibo, Richards, Rios~Diaz, Ripepi, Riva, Rix, Rixon, Robichon, Robin, Robin, Roelens, Rogues, Rohrbasser, {Romero-G{\'o}mez}, Rowell, Royer, Ruz~Mieres, Rybicki, Sadowski, S{\'a}ez~N{\'u}{\~n}ez, Sagrist{\`a}~Sell{\'e}s, Sahlmann, Salguero, Samaras, Sanchez~Gimenez, Sanna, Santove{\~n}a, Sarasso, Schultheis, Sciacca, Segol, Segovia, S{\'e}gransan, Semeux, Shahaf, Siddiqui, Siebert, Siltala, Silvelo, Slezak, Slezak, Smart, Snaith, Solano, Solitro, Souami, Souchay, Spagna, Spina, Spoto, Steele, Steidelm{\"u}ller, Stephenson, S{\"u}veges, Surdej, Szabados, {Szegedi-Elek}, Taris, Taylor, Teixeira, Tolomei, Tonello, Torra, Torra, Torralba~Elipe, Trabucchi, Tsounis, Turon, Ulla, Unger, Vaillant, {van Dillen}, {van Reeven}, Vanel, Vecchiato, Viala, Vicente, Voutsinas, Weiler, Wevers, Wyrzykowski, Yoldas, Yvard, Zhao, Zorec, Zucker, \& Zwitter}]{GaiaCollaboration2023}
{Gaia Collaboration}, Vallenari, A., Brown, A. G.~A., {et~al.} 2023, A\&A, 674, A1

\bibitem[{Galland {et~al.}(2005)Galland, Lagrange, Udry, Chelli, Pepe, Queloz, Beuzit, \& Mayor}]{Galland2005}
Galland, F., Lagrange, A.-M., Udry, S., {et~al.} 2005, A\&A, 443, 337

\bibitem[{Gehan {et~al.}(2024)Gehan, {Godoy-Rivera}, \& Gaulme}]{Gehan2024}
Gehan, C., {Godoy-Rivera}, D., \& Gaulme, P. 2024, A\&A, 686, A93

\bibitem[{Ginsburg {et~al.}(2019)Ginsburg, Sip{\H o}cz, Brasseur, Cowperthwaite, Craig, Deil, Guillochon, Guzman, Liedtke, Lian~Lim, Lockhart, Mommert, Morris, Norman, Parikh, Persson, Robitaille, Segovia, Singer, Tollerud, {de Val-Borro}, Valtchanov, Woillez, {Astroquery Collaboration}, \& {a subset of astropy Collaboration}}]{Ginsburg2019}
Ginsburg, A., Sip{\H o}cz, B.~M., Brasseur, C.~E., {et~al.} 2019, AJ, 157, 98

\bibitem[{Goldberg {et~al.}(2024)Goldberg, Joyce, \& Moln{\'a}r}]{Goldberg2024}
Goldberg, J.~A., Joyce, M., \& Moln{\'a}r, L. 2024, ApJ, 977, 35

\bibitem[{Golovin {et~al.}(2023)Golovin, Reffert, Just, Jordan, Vani, \& Jahrei{\ss}}]{Golovin2023}
Golovin, A., Reffert, S., Just, A., {et~al.} 2023, A\&A, 670, A19

\bibitem[{Grundahl {et~al.}(2017)Grundahl, Andersen, {Christensen-Dalsgaard}, Antoci, Kjeldsen, Handberg, Houdek, Bedding, Pall{\'e}, {Jessen-Hansen}, Aguirre, White, Frandsen, Albrecht, Andersen, Arentoft, Brogaard, Chaplin, Harps{\o}e, J{\o}rgensen, Karovicova, Karoff, Rasmussen, Lund, Lundkvist, Skottfelt, Norup~S{\o}rensen, Tronsgaard, \& Weiss}]{Grundahl2017}
Grundahl, F., Andersen, M.~F., {Christensen-Dalsgaard}, J., {et~al.} 2017, ApJ, 836, 142

\bibitem[{Harris {et~al.}(2020)Harris, Millman, {van der Walt}, Gommers, Virtanen, Cournapeau, Wieser, Taylor, Berg, Smith, Kern, Picus, Hoyer, {van Kerkwijk}, Brett, Haldane, {del R{\'i}o}, Wiebe, Peterson, {G{\'e}rard-Marchant}, Sheppard, Reddy, Weckesser, Abbasi, Gohlke, \& Oliphant}]{Harris2020}
Harris, C.~R., Millman, K.~J., {van der Walt}, S.~J., {et~al.} 2020, Nat, 585, 357

\bibitem[{Hatzes(1996)}]{Hatzes1996}
Hatzes, A.~P. 1996, PASP, 108, 839

\bibitem[{Hatzes(2013)}]{Hatzes2013}
Hatzes, A.~P. 2013, Astron. Nachr., 334, 616

\bibitem[{Hatzes \& Cochran(1993)}]{Hatzes1993}
Hatzes, A.~P. \& Cochran, W.~D. 1993, ApJ, 413, 339

\bibitem[{Hatzes \& Cochran(1994)}]{Hatzes1994}
Hatzes, A.~P. \& Cochran, W.~D. 1994, ApJ, 422, 366

\bibitem[{{Hatzes} \& {Cochran}(1998)}]{Hatzes1998a}
{Hatzes}, A.~P. \& {Cochran}, W.~D. 1998, in ASP Conf. Ser., Vol. 154, 311

\bibitem[{Hatzes {et~al.}(2006)Hatzes, Cochran, Endl, Guenther, Saar, Walker, Yang, Hartmann, Esposito, Paulson, \& D{\"o}llinger}]{Hatzes2006}
Hatzes, A.~P., Cochran, W.~D., Endl, M., {et~al.} 2006, A\&A, 457, 335

\bibitem[{Hatzes {et~al.}(2018)Hatzes, Endl, Cochran, MacQueen, Han, Lee, Kim, Mkrtichian, D{\"o}llinger, Hartmann, Karjalainen, \& Dreizler}]{Hatzes2018}
Hatzes, A.~P., Endl, M., Cochran, W.~D., {et~al.} 2018, AJ, 155, 120

\bibitem[{Heeren {et~al.}(2021)Heeren, Reffert, Trifonov, Wong, Lee, {Lillo-Box}, Quirrenbach, Arentoft, Albrecht, Grundahl, Andersen, Antoci, \& Pall{\'e}}]{Heeren2021}
Heeren, P., Reffert, S., Trifonov, T., {et~al.} 2021, A\&A, 647, A160

\bibitem[{Heeren {et~al.}(2023)Heeren, Tronsgaard, Grundahl, Reffert, Quirrenbach, \& Pall{\'e}}]{Heeren2023}
Heeren, P., Tronsgaard, R., Grundahl, F., {et~al.} 2023, A\&A, 674, A164

\bibitem[{Hekker \& Mel{\'e}ndez(2007)}]{Hekker2007}
Hekker, S. \& Mel{\'e}ndez, J. 2007, A\&A, 475, 1003

\bibitem[{Hekker {et~al.}(2005)Hekker, Reffert, \& Quirrenbach}]{Hekker2005}
Hekker, S., Reffert, S., \& Quirrenbach, A. 2005, arXiv e-prints [\eprint[arXiv]{astro-ph/0510337}]

\bibitem[{Hekker {et~al.}(2006)Hekker, Reffert, Quirrenbach, Mitchell, Fischer, Marcy, \& Butler}]{Hekker2006a}
Hekker, S., Reffert, S., Quirrenbach, A., {et~al.} 2006, A\&A, 454, 943

\bibitem[{Hekker {et~al.}(2008)Hekker, Snellen, Aerts, Quirrenbach, Reffert, \& Mitchell}]{Hekker2008}
Hekker, S., Snellen, I. A.~G., Aerts, C., {et~al.} 2008, A\&A, 480, 215

\bibitem[{Higson {et~al.}(2019)Higson, Handley, Hobson, \& Lasenby}]{Higson2019}
Higson, E., Handley, W., Hobson, M., \& Lasenby, A. 2019, Stat Comput, 29, 891

\bibitem[{Hill {et~al.}(2021)Hill, Kane, Campante, Li, Dalba, Brandt, White, Pope, Stassun, Fulton, Corsaro, Li, Ong, Bedding, Bossini, Buzasi, Chaplin, Cunha, Garc{\'i}a, Breton, Hon, Huber, Jiang, Kayhan, Kuszlewicz, Mathur, Serenelli, \& Stello}]{Hill2021}
Hill, M.~L., Kane, S.~R., Campante, T.~L., {et~al.} 2021, AJ, 162, 211

\bibitem[{Huang {et~al.}(2024)Huang, He, Bai, Yuan, Yang, Zhou, Dong, Wang, He, Zhang, Chu, Zhao, Zhang, \& Zhang}]{Huang2024}
Huang, X., He, Y., Bai, Z., {et~al.} 2024, ApJS, 272, 6

\bibitem[{Hu{\'e}lamo {et~al.}(2008)Hu{\'e}lamo, Figueira, Bonfils, Santos, Pepe, Gillon, Azevedo, Barman, Fern{\'a}ndez, Di~Folco, Guenther, Lovis, Melo, Queloz, \& Udry}]{Huelamo2008}
Hu{\'e}lamo, N., Figueira, P., Bonfils, X., {et~al.} 2008, A\&A, 489, L9

\bibitem[{Hunter(2007)}]{Hunter2007}
Hunter, J.~D. 2007, Comput. Sci. Eng., 9, 90

\bibitem[{Johnson {et~al.}(2007)Johnson, Fischer, Marcy, Wright, Driscoll, Butler, Hekker, Reffert, \& Vogt}]{Johnson2007}
Johnson, J.~A., Fischer, D.~A., Marcy, G.~W., {et~al.} 2007, ApJ, 665, 785

\bibitem[{Kane {et~al.}(2010)Kane, Reffert, Henry, Fischer, Schwab, Clubb, \& Bergmann}]{Kane2010}
Kane, S.~R., Reffert, S., Henry, G.~W., {et~al.} 2010, ApJ, 720, 1644

\bibitem[{Kemmer {et~al.}(2025)Kemmer, Lafarga, Fuhrmeister, Sch{\"o}fer, Shan, Quirrenbach, Jeffers, Amado, Caballero, Reiners, Ribas, B{\'e}jar, Sordo, Hatzes, Henning, Hermelo, Kaminski, Montes, Morales, \& Reffert}]{Kemmer2024}
Kemmer, J., Lafarga, M., Fuhrmeister, B., {et~al.} 2025, A\&A (in press)

\bibitem[{Kjeldsen \& Bedding(1995)}]{Kjeldsen1995}
Kjeldsen, H. \& Bedding, T.~R. 1995, A\&A, 293, 87

\bibitem[{Kjeldsen \& Bedding(2011)}]{Kjeldsen2011}
Kjeldsen, H. \& Bedding, T.~R. 2011, A\&A, 529, L8

\bibitem[{Koposov {et~al.}(2024)Koposov, Speagle, Barbary, Ashton, Bennett, Buchner, Scheffler, Cook, Talbot, Guillochon, Cubillos, Ramos, Dartiailh, Ilya, Tollerud, Lang, Johnson, {jtmendel}, Higson, Vandal, Daylan, Angus, {patelR}, Cargile, Sheehan, Pitkin, Kirk, Leja, {joezuntz}, \& Goldstein}]{Koposov2024}
Koposov, S., Speagle, J., Barbary, K., {et~al.} 2024, Joshspeagle/Dynesty: V2.1.4, Zenodo

\bibitem[{K{\"u}rster {et~al.}(2003)K{\"u}rster, Endl, Rouesnel, Els, Kaufer, Brillant, Hatzes, Saar, \& Cochran}]{Kurster2003}
K{\"u}rster, M., Endl, M., Rouesnel, F., {et~al.} 2003, A\&A, 403, 1077

\bibitem[{Lafarga {et~al.}(2020)Lafarga, Ribas, Lovis, Perger, Zechmeister, Bauer, K{\"u}rster, {Cort{\'e}s-Contreras}, Morales, Herrero, Rosich, Baroch, Reiners, Caballer, Quirrenbach, Amado, Alacid, B{\'e}jar, Dreizler, Hatzes, Henning, Jeffers, Kaminski, Montes, Pedraz, {Rodr{\'i}guez-L{\'o}pez}, \& Schmitt}]{Lafarga2020}
Lafarga, M., Ribas, I., Lovis, C., {et~al.} 2020, A\&A, 636, A36

\bibitem[{Lagrange {et~al.}(2009)Lagrange, Desort, Galland, Udry, \& Mayor}]{Lagrange2009}
Lagrange, A.-M., Desort, M., Galland, F., Udry, S., \& Mayor, M. 2009, A\&A, 495, 335

\bibitem[{Laney {et~al.}(2012)Laney, Joner, \& Pietrzy{\'n}ski}]{Laney2012}
Laney, C.~D., Joner, M.~D., \& Pietrzy{\'n}ski, G. 2012, MNRAS, 419, 1637

\bibitem[{L{\`e}bre {et~al.}(2006)L{\`e}bre, {de Laverny}, Do~Nascimento, \& {de Medeiros}}]{Lebre2006}
L{\`e}bre, A., {de Laverny}, P., Do~Nascimento, Jr., J.~D., \& {de Medeiros}, J.~R. 2006, A\&A, 450, 1173

\bibitem[{Lee {et~al.}(2014)Lee, Han, Park, Mkrtichian, Jeong, Kim, \& Valyavin}]{Lee2014a}
Lee, B.-C., Han, I., Park, M.-G., {et~al.} 2014, J. Korean Astron. Soc., 47, 69

\bibitem[{Lee {et~al.}(2023)Lee, Koo, Choi, Bang, Lim, Park, \& Jeong}]{Lee2023a}
Lee, B.-C., Koo, J.-R., Choi, Y.-H., {et~al.} 2023, J. Korean Astron. Soc., 56, 277

\bibitem[{Li {et~al.}(2022)Li, Deheuvels, Ballot, \& Ligni{\`e}res}]{Li2022}
Li, G., Deheuvels, S., Ballot, J., \& Ligni{\`e}res, F. 2022, Nat, 610, 43

\bibitem[{Li {et~al.}(2023)Li, Deheuvels, Li, Ballot, \& Ligni{\`e}res}]{Li2023}
Li, G., Deheuvels, S., Li, T., Ballot, J., \& Ligni{\`e}res, F. 2023, A\&A, 680, A26

\bibitem[{Liu {et~al.}(2008)Liu, Sato, Zhao, Noguchi, Wang, Kambe, Ando, Izumiura, Chen, Okada, Toyota, Omiya, Masuda, Takeda, Murata, Itoh, Yoshida, Kokubo, \& Ida}]{Liu2008}
Liu, Y.~J., Sato, B., Zhao, G., {et~al.} 2008, ApJ, 672, 553

\bibitem[{Liu {et~al.}(2014)Liu, Tan, Wang, Zhao, Sato, Takeda, \& Li}]{Liu2014}
Liu, Y.~J., Tan, K.~F., Wang, L., {et~al.} 2014, ApJ, 785, 94

\bibitem[{Lovis \& Mayor(2007)}]{Lovis2007}
Lovis, C. \& Mayor, M. 2007, A\&A, 472, 657

\bibitem[{Luque {et~al.}(2019)Luque, Trifonov, Reffert, Quirrenbach, Lee, Albrecht, Fredslund~Andersen, Antoci, Grundahl, Schwab, \& Wolthoff}]{Luque2019}
Luque, R., Trifonov, T., Reffert, S., {et~al.} 2019, A\&A, 631, A136

\bibitem[{Malla {et~al.}(2024)Malla, Stello, Montet, Huber, Hon, Bedding, Reyes, \& Hey}]{Malla2024}
Malla, S.~P., Stello, D., Montet, B.~T., {et~al.} 2024, MNRAS, 534, 1775

\bibitem[{Martin {et~al.}(2017)Martin, Fuhrmeister, Mittag, Schmidt, Hempelmann, {Gonz{\'a}lez-P{\'e}rez}, \& Schmitt}]{Martin2017}
Martin, J., Fuhrmeister, B., Mittag, M., {et~al.} 2017, A\&A, 605, A113

\bibitem[{{Mart{\'i}nez-Arn{\'a}iz} {et~al.}(2011){Mart{\'i}nez-Arn{\'a}iz}, {L{\'o}pez-Santiago}, {Crespo-Chac{\'o}n}, \& Montes}]{Martinez-Arnaiz2011}
{Mart{\'i}nez-Arn{\'a}iz}, R., {L{\'o}pez-Santiago}, J., {Crespo-Chac{\'o}n}, I., \& Montes, D. 2011, MNRAS, 414, 2629

\bibitem[{Mayor {et~al.}(2003)Mayor, Pepe, Queloz, Bouchy, Rupprecht, Lo~Curto, Avila, Benz, Bertaux, Bonfils, Dall, Dekker, Delabre, Eckert, Fleury, Gilliotte, Gojak, Guzman, Kohler, Lizon, Longinotti, Lovis, Megevand, Pasquini, Reyes, Sivan, Sosnowska, Soto, Udry, {van Kesteren}, Weber, \& Weilenmann}]{Mayor2003}
Mayor, M., Pepe, F., Queloz, D., {et~al.} 2003, Msngr, 114, 20

\bibitem[{{McKerns} {et~al.}(2012){McKerns}, {Strand}, {Sullivan}, {Fang}, \& {Aivazis}}]{McKerns2012}
{McKerns}, M.~M., {Strand}, L., {Sullivan}, T., {Fang}, A., \& {Aivazis}, M. A.~G. 2012, arXiv e-prints, arXiv:1202.1056

\bibitem[{McKinney(2010)}]{McKinney2010}
McKinney, W. 2010, in Proc. 9th {{Python Sci}}. {{Conf}}., ed. S.~{van der Walt} \& {Jarrod Millman}, 56--61

\bibitem[{Mitchell {et~al.}(2013)Mitchell, Reffert, Trifonov, Quirrenbach, \& Fischer}]{Mitchell2013}
Mitchell, D.~S., Reffert, S., Trifonov, T., Quirrenbach, A., \& Fischer, D.~A. 2013, A\&A, 555, A87

\bibitem[{Mortier \& Collier~Cameron(2017)}]{Mortier2017}
Mortier, A. \& Collier~Cameron, A. 2017, A\&A, 601, A110

\bibitem[{Mortier {et~al.}(2015)Mortier, Faria, Correia, Santerne, \& Santos}]{Mortier2015}
Mortier, A., Faria, J.~P., Correia, C.~M., Santerne, A., \& Santos, N.~C. 2015, A\&A, 573, A101

\bibitem[{M{\"u}ller(2019)}]{Muller2019}
M{\"u}ller, R.~G. 2019, Bachelor's thesis, University of Heidelberg

\bibitem[{Nagel {et~al.}(2023)Nagel, Czesla, Kaminski, Zechmeister, {Tal-Or}, Schmitt, Reiners, Quirrenbach, Garc{\'i}a~L{\'o}pez, Caballero, Ribas, Amado, B{\'e}jar, {Cort{\'e}s-Contreras}, Dreizler, Hatzes, Henning, Jeffers, K{\"u}rster, Lafarga, {L{\'o}pez-Puertas}, Montes, Morales, Pedraz, \& Schweitzer}]{Nagel2023}
Nagel, E., Czesla, S., Kaminski, A., {et~al.} 2023, A\&A, 680, A73

\bibitem[{Newville {et~al.}(2021)Newville, Otten, Nelson, Ingargiola, Stensitzki, Allan, Fox, Carter, {Micha{\l}}, Osborn, Pustakhod, {lneuhaus}, Weigand, {Glenn}, Deil, {Mark}, Hansen, Pasquevich, Foks, Zobrist, Frost, Beelen, {Stuermer}, {azelcer}, Hannum, Polloreno, Nielsen, Caldwell, Almarza, \& Persaud}]{Newville2021}
Newville, M., Otten, R., Nelson, A., {et~al.} 2021, Lmfit/Lmfit-Py: 1.0.3, Zenodo

\bibitem[{Ortiz {et~al.}(2016)Ortiz, Reffert, Trifonov, Quirrenbach, Mitchell, Nowak, Buenzli, Zimmerman, Bonnefoy, Skemer, Defr{\`e}re, Lee, Fischer, \& Hinz}]{Ortiz2016}
Ortiz, M., Reffert, S., Trifonov, T., {et~al.} 2016, A\&A, 595, A55

\bibitem[{pandas~development {team}(2024)}]{pandas2024}
pandas~development {team}, T. 2024, Pandas-Dev/Pandas: {{Pandas}}, Zenodo

\bibitem[{Pedregosa {et~al.}(2011)Pedregosa, Varoquaux, Gramfort, Michel, Thirion, Grisel, Blondel, Prettenhofer, Weiss, Dubourg, Vanderplas, Passos, Cournapeau, Brucher, Perrot, \& Duchesnay}]{Pedregosa2011}
Pedregosa, F., Varoquaux, G., Gramfort, A., {et~al.} 2011, J. Mach. Learn. Res., 12, 2825

\bibitem[{Percy {et~al.}(2001)Percy, Wilson, \& Henry}]{Percy2001}
Percy, J.~R., Wilson, J.~B., \& Henry, G.~W. 2001, PASP, 113, 983

\bibitem[{Phipson \& Smyth(2010)}]{Phipson2010}
Phipson, B. \& Smyth, G.~K. 2010, Stat. Appl. Genet. Mol. Biol., 9

\bibitem[{Quirrenbach {et~al.}(2014)Quirrenbach, Amado, Caballero, Mundt, Reiners, Ribas, Seifert, Abril, Aceituno, {Alonso-Floriano}, {Ammler-von Eiff}, Antona~Jim{\'e}nez, {Anwand-Heerwart}, Azzaro, Bauer, Barrado, Becerril, B{\'e}jar, Ben{\'i}tez, Berdi{\~n}as, C{\'a}rdenas, Casal, Claret, Colom{\'e}, {Cort{\'e}s-Contreras}, Czesla, Doellinger, Dreizler, Feiz, Fern{\'a}ndez, Galad{\'i}, {G{\'a}lvez-Ortiz}, {Garc{\'i}a-Piquer}, {Garc{\'i}a-Vargas}, Garrido, Gesa, G{\'o}mez~Galera, Gonz{\'a}lez~{\'A}lvarez, Gonz{\'a}lez~Hern{\'a}ndez, Gr{\"o}zinger, Gu{\`a}rdia, Guenther, {de Guindos}, {Guti{\'e}rrez-Soto}, Hagen, Hatzes, Hauschildt, Helmling, Henning, Hermann, Hern{\'a}ndez~Casta{\~n}o, Herrero, Hidalgo, Holgado, Huber, Huber, Jeffers, Joergens, {de Juan}, Kehr, Klein, K{\"u}rster, Lamert, Lalitha, Laun, Lemke, Lenzen, {L{\'o}pez del Fresno}, L{\'o}pez~Mart{\'i}, {L{\'o}pez-Santiago}, Mall, Mandel, Mart{\'i}n, {Mart{\'i}n-Ruiz}, {Mart{\'i}nez-Rodr{\'i}guez}, Marvin, Mathar, Mirabet, Montes, Morales~Mu{\~n}oz, Moya, Naranjo, Ofir, Oreiro, Pall{\'e}, Panduro, Passegger, {P{\'e}rez-Calpena}, P{\'e}rez~Medialdea, Perger, Pluto, Ram{\'o}n, Rebolo, Redondo, Reffert, Reinhardt, Rhode, Rix, Rodler, Rodr{\'i}guez, {Rodr{\'i}guez-L{\'o}pez}, {Rodr{\'i}guez-P{\'e}rez}, Rohloff, Rosich, {S{\'a}nchez-Blanco}, S{\'a}nchez~Carrasco, {Sanz-Forcada}, Sarmiento, Sch{\"a}fer, Schiller, Schmidt, Schmitt, Solano, Stahl, Storz, St{\"u}rmer, Su{\'a}rez, Ulbrich, Veredas, Wagner, Winkler, Zapatero~Osorio, Zechmeister, {Abell{\'a}n de Paco}, {Anglada-Escud{\'e}}, {del Burgo}, Klutsch, Lizon, {L{\'o}pez-Morales}, Morales, Perryman, Tulloch, \& Xu}]{Quirrenbach2014}
Quirrenbach, A., Amado, P.~J., Caballero, J.~A., {et~al.} 2014, SPIE Conf. Ser., 9147, 91471F

\bibitem[{Quirrenbach {et~al.}(2011)Quirrenbach, Reffert, \& Bergmann}]{Quirrenbach2011}
Quirrenbach, A., Reffert, S., \& Bergmann, C. 2011, in {{AIP Conf}}. {{Ser}}., Vol. 1331, Planet. {{Syst}}. {{Main Seq}}., ed. S.~Schuh, H.~Drechsel, \& U.~Heber (AIP), 102--109

\bibitem[{Quirrenbach {et~al.}(2019)Quirrenbach, Trifonov, Lee, \& Reffert}]{Quirrenbach2019}
Quirrenbach, A., Trifonov, T., Lee, M.~H., \& Reffert, S. 2019, A\&A, 624, A18

\bibitem[{Reffert {et~al.}(2015)Reffert, Bergmann, Quirrenbach, Trifonov, \& K{\"u}nstler}]{Reffert2015}
Reffert, S., Bergmann, C., Quirrenbach, A., Trifonov, T., \& K{\"u}nstler, A. 2015, A\&A, 574, A116

\bibitem[{Reffert {et~al.}(2006)Reffert, Quirrenbach, Mitchell, Albrecht, Hekker, Fischer, Marcy, \& Butler}]{Reffert2006}
Reffert, S., Quirrenbach, A., Mitchell, D.~S., {et~al.} 2006, ApJ, 652, 661

\bibitem[{Reichert {et~al.}(2019)Reichert, Reffert, Stock, Trifonov, \& Quirrenbach}]{Reichert2019}
Reichert, K., Reffert, S., Stock, S., Trifonov, T., \& Quirrenbach, A. 2019, A\&A, 625, A22

\bibitem[{Reiners {et~al.}(2010)Reiners, Bean, Huber, Dreizler, Seifahrt, \& Czesla}]{Reiners2010}
Reiners, A., Bean, J.~L., Huber, K.~F., {et~al.} 2010, ApJ, 710, 432

\bibitem[{Reiners {et~al.}(2018)Reiners, Zechmeister, Caballero, Ribas, Morales, Jeffers, Sch{\"o}fer, {Tal-Or}, Quirrenbach, Amado, Kaminski, Seifert, Abril, Aceituno, {Alonso-Floriano}, {Ammler-von Eiff}, Antona, {Anglada-Escud{\'e}}, {Anwand-Heerwart}, {Arroyo-Torres}, Azzaro, Baroch, Barrado, Bauer, Becerril, B{\'e}jar, Ben{\'i}tez, Berdinas, Bergond, Bl{\"u}mcke, Brinkm{\"o}ller, {del Burgo}, Cano, C{\'a}rdenas~V{\'a}zquez, Casal, Cifuentes, Claret, Colom{\'e}, {Cort{\'e}s-Contreras}, Czesla, {D{\'i}ez-Alonso}, Dreizler, Feiz, Fern{\'a}ndez, Ferro, Fuhrmeister, {Galad{\'i}-Enr{\'i}quez}, {Garcia-Piquer}, Garc{\'i}a~Vargas, Gesa, G{\'o}mez~Galera, Gonz{\'a}lez~Hern{\'a}ndez, {Gonz{\'a}lez-Peinado}, Gr{\"o}zinger, Grohnert, Gu{\`a}rdia, Guenther, Guijarro, {de Guindos}, {Guti{\'e}rrez-Soto}, Hagen, Hatzes, Hauschildt, Hedrosa, Helmling, Henning, Hermelo, Hern{\'a}ndez~Arab{\'i}, Hern{\'a}ndez~Casta{\~n}o, Hern{\'a}ndez~Hernando, Herrero, Huber, Huke, Johnson, {de Juan}, Kim, Klein, Kl{\"u}ter, Klutsch, K{\"u}rster, Lafarga, Lamert, Lamp{\'o}n, Lara, Laun, Lemke, Lenzen, Launhardt, {L{\'o}pez del Fresno}, {L{\'o}pez-Gonz{\'a}lez}, {L{\'o}pez-Puertas}, L{\'o}pez~Salas, {L{\'o}pez-Santiago}, Luque, Mag{\'a}n~Madinabeitia, Mall, Mancini, Mandel, Marfil, Mar{\'i}n~Molina, Maroto~Fern{\'a}ndez, Mart{\'i}n, {Mart{\'i}n-Ruiz}, Marvin, Mathar, Mirabet, Montes, {Moreno-Raya}, Moya, Mundt, Nagel, Naranjo, Nortmann, Nowak, Ofir, Oreiro, Pall{\'e}, Panduro, Pascual, Passegger, Pavlov, Pedraz, {P{\'e}rez-Calpena}, P{\'e}rez~Medialdea, Perger, Perryman, Pluto, Rabaza, Ram{\'o}n, Rebolo, Redondo, Reffert, Reinhart, Rhode, Rix, Rodler, Rodr{\'i}guez, {Rodr{\'i}guez-L{\'o}pez}, Rodr{\'i}guez~Trinidad, Rohloff, Rosich, Sadegi, {S{\'a}nchez-Blanco}, S{\'a}nchez~Carrasco, {S{\'a}nchez-L{\'o}pez}, {Sanz-Forcada}, Sarkis, Sarmiento, Sch{\"a}fer, Schmitt, Schiller, Schweitzer, Solano, Stahl, Strachan, St{\"u}rmer, Su{\'a}rez, Tabernero, Tala, Trifonov, Tulloch, Ulbrich, Veredas, Vico~Linares, Vilardell, Wagner, Winkler, Wolthoff, Xu, Yan, \& Zapatero~Osorio}]{Reiners2018}
Reiners, A., Zechmeister, M., Caballero, J.~A., {et~al.} 2018, A\&A, 612, A49

\bibitem[{Ribas {et~al.}(2023)Ribas, Reiners, Zechmeister, Caballero, Morales, Sabotta, Baroch, Amado, Quirrenbach, Abril, Aceituno, {Anglada-Escud{\'e}}, Azzaro, Barrado, B{\'e}jar, {Ben{\'i}tez de Haro}, Bergond, Bluhm, Calvo~Ortega, Cardona~Guill{\'e}n, Chaturvedi, Cifuentes, Colom{\'e}, Cont, {Cort{\'e}s-Contreras}, Czesla, {D{\'i}ez-Alonso}, Dreizler, {Duque-Arribas}, Espinoza, Fern{\'a}ndez, Fuhrmeister, {Galad{\'i}-Enr{\'i}quez}, {Garc{\'i}a-L{\'o}pez}, {Gonz{\'a}lez-{\'A}lvarez}, Gonz{\'a}lez~Hern{\'a}ndez, Guenther, {de Guindos}, Hatzes, Henning, Herrero, Hintz, Huelmo, Jeffers, Johnson, {de Juan}, Kaminski, Kemmer, Khaimova, Khalafinejad, Kossakowski, K{\"u}rster, Labarga, Lafarga, Lalitha, Lamp{\'o}n, {Lillo-Box}, Lodieu, L{\'o}pez~Gonz{\'a}lez, {L{\'o}pez-Puertas}, Luque, Mag{\'a}n, Mancini, Marfil, Mart{\'i}n, {Mart{\'i}n-Ruiz}, Molaverdikhani, Montes, Nagel, Nortmann, Nowak, Pall{\'e}, Passegger, Pavlov, Pedraz, Perdelwitz, Perger, {Ram{\'o}n-Ballesta}, Reffert, Revilla, Rodr{\'i}guez, {Rodr{\'i}guez-L{\'o}pez}, Sadegi, S{\'a}nchez~Carrasco, {S{\'a}nchez-L{\'o}pez}, {Sanz-Forcada}, Sch{\"a}fer, Schlecker, Schmitt, Sch{\"o}fer, Schweitzer, Seifert, Shan, Skrzypinski, Solano, Stahl, Stangret, Stock, St{\"u}rmer, Tabernero, {Tal-Or}, Trifonov, Vanaverbeke, Yan, \& Zapatero~Osorio}]{Ribas2023}
Ribas, I., Reiners, A., Zechmeister, M., {et~al.} 2023, A\&A, 670, A139

\bibitem[{Robertson {et~al.}(2014)Robertson, Mahadevan, Endl, \& Roy}]{Robertson2014}
Robertson, P., Mahadevan, S., Endl, M., \& Roy, A. 2014, Sci, 345, 440

\bibitem[{Rolo {et~al.}(2024)Rolo, Delgado~Mena, Tsantaki, \& {Gomes da Silva}}]{Rolo2024}
Rolo, I., Delgado~Mena, E., Tsantaki, M., \& {Gomes da Silva}, J. 2024, A\&A, 688, A68

\bibitem[{Saio {et~al.}(2015)Saio, Wood, Takayama, \& Ita}]{Saio2015}
Saio, H., Wood, P.~R., Takayama, M., \& Ita, Y. 2015, MNRAS, 452, 3863

\bibitem[{Sarkis {et~al.}(2018)Sarkis, Henning, K{\"u}rster, Trifonov, Zechmeister, {Tal-Or}, {Anglada-Escud{\'e}}, Hatzes, Lafarga, Dreizler, Ribas, Caballero, Reiners, Mallonn, Morales, Kaminski, Aceituno, Amado, B{\'e}jar, Hagen, Jeffers, Quirrenbach, Launhardt, Marvin, \& Montes}]{Sarkis2018}
Sarkis, P., Henning, T., K{\"u}rster, M., {et~al.} 2018, AJ, 155, 257

\bibitem[{Sato {et~al.}(2003)Sato, Ando, Kambe, Takeda, Izumiura, Masuda, Watanabe, Noguchi, Wada, Okada, Koyano, Maehara, Norimoto, Okada, Shimizu, Uraguchi, Yanagisawa, \& Yoshida}]{Sato2003}
Sato, B., Ando, H., Kambe, E., {et~al.} 2003, ApJ, 597, L157

\bibitem[{Sato {et~al.}(2007)Sato, Izumiura, Toyota, Kambe, Takeda, Masuda, Omiya, Murata, Itoh, Ando, Yoshida, Ikoma, Kokubo, \& Ida}]{Sato2007}
Sato, B., Izumiura, H., Toyota, E., {et~al.} 2007, ApJ, 661, 527

\bibitem[{{Sch{\"a}fer} {et~al.}(2018){Sch{\"a}fer}, {Guenther}, {Reiners}, {Winkler}, {Pluto}, \& {Schiller}}]{Schafer2018}
{Sch{\"a}fer}, S., {Guenther}, E.~W., {Reiners}, A., {et~al.} 2018, in SPIE Conf. Ser., Vol. 10702, Ground-based and Airborne Instrumentation for Astronomy VII, ed. C.~J. {Evans}, L.~{Simard}, \& H.~{Takami}, 1070276

\bibitem[{Sch{\"o}fer {et~al.}(2019)Sch{\"o}fer, Jeffers, Reiners, Shulyak, Fuhrmeister, Johnson, Zechmeister, Ribas, Quirrenbach, Amado, Caballero, {Anglada-Escud{\'e}}, Bauer, B{\'e}jar, {Cort{\'e}s-Contreras}, Dreizler, Guenther, Kaminski, K{\"u}rster, Lafarga, Montes, Morales, Pedraz, \& {Tal-Or}}]{Schofer2019}
Sch{\"o}fer, P., Jeffers, S.~V., Reiners, A., {et~al.} 2019, A\&A, 623, A44

\bibitem[{Schwab(2010)}]{Schwab2010}
Schwab, C. 2010, PhD thesis, University of Heidelberg

\bibitem[{{Skilling}(2004)}]{Skilling2004}
{Skilling}, J. 2004, in AIP Conf. Ser., Vol. 735, Bayesian Inference and Maximum Entropy Methods in Science and Engineering: 24th International Workshop on Bayesian Inference and Maximum Entropy Methods in Science and Engineering, ed. R.~{Fischer}, R.~{Preuss}, \& U.~V. {Toussaint} (AIP), 395--405

\bibitem[{Skilling(2006)}]{Skilling2006}
Skilling, J. 2006, Bayesian Anal., 1, 833

\bibitem[{Sneden {et~al.}(2022)Sneden, Af{\c s}ar, Bozkurt, Adam{\'o}w, Mallick, Reddy, Janowiecki, Mahadevan, Bowler, Hawkins, Lind, Dupree, Ninan, Nagarajan, Topcu, Froning, Bender, Terrien, Ramsey, \& Mace}]{Sneden2022}
Sneden, C., Af{\c s}ar, M., Bozkurt, Z., {et~al.} 2022, ApJ, 940, 12

\bibitem[{Soszy{\'n}ski {et~al.}(2021)Soszy{\'n}ski, Olechowska, Ratajczak, Iwanek, Skowron, Mr{\'o}z, Pietrukowicz, Udalski, Szyma{\'n}ski, Skowron, Gromadzki, Poleski, Koz{\l}owski, Wrona, Ulaczyk, \& Rybicki}]{Soszynski2021}
Soszy{\'n}ski, I., Olechowska, A., Ratajczak, M., {et~al.} 2021, ApJL, 911, L22

\bibitem[{Soszy{\'n}ski \& Udalski(2014)}]{Soszynski2014}
Soszy{\'n}ski, I. \& Udalski, A. 2014, ApJ, 788, 13

\bibitem[{Soszy{\'n}ski {et~al.}(2009)Soszy{\'n}ski, Udalski, Szyma{\'n}ski, Kubiak, Pietrzy{\'n}ski, Wyrzykowski, Szewczyk, Ulaczyk, \& Poleski}]{Soszynski2009}
Soszy{\'n}ski, I., Udalski, A., Szyma{\'n}ski, M.~K., {et~al.} 2009, Acta Astron., 59, 239

\bibitem[{Spaeth {et~al.}(2024)Spaeth, Reffert, Hunt, Kaminski, \& Quirrenbach}]{Spaeth2024}
Spaeth, D., Reffert, S., Hunt, E.~L., Kaminski, A., \& Quirrenbach, A. 2024, A\&A, 689, A91

\bibitem[{Speagle(2020)}]{Speagle2020}
Speagle, J.~S. 2020, MNRAS, 493, 3132

\bibitem[{Staudt(2020)}]{Staudt2020}
Staudt, P. 2020, Bachelor's thesis, University of Heidelberg

\bibitem[{Stock {et~al.}(2023)Stock, Kemmer, Kossakowski, Sabotta, Reffert, \& Quirrenbach}]{Stock2023}
Stock, S., Kemmer, J., Kossakowski, D., {et~al.} 2023, A\&A, 674, A108

\bibitem[{Stock {et~al.}(2018)Stock, Reffert, \& Quirrenbach}]{Stock2018}
Stock, S., Reffert, S., \& Quirrenbach, A. 2018, A\&A, 616, A33

\bibitem[{Stoehr {et~al.}(2008)Stoehr, White, Smith, Kamp, Thompson, Durand, Freudling, Fraquelli, Haase, Hook, Kimball, K{\"u}mmel, Levay, Lombardi, Micol, \& Rogers}]{Stoehr2008}
Stoehr, F., White, R., Smith, M., {et~al.} 2008, in Astronomical Society of the Pacific Conference Series, Vol. 394, Astron. {{Data Anal}}. {{Softw}}. {{Syst}}. {{XVII}}, ed. R.~W. Argyle, P.~S. Bunclark, \& J.~R. Lewis, 505

\bibitem[{Takarada {et~al.}(2018)Takarada, Sato, Omiya, Harakawa, Nagasawa, Izumiura, Kambe, Takeda, Yoshida, Itoh, Ando, Kokubo, \& Ida}]{Takarada2018}
Takarada, T., Sato, B., Omiya, M., {et~al.} 2018, PASJ, 70, 59

\bibitem[{Tala {et~al.}(2016)Tala, Heeren, Grill, Harris, St{\"u}rmer, Schwab, Gutcke, Reffert, Quirrenbach, Seifert, Mandel, Geuer, Sch{\"a}ffner, Thimm, Seeman, Tietz, \& Wagner}]{Tala2016}
Tala, M., Heeren, P., Grill, M., {et~al.} 2016, in SPIE Conf. Ser., Vol. 9908, Ground-{{Based Airborne Instrum}}. {{Astron}}. {{VI}}, ed. C.~J. Evans, L.~Simard, \& H.~Takami, 99086O

\bibitem[{Tala~Pinto {et~al.}(2020)Tala~Pinto, Reffert, Quirrenbach, Stock, Trifonov, \& Mitchell}]{TalaPinto2020}
Tala~Pinto, M., Reffert, S., Quirrenbach, A., {et~al.} 2020, A\&A, 644, A1

\bibitem[{Teng {et~al.}(2023{\natexlab{a}})Teng, Sato, Kunitomo, Takarada, Omiya, Harakawa, Xiao, Liu, Izumiura, Kambe, Yoshida, Itoh, Ando, Kokubo, \& Ida}]{Teng2023}
Teng, H.-Y., Sato, B., Kunitomo, M., {et~al.} 2023{\natexlab{a}}, PASJ, 75, 169

\bibitem[{Teng {et~al.}(2023{\natexlab{b}})Teng, Sato, Kuzuhara, Takarada, Omiya, Harakawa, Izumiura, Kambe, Yilmaz, Bikmaev, Selam, Brandt, Xiao, Yoshida, Itoh, Ando, Kokubo, \& Ida}]{Teng2023a}
Teng, H.-Y., Sato, B., Kuzuhara, M., {et~al.} 2023{\natexlab{b}}, PASJ, 75, 1030

\bibitem[{Trifonov(2019)}]{Trifonov2019a}
Trifonov, T. 2019, The {{Exo-Striker}}: {{Transit}} and Radial Velocity Interactive Fitting Tool for Orbital Analysis and {{N-body}} Simulations, Astrophysics Source Code Library, record ascl:1906.004

\bibitem[{Trifonov {et~al.}(2014)Trifonov, Reffert, Tan, Lee, \& Quirrenbach}]{Trifonov2014}
Trifonov, T., Reffert, S., Tan, X., Lee, M.~H., \& Quirrenbach, A. 2014, A\&A, 568, A64

\bibitem[{Trifonov {et~al.}(2015)Trifonov, Reffert, Zechmeister, Reiners, \& Quirrenbach}]{Trifonov2015}
Trifonov, T., Reffert, S., Zechmeister, M., Reiners, A., \& Quirrenbach, A. 2015, A\&A, 582, A54

\bibitem[{Trotta(2008)}]{Trotta2008}
Trotta, R. 2008, Contemporary Physics, 49, 71

\bibitem[{Udry \& Santos(2007)}]{Udry2007}
Udry, S. \& Santos, N.~C. 2007, ARAA, 45, 397

\bibitem[{VanderPlas(2018)}]{VanderPlas2018}
VanderPlas, J.~T. 2018, ApJS, 236, 16

\bibitem[{Virtanen {et~al.}(2020)Virtanen, Gommers, Oliphant, Haberland, Reddy, Cournapeau, Burovski, Peterson, Weckesser, Bright, {van der Walt}, Brett, Wilson, Millman, Mayorov, Nelson, Jones, Kern, Larson, Carey, Polat, Feng, Moore, VanderPlas, Laxalde, Perktold, Cimrman, Henriksen, Quintero, Harris, Archibald, Ribeiro, Pedregosa, {van Mulbregt}, \& {SciPy 1.0 Contributors}}]{Virtanen2020}
Virtanen, P., Gommers, R., Oliphant, T.~E., {et~al.} 2020, Nat. Methods, 17, 261

\bibitem[{Vogt(1987)}]{Vogt1987}
Vogt, S.~S. 1987, PASP, 99, 1214

\bibitem[{Von~Stauffenberg {et~al.}(2024)Von~Stauffenberg, Trifonov, Quirrenbach, Reffert, Kaminski, Dreizler, Ribas, Reiners, K{\"u}rster, Twicken, Rapetti, Caballero, Amado, B{\'e}jar, Cifuentes, G{\'o}ngora, Hatzes, Henning, Montes, Morales, \& Schweitzer}]{VonStauffenberg2024}
Von~Stauffenberg, A., Trifonov, T., Quirrenbach, A., {et~al.} 2024, A\&A, 688, A112

\bibitem[{Waskom(2021)}]{Waskom2021}
Waskom, M.~L. 2021, J. Open Source Softw., 6, 3021

\bibitem[{Wittenmyer {et~al.}(2011)Wittenmyer, Endl, Wang, Johnson, Tinney, \& O'Toole}]{Wittenmyer2011}
Wittenmyer, R.~A., Endl, M., Wang, L., {et~al.} 2011, ApJ, 743, 184

\bibitem[{Wolthoff {et~al.}(2022)Wolthoff, Reffert, Quirrenbach, Jones, Wittenmyer, \& Jenkins}]{Wolthoff2022}
Wolthoff, V., Reffert, S., Quirrenbach, A., {et~al.} 2022, A\&A, 661, A63

\bibitem[{Wood {et~al.}(1999)Wood, Alcock, Allsman, Alves, Axelrod, Becker, Bennett, Cook, Drake, Freeman, Griest, King, Lehner, Marshall, Minniti, Peterson, Pratt, Quinn, Stubbs, Sutherland, Tomaney, Vandehei, \& Welch}]{Wood1999}
Wood, P.~R., Alcock, C., Allsman, R.~A., {et~al.} 1999, in Asymptot. {{Giant Branch Stars}}, ed. T.~Le~Bertre, A.~Lebre, \& C.~Waelkens, Vol. 191, 151

\bibitem[{Zaire {et~al.}(2024)Zaire, Donati, Alencar, Bouvier, Moutou, Bellotti, Carmona, Petit, K{\'o}sp{\'a}l, Shang, Grankin, Manara, Alecian, Gregory, Fouqu{\'e}, \& {the SLS consortium}}]{Zaire2024}
Zaire, B., Donati, J.~F., Alencar, S.~P., {et~al.} 2024, MNRAS, 533, 2893

\bibitem[{Zechmeister {et~al.}(2019)Zechmeister, Dreizler, Ribas, Reiners, Caballero, Bauer, B{\'e}jar, {Gonz{\'a}lez-Cuesta}, Herrero, Lalitha, {L{\'o}pez-Gonz{\'a}lez}, Luque, Morales, Pall{\'e}, Rodr{\'i}guez, Rodr{\'i}guez~L{\'o}pez, {Tal-Or}, {Anglada-Escud{\'e}}, Quirrenbach, Amado, Abril, Aceituno, Aceituno, {Alonso-Floriano}, {Ammler-von Eiff}, Antona~Jim{\'e}nez, {Anwand-Heerwart}, {Arroyo-Torres}, Azzaro, Baroch, Barrado, Becerril, Ben{\'i}tez, Berdi{\~n}as, Bergond, Bluhm, Brinkm{\"o}ller, Del~Burgo, Calvo~Ortega, Cano, Cardona~Guill{\'e}n, Carro, C{\'a}rdenas~V{\'a}zquez, Casal, {Casasayas-Barris}, Casanova, Chaturvedi, Cifuentes, Claret, Colom{\'e}, {Cort{\'e}s-Contreras}, Czesla, {D{\'i}ez-Alonso}, Dorda, Fern{\'a}ndez, {Fern{\'a}ndez-Mart{\'i}n}, Fuhrmeister, Fukui, {Galad{\'i}-Enr{\'i}quez}, Gallardo~Cava, Garcia De La~Fuente, {Garcia-Piquer}, Garc{\'i}a~Vargas, Gesa, G{\'o}ngora~Rueda, {Gonz{\'a}lez-{\'A}lvarez}, Gonz{\'a}lez~Hern{\'a}ndez, {Gonz{\'a}lez-Peinado}, Gr{\"o}zinger, Gu{\`a}rdia, Guijarro, De~Guindos, Hatzes, Hauschildt, Hedrosa, Helmling, Henning, Hermelo, Hern{\'a}ndez~Arabi, Hern{\'a}ndez~Casta{\~n}o, Hern{\'a}ndez~Otero, Hintz, Huke, Huber, Jeffers, Johnson, De~Juan, Kaminski, Kemmer, Kim, Klahr, Klein, Kl{\"u}ter, Klutsch, Kossakowski, K{\"u}rster, Labarga, Lafarga, Llamas, Lamp{\'o}n, Lara, Launhardt, L{\'a}zaro, Lodieu, L{\'o}pez Del~Fresno, {L{\'o}pez-Puertas}, L{\'o}pez~Salas, {L{\'o}pez-Santiago}, Mag{\'a}n~Madinabeitia, Mall, Mancini, Mandel, Marfil, Mar{\'i}n~Molina, Maroto~Fern{\'a}ndez, Mart{\'i}n, {Mart{\'i}n-Fern{\'a}ndez}, {Mart{\'i}n-Ruiz}, Marvin, Mirabet, {Monta{\~n}{\'e}s-Rodr{\'i}guez}, Montes, {Moreno-Raya}, Nagel, Naranjo, Narita, Nortmann, Nowak, Ofir, Oshagh, Panduro, Parviainen, Pascual, Passegger, Pavlov, Pedraz, {P{\'e}rez-Calpena}, P{\'e}rez~Medialdea, Perger, Perryman, Rabaza, Ram{\'o}n~Ballesta, Rebolo, Redondo, Reffert, Reinhardt, Rhode, Rix, Rodler, Rodr{\'i}guez~Trinidad, Rosich, Sadegi, {S{\'a}nchez-Blanco}, S{\'a}nchez~Carrasco, {S{\'a}nchez-L{\'o}pez}, {Sanz-Forcada}, Sarkis, Sarmiento, Sch{\"a}fer, Schmitt, Sch{\"o}fer, Schweitzer, Seifert, Shulyak, Solano, Sota, Stahl, Stock, Strachan, Stuber, St{\"u}rmer, Su{\'a}rez, Tabernero, Tala~Pinto, Trifonov, Veredas, Vico~Linares, Vilardell, Wagner, Wolthoff, Xu, Yan, \& Zapatero~Osorio}]{Zechmeister2019}
Zechmeister, M., Dreizler, S., Ribas, I., {et~al.} 2019, A\&A, 627, A49

\bibitem[{Zechmeister \& K{\"u}rster(2009)}]{Zechmeister2009}
Zechmeister, M. \& K{\"u}rster, M. 2009, A\&A, 496, 577

\bibitem[{Zechmeister {et~al.}(2008)Zechmeister, Reffert, Hatzes, Endl, \& Quirrenbach}]{Zechmeister2008}
Zechmeister, M., Reffert, S., Hatzes, A.~P., Endl, M., \& Quirrenbach, A. 2008, A\&A, 491, 531

\bibitem[{Zechmeister {et~al.}(2018)Zechmeister, Reiners, Amado, Azzaro, Bauer, B{\'e}jar, Caballero, Guenther, Hagen, Jeffers, Kaminski, K{\"u}rster, Launhardt, Montes, Morales, Quirrenbach, Reffert, Ribas, Seifert, {Tal-Or}, \& Wolthoff}]{Zechmeister2018}
Zechmeister, M., Reiners, A., Amado, P.~J., {et~al.} 2018, A\&A, 609, A12

\bibitem[{Zhou {et~al.}(2023)Zhou, Latham, Quinn, Bieryla, Vanderburg, Berlind, Calkins, \& Esquerdo}]{Zhou2023}
Zhou, Q., Latham, D.~W., Quinn, S.~N., {et~al.} 2023, AJ, 166, 160

\end{thebibliography}

\begin{appendix}
\onecolumn
\section{Results of the activity periodogram analysis}

\begin{figure*}[h!]
    \centering
    \includegraphics[scale=1.0]{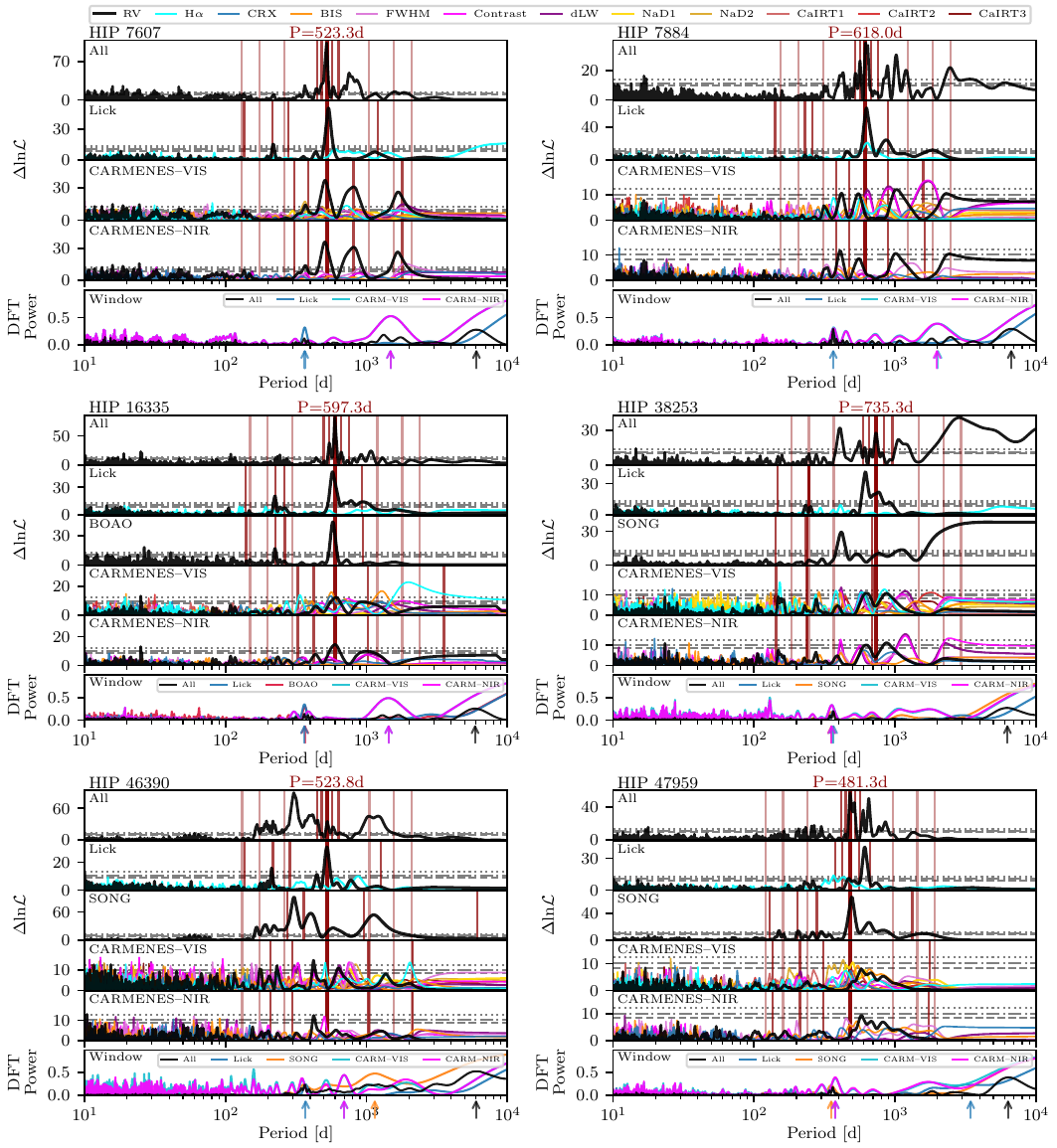}
    \caption{Maximum likelihood periodograms for six sample stars. The top panel for each star portrays the combined RV MLP. In the remaining panels, we show the MLPs of the RVs (black) and activity indicators (see legend) for each instrument. The FAPs of $5\%$ (dashed), $1\%$ (dash-dotted), and $0.1\%$ (dotted) are plotted as gray horizontal lines. The bottom panel shows the instruments' window functions derived as discrete Fourier transforms. The thick, red, vertical line indicates the period of the Keplerian fit (see \autoref{tab:nest_samp_params}). The alias periods, corresponding to the most significant peak in the respective window function, are shown as thinner lines. The positions of these window peaks are marked with colored arrows in the bottom panels. Notably, the CARMENES VIS and NIR window functions largely overlap. Additionally, we display the harmonics and sub-harmonics of the period of the Keplerian fit as red lines with lower opacity.}
    \label{fig:mlp_part1}
\end{figure*}

\begin{figure*}[h!]
    \centering
\includegraphics{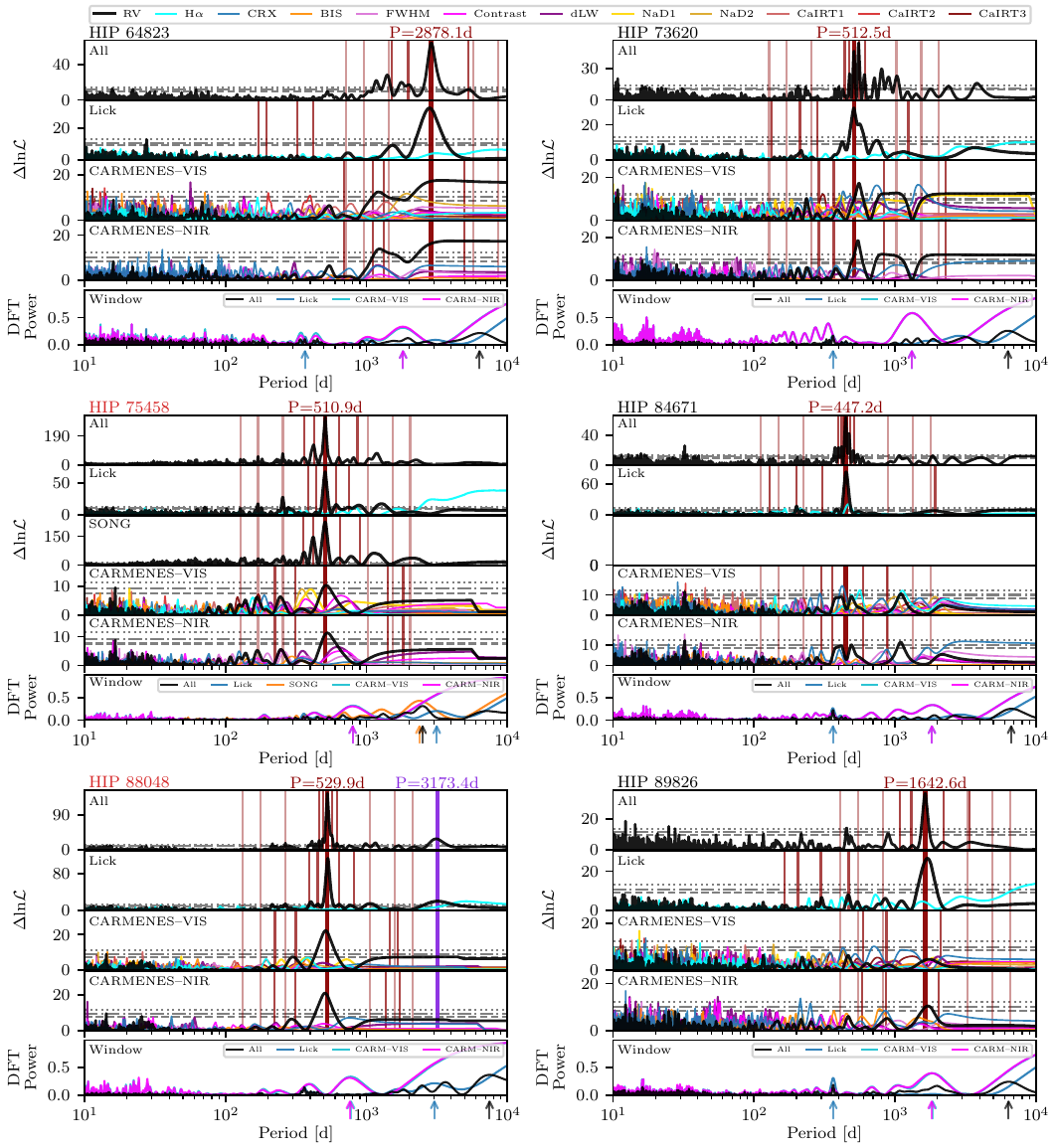}
    \caption{Maximum likelihood periodograms for the remaining six stars. The same description as in \autoref{fig:mlp_part1} applies. The FAPs of $5\%$ (dashed), $1\%$ (dash-dotted), and $0.1\%$ (dotted) are plotted as gray horizontal lines. The identifiers of the two planet hosts, HIP~75458 and HIP~88048, are marked in red. For HIP~80488, we mark the second significant periodicity, corresponding to the outer brown-dwarf companion, as the violet, vertical line. We omit plotting its aliases and harmonics for visual clarity.}
    \label{fig:mlp_part2}
\end{figure*}

\FloatBarrier

\clearpage
\section{Results of the DBSCAN clustering algorithm}

\begin{figure*}[h!]
    \centering
    \includegraphics{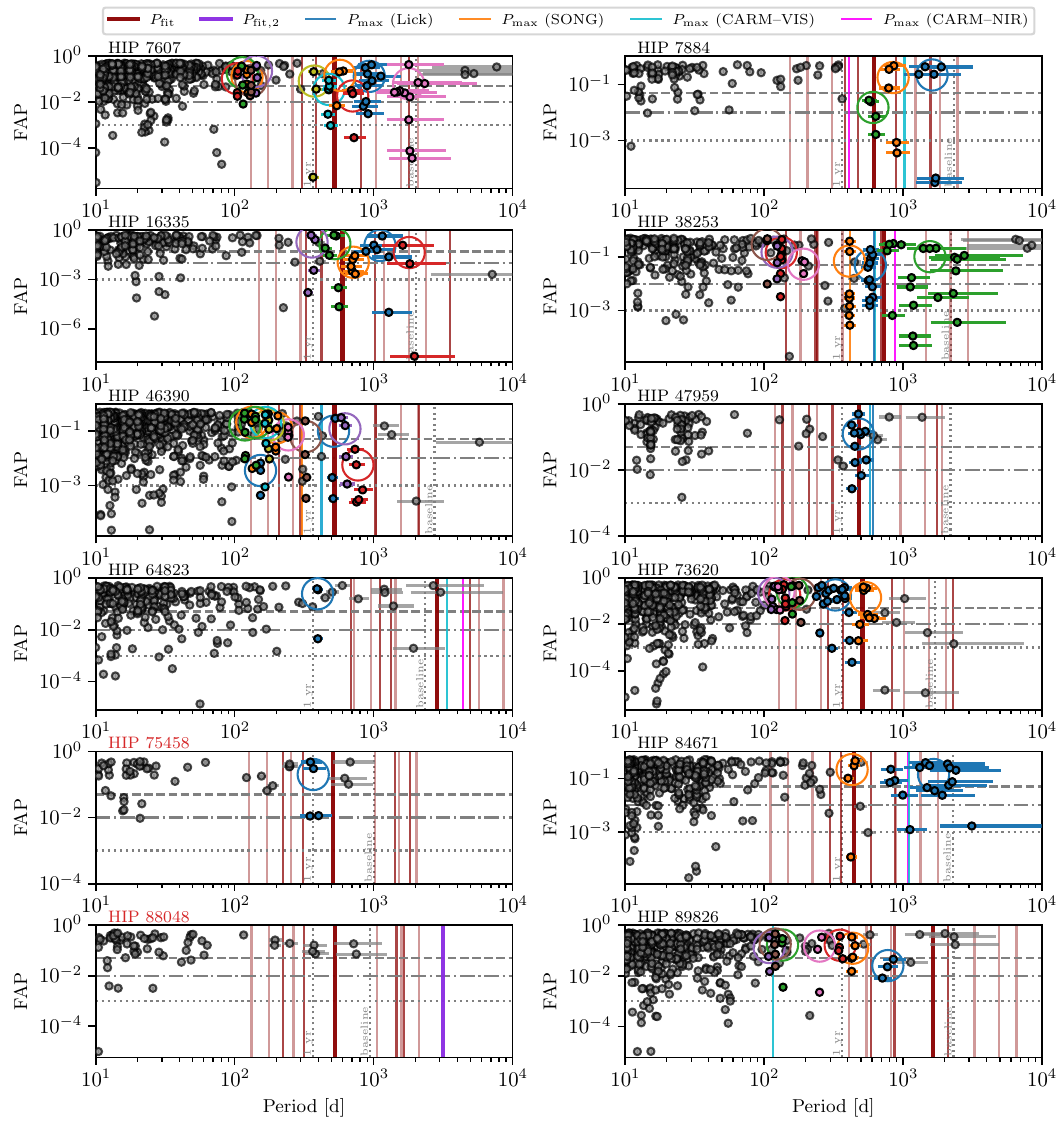}
    \caption{Results of the DBSCAN clustering algorithm. In each panel, all identified peaks in the MLPs of the CARMENES activity indicators are shown as filled circles. Arbitrarily colored circles represent activity peaks assigned to clusters by DBSCAN, while gray circles indicate peaks that are not part of any cluster. The larger, empty circles indicate the center periods and FAPs of the clusters. We further overplot the FAP levels of $5\%$ (dashed), $1\%$ (dash-dotted), and $0.1\%$ (dotted) as horizontal gray lines, and the fitted RV period (thick), its aliases (thin, based on the VIS window function), and harmonics (lower opacity) as vertical red lines.
    Furthermore, for stars where individual instruments exhibit a strongest period that deviates from the fitted period by at least $10\%$ (restricted to $100\,\mathrm{d} < P_\mathrm{max} < 5000\,\mathrm{d}$), we overplot the strongest RV period detected in the MLP of the respective instrument using vertical, colored lines specified in the figure legend. The identifiers of the two planet hosts, HIP~75458 and HIP~88048, are marked in red. For the latter, we also mark the periodicity of the outer orbital companion in violet.}
    \label{fig:cluster}
\end{figure*}

\FloatBarrier
\clearpage
\section{Priors used for the Keplerian modeling}
\FloatBarrier
\begin{table*}[h]
\caption{Priors used for the Keplerian models in \autoref{sec:results:keplerian_modelling}.}
\label{tab:Kepler_priors}
\resizebox{\textwidth}{!}{
\begin{tabular}{lllllllllllll}
\hline
HIP & \phantom{(}$P$ & \phantom{(}$K$ & \phantom{(}$e$ & \phantom{(}$\omega$ & \phantom{(}$M_0$ &  \phantom{(}$\sigma_\mathrm{jit}$ & \phantom{(}$\mathrm{RV_{off}}$\\ 
& (d) &  ($\mathrm{m\,s^{-1}}$) & &  ($\deg$) &  ($\deg$) &  ($\mathrm{m\,s^{-1}}$) &  $(\mathrm{m\,s^{-1}}$) \\
\hline
Default & $\mathcal{U}(100, 2000)$ & $\mathcal{U}(1, 1000)$ & $\mathcal{U}(0, 0.9)$ & $\mathcal{U}(0, 540)$ & $\mathcal{U}(0, 540)$ & $\mathcal{U}(0, 300)$ & $\mathcal{U}(-5000, 5000)$\\
46390 & $\mathcal{U}(100, 1000)$ \\
47959 & & & & & & & $\mathcal{U}(-50000, 50000)$\\
64823 & $\mathcal{U}(100, 5000)$ \\
75458 b & $\mathcal{U}(100, 5000)$ & & & & & & $\mathcal{U}(-30000, 30000)$ \\
75458 c & $\mathcal{U}(100, 50000)$ & & & & & & $\mathcal{U}(-30000, 30000)$ \\
88048 b & $\mathcal{U}(100, 5000)$ \\
88048 c & $\mathcal{U}(100, 5000)$ \\
89826 & $\mathcal{U}(100, 5000)$ \\
\hline 
\end{tabular}}
\tablefoot{We give the default priors, as well as deviations from it for particular stars. $\mathcal{U}$ denotes the uniform distribution. }
\end{table*}

\begin{table*}[h]
\caption{Priors used for the modeling of HIP~64823 in \autoref{sec:hip64823}.}
\label{tab:HIP64823_priors}
\begin{tabular}{llll}
     \hline
     Parameter & Unit & 2P & 1P+GP \\
     \hline
     $P_\mathrm{b}$ & ($\mathrm{d}$) & $\mathcal{U}(300, 500)$ & $\mathcal{U}(100, 5000)$\\
     $K_\mathrm{b}$ & ($\mathrm{m\,s^{-1}}$) & $\mathcal{U}(1, 1000)$ &  $\mathcal{U}(1, 1000)$\\
     $e_\mathrm{b}$ & & $\mathcal{U}(0, 0.7)$ & $\mathcal{U}(0, 0.9)$\\
     $\omega_\mathrm{b}$ & ($\deg$) & $\mathcal{U}(0, 540)$ & $\mathcal{U}(0, 540)$\\
     $M_{0,\mathrm{b}}$ & ($\deg$) & $\mathcal{U}(0, 540)$ & $\mathcal{U}(0, 540)$\\
     $P_\mathrm{c}$ & ($\mathrm{d}$) & $\mathcal{U}(100, 5000)$\\
     $K_\mathrm{c}$ & ($\mathrm{m\,s^{-1}}$) & $\mathcal{U}(1, 1000)$\\
     $e_\mathrm{c}$ & & $\mathcal{U}(0, 0.9)$ \\
     $\omega_\mathrm{c}$ & ($\deg$) & $\mathcal{U}(0, 540)$ \\
     $M_{0,\mathrm{c}}$ & ($\deg$) & $\mathcal{U}(0, 540)$ \\
     $\sigma_\mathrm{jit}$ & ($\mathrm{m\,s^{-1}}$) & $\mathcal{U}(0, 300)$ & $\mathcal{U}(0, 300)$ \\
     $\mathrm{RV_{off}}$ & ($\mathrm{m\,s^{-1}}$) & $\mathcal{U}(-5000, 5000)$ & $\mathcal{U}(-5000, 5000)$ \\
     $B_\mathrm{GP}$ & ($(\mathrm{m\,s^{-1}})^{2}$) & &  $\mathcal{U}(100, 10000)$\\
     $L_\mathrm{GP}$ & ($\mathrm{d}$) & & $\mathcal{U}(450, 10000)$\\
     $P_\mathrm{GP}$ & ($\mathrm{d}$) & & $\mathcal{U}(10, 800)$ \\
     $C_\mathrm{GP}$ & & & $\mathcal{U}(0, 1)$ \\
     \hline
\end{tabular}
\tablefoot{$\mathcal{U}$ denotes the uniform distribution. }
\end{table*}

\FloatBarrier
\clearpage

\section{Corner plot of the joint 1P+GP model for HIP 64823}
\begin{figure*}[h]
    \centering
    \includegraphics[scale=0.25]{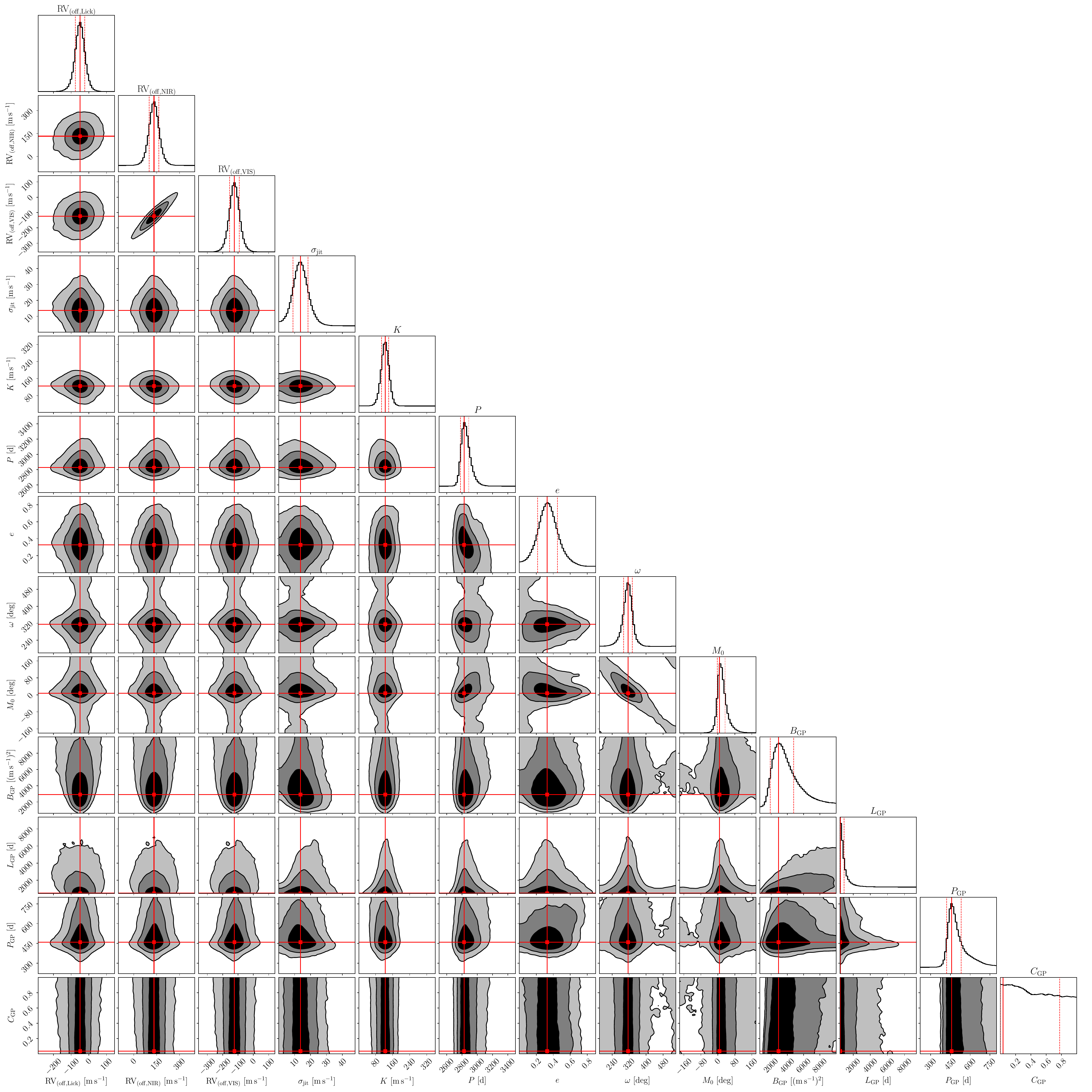}
    \caption{Corner plot of the posterior samples of the joint 1P+GP model for HIP 64283. The mode of each parameter, marked as the continuous, red line, was adopted as the best fitting parameter in \autoref{tab:64283_params}. The red dashed lines mark the interval that contains 68\% of the posterior samples. 
     We note that the posteriors of the planet model are well constrained, while the posteriors of the GP model (especially the factor $C$) are relatively broad and thus allow for a wide range of models.}
    \label{fig:64823_corner}
\end{figure*}
\end{appendix}

\end{document}